\documentclass[a4paper,11pt]{article}
\usepackage{jinstpub} % for details on the use of the package, please see the JINST-author-manual
\usepackage{lineno}
\usepackage{hyperref}
\usepackage[table]{xcolor}
\usepackage{booktabs}
\usepackage{multirow}
\usepackage{siunitx}
\usepackage[capitalise]{cleveref}
\usepackage{subcaption} 
\usepackage{float}
\usepackage{adjustbox}
\newcommand{\prometheus}{\texttt{P\footnotesize{ROMETHEUS}}}
\newcommand{\fennel}{\texttt{fennel}}
\newcommand{\hyperion}{\texttt{hyperion}}
\newcommand{\geant}{\texttt{GEANT4}}
\newcommand{\ppc}{\texttt{PPC}}

\newcommand{\cluster}{\texttt{Cluster}}
\newcommand{\Triangle}{\texttt{Triangle}}
\newcommand{\flowers}{\texttt{Flower S}}
\newcommand{\flowerl}{\texttt{Flower L}}
\newcommand{\flowerxl}{\texttt{Flower XL}}
\newcommand{\hexagon}{\texttt{Hexagon}}
\newcommand{\hexagonice}{\texttt{Hexagon Ice LE}}

\newcommand{\particlenet}{\textsc{ParticleNeT}}
\newcommand{\dynedge}{\textsc{DynEdge}}
\newcommand{\deepice}{\textsc{DeepIce}}
\newcommand{\grit}{\textsc{GRIT}}

\title{NuBench: An Open Benchmark for Deep Learning–Based Event Reconstruction in Neutrino Telescopes}

\author[a]{Rasmus {\O}rsøe}
\author[b,c,d]{Stephan Meighen-Berger}
\author[e]{Jeffrey Lazar}
\author[f]{Jorge Prado}
\author[g]{Iván Mozún-Mateo}
\author[h]{Aske Rosted}
\author[i]{Philip Weigel}
\author{Arturo Llorente Anaya}

\affiliation[a]{Physik-department, Technische Universit\"{a}t M\"{u}nchen, D-85748 Garching, Germany}
\affiliation[b]{School of Physics, The University of Melbourne, Victoria 3010, Australia}
\affiliation[c]{Center for Cosmology and AstroParticle Physics (CCAPP), Ohio State University, Columbus, OH 43210, USA}
\affiliation[d]{University of Iowa, 30 N Dubuque St, Iowa City, IA, 52242, United States of America}
\affiliation[e]{Université Catholique de Louvain, Pl. de l'Universit\'{e} 1, 1348 Ottignies-Louvain-la-Neuve}
\affiliation[f]{IFIC - Instituto de F\'{i}sica Corpuscular (CSIC - Universitat de Val\'{e}ncia), c/Catedr\'{a}tico Jos\'{e} Beltr\'{a}n, 2, 46980 Paterna, Valencia, Spain}
\affiliation[g]{LPC CAEN, Normandie Univ, ENSICAEN, UNICAEN, CNRS/IN2P3, 6 boulevard Mar\'{e}chal Juin, Caen, 14050 France}
\affiliation[h]{Dept. of Physics and The International Center for Hadron Astrophysics, Chiba University, Chiba 263-8522, Japan}
\affiliation[i]{Dept. of Physics, Massachusetts Institute of Technology, Cambridge, MA 02139, USA}
\emailAdd{rasmus.orsoe@tum.de}

\abstract{Neutrino telescopes are large-scale detectors designed to observe Cherenkov radiation produced from neutrino interactions in water or ice. They exist to identify extraterrestrial neutrino sources and to probe fundamental questions pertaining to the elusive neutrino itself.  A central challenge common across neutrino telescopes is to solve a series of inverse problems known as event reconstruction, which seeks to resolve properties of the incident neutrino, based on the detected Cherenkov light. In recent times, significant efforts have been made in adapting advances from deep learning research to event reconstruction, as such techniques provide several benefits over traditional methods. While a large degree of similarity in reconstruction needs and low-level data exists, cross-experimental collaboration has been hindered by a lack of diverse open-source datasets for comparing methods.

We present NuBench, an open benchmark for deep learning–based event reconstruction in neutrino telescopes. NuBench comprises seven large-scale simulated datasets containing nearly 130 million charged- and neutral-current muon-neutrino interactions spanning 10 GeV to 100 TeV, generated across six detector geometries inspired by existing and proposed experiments. These datasets provide pulse- and event-level information suitable for developing and comparing machine-learning reconstruction methods in both water and ice environments. Using NuBench, we evaluate four reconstruction algorithms—ParticleNeT and DynEdge, both actively used within the KM3NeT and IceCube collaborations, respectively, along with GRIT and DeepIce—on up to five core tasks: energy and direction reconstruction, topology classification, interaction vertex prediction, and inelasticity estimation. Datasets, predictions and model artifacts are available \href{https://github.com/graphnet-team/NuBench}{here}}


\arxivnumber{1234.56789} % Only if you have one

\begin{document}

\maketitle
\flushbottom

\section{Introduction}
In recent decades, a class of experiments known as neutrino telescopes has emerged as cutting-edge scientific instruments capable of detecting neutrinos of extraterrestrial origin and probing fundamental questions regarding the nature of the neutrino. In order to shield the experiments from the dominant atmospheric muon background, neutrino telescopes are constructed in deep subsurface locations that offer a transparent detection medium. To compensate for the extremely low neutrino interaction cross section and the rarity of extraterrestrial neutrino events, these detectors may span volumes on the order of a cubic kilometer, making them the largest human-made structures by volume.

Today, three large-scale neutrino telescopes exist at various stages of completion, detection media, and geometric design. The IceCube Neutrino Observatory~\cite{icecube86}, completed in 2011, was installed deep within the Antarctic ice and has been operational for more than a decade. The Baikal-GVD telescope~\cite{baikal_gvd}  is currently being constructed in Lake Baikal, the world’s deepest freshwater lake. Finally, KM3NeT, which is designed to host two detectors (ORCA ~\cite{km3net_orca_osc} and ARCA ~\cite{arca_astronomy}), is under construction at two sites on the floor of the Mediterranean Sea. Collectively, the existing neutrino telescopes have made significant contributions to a wide range of fields, including astronomy~\cite{IceCube:2013low, IceCube:2013cdw, IceCube:2014stg, Baikal-GVD:2022fis, Naab:2023xcz, IceCube:2024fxo, IceCube:2022der, IceCube:2023ame, KM3NeT:2025npi}, neutrino physics~\cite{KM3NeT:2024ecf, IceCubeCollaboration:2024dxk, IceCubeCollaboration:2024nle, IceCubeCollaboration:2022tso}, and particle physics~\cite{IceCube:2021rpz, IceCube:2021xzo, IceCube:2021tdn, ICECUBE:2023gdv, KM3NeT:2025mfl}. Among these, IceCube has played a particularly pivotal role, enabled by its sustained operation of its cubic-kilometer detector volume for over a decade. 

Following the demonstrated potential of current-generation neutrino telescopes, a global effort to build next-generation detectors with enhanced instrumentation and optimized geometries is unfolding. Alongside the ongoing construction of ORCA and ARCA, and the imminent extension of IceCube \cite{icecube_upgrade}, several new telescopes -- such as P-ONE~\cite{pone}, TRIDENT~\cite{TRIDENT:2022hql}, NEON~\cite{neon_detector}, and HUNT~\cite{hunt_detector} -- have been proposed, suggesting a near future in which a diverse array of large-scale detectors may coexist in complement. 

While current and next-generation telescopes may differ significantly in detection medium, instrumentation, and geometry, they share the same detection principle and overall methodology.
Each neutrino telescope is composed of one or more arrays of vertical lines equipped with optical modules (OMs), containing photomultiplier tubes (PMTs) and/or silicon photomultipliers (SiPMs) that detect the faint Cherenkov radiation produced by neutrino interactions.
As a result, the low-level observations in these telescopes---Cherenkov light detected by individual OMs at different times---share a common data structure, consisting of irregular, spatially distributed time series, known as an \textit{event}, and are illustrated in \cref{fig:event_display}. 

Central to the operation of current and next-generation telescopes is the solution of a collection of inverse problems known as \textit{event reconstruction}, which aims to infer the direction, energy, and other properties of the incident neutrino. Reconstruction algorithms define the methods used to infer neutrino event properties and remain an area of continuous improvement.
These were traditionally addressed via maximum likelihood estimation (MLE).
In recent years, deep learning–based approaches have gained prominence, providing key contributions to central results such as the detection of neutrino emission from the Galactic Plane~\cite{icecube_galactic_plane} and strong evidence of neutrino emission from NGC-1068~\cite{icecube_ngc1068}.

The similarity in low-level observations, shared reconstruction needs, and the increasing adoption of deep-learning-based methods provide a strong foundation for cross-experimental collaboration on common reconstruction challenges. However, such collaboration requires the availability of high-quality, diverse, and openly accessible datasets to enable meaningful benchmarking and method development —resources that are currently scarce.

\subsection*{Related work \& Our Contributions}
A single large dataset was released as part of an open-data challenge by IceCube in 2023~\cite{kaggle_competition, kaggle_3solutions}, containing around 140 million simulated neutrino interactions spanning energies from 100~\si{GeV} to 100~\si{PeV}, across all flavours and arrival directions. The challenge clearly demonstrated that individuals outside the field can develop competitive algorithms using open-source datasets, and some of the winning solutions are now being explored for scientific use within IceCube~\cite{icrc_gen2_kaggle_usage}. However, the dataset is specific to the IceCube detector, includes significant contamination from coincident atmospheric muons, and is limited in scope to direction reconstruction, which is one of several common reconstruction tasks in neutrino telescopes. 
\newline
\newline
This work introduces a collection of datasets comprising nearly 130 million simulated charged-current (CC) and neutral-current (NC) muon neutrino interactions ($\nu_\mu^\text{CC}$ and $\nu_\mu^\text{NC}$) with energies ranging from 10~\si{GeV} to 100~\si{TeV}, simulated across six distinct detector geometries that resemble existing or proposed neutrino telescopes. Using the open-source simulation tool \prometheus{}~\cite{prometheus}, each geometry is simulated in water, with one additionally simulated in ice, yielding a total of seven datasets. These datasets include rich event-level ground-truth information, making them well-suited for benchmarking and comparing reconstruction methods across a wide range of problems of shared interest in the field. Five reconstruction and classification tasks of common interest are described in detail, including their relevance to physics analyses and the conventional metrics used to quantify reconstruction performance, with the aim of providing a practical benchmarking resource for the field. Finally, we use these datasets to provide a comprehensive comparison of four modern reconstruction architectures: two graph neural networks currently used within KM3NeT and IceCube—\particlenet{}~\cite{Qu_2020,stefan_phd} and \dynedge{}~\cite{icecube_dynedge}, respectively; a transformer-based model, \deepice{}~\cite{kaggle_3solutions}, one of the winning solutions of the open-data challenge; and a new hybrid algorithm, \grit{}~\cite{ma2023graph}, which combines graph representations with attention-based mechanisms. The models have been implemented in \textsc{GraphNeT}~\cite{graphnet_1, graphnet_2}--an open-source deep learning library for neutrino telescopes--ensuring they are readily available for future use and comparative studies. 
\newline
\newline
This work is structured as follows. In \cref{sec:neutrino_events}, we further define the notion of neutrino events and describe the five commonly sought attributes in neutrino event reconstruction, followed by details regarding MLE-based reconstruction methods in \cref{sec:likelihood_reconstructions}, and deep learning–based reconstruction methods in \cref{sec:ml_reconstruction}. In \cref{sec:datasets} we introduce the datasets, and in \cref{sec:compare}, we present comparisons of the four algorithms across these five tasks and all datasets. Technical details regarding dataset generation and model implementation are provided in \cref{sec:app_datasets} and \cref{sec:models}, respectively.

\section{Neutrino Events \& Reconstruction}
\label{sec:reconstruction}
\label{sec:neutrino_events}
Neutrino telescopes do not detect neutrinos directly, but instead register Cherenkov radiation emitted by relativistic charged particles produced when neutrinos interact within the transparent detection medium. Only a fraction of the neutrino’s initial energy is ultimately released as Cherenkov radiation, and, owing to the sparse density of optical instrumentation, only a fraction of this light is registered by the OMs of the detector array. When a Cherenkov photon strikes the photocathode of a PMT, photoelectrons (p.e.) are released and accelerated toward the anode, where they are multiplied into a cascade through successive dynode stages. Once sufficient charge has accumulated at the anode, the analogue signal is processed by the onboard electronics of the OMs. Both the charge threshold and the subsequent signal processing vary between telescopes. In IceCube, for example, the analogue signal is digitized into waveforms, which are then processed with a wave-unfolding algorithm to estimate the arrival times and charges of individual photons~\cite{icecube_daq}. These processed waveforms are referred to as pulses of Cherenkov radiation in the following. Because PMTs rely on photoelectrons traveling from the photocathode to the anode, they have a finite timing resolution. Coincident photons may not be individually resolved, but the total induced charge can serve as an indicator of their presence. In other telescopes, such as KM3NeT, the arrival time of Cherenkov photons is instead defined as the instant the PMT signal crosses the charge threshold, and the Time-over-Threshold (ToT) is recorded in place of a charge estimate~\cite{k3mnet_daq}. The datasets presented in this work simulate OM responses in the form used by IceCube.
\begin{figure}[h!]
    \centering
   \includegraphics[width=\textwidth]{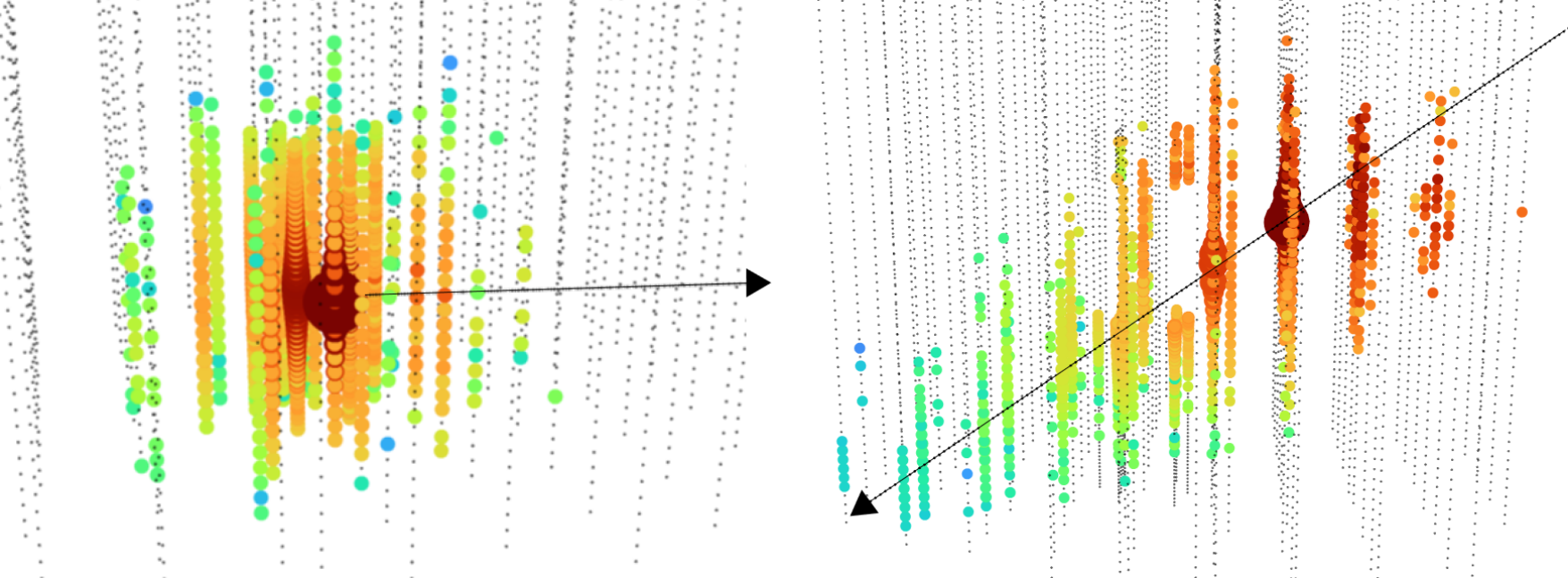}
    \caption{Illustrations of the two neutrino event morphologies: Cascade (left) and Track (right). Each dot represents an optical module, and the color indicates the arrival time of the pulses, ranging from red (early) to cyan (late). The size of the dots is adjusted to be proportional to the observed charge. Grey dots represent modules that did not observe pulses during the interaction.}
    \label{fig:event_display}
\end{figure}

Superimposed on this faint neutrino signal are backgrounds from coincident atmospheric muons and from intrinsic detector noise, including radioactive decays within the PMT housings, the OM glass, and the detection medium itself. In water-based experiments, bioluminescence constitutes an additional significant source of background~\cite{km3net_bio_patterns, gvd_bio, pone_bio}. During data taking, the continuous stream from the detector array is monitored, and, as in other particle physics experiments~\cite{trigger_systems_review_lhc}, event triggers are employed to suppress detector readouts dominated by stochastic noise. Because stochastic noise leads to uncorrelated observations at the OMs, trigger conditions typically employ a notion of local coincidence to identify causally connected signals within a trigger window on the order of microseconds ~\cite{icecube_daq}. The collection of pulses that satisfy these trigger conditions constitutes a single \textit{event}, which may be induced by either stochastic noise (accidental triggering), atmospheric muons or neutrino interactions. In this work, all events are caused by $\nu_\mu^\text{NC}$ or $\nu_\mu^\text{CC}$ interactions without contamination from coincident atmospheric muons, and employ a simplified trigger condition elaborated upon in \cref{sec:detector_response_sim}.

\subsection*{Neutrino Event Morphologies \& Containment}
In neutrino telescopes, most events can be classified into two primary morphologies -- \textit{tracks} and \textit{cascades} -- which are closely correlated with the flavor of the interacting neutrino. Tracks are produced predominantly by muons, either from cosmic-ray air showers (background) or from muon-neutrino charged-current interactions. Cascades appear as approximately spherical light patterns and arise from electron-neutrino charged-current interactions as well as from neutral-current interactions of all flavors. These interactions produce hadrons and, in the case of charged-current electron neutrinos, an electron, which deposits its energy over a length scale of roughly ten meters~\cite{Radel:2012ij}, typically much smaller than the interstring spacing of neutrino telescopes. While tau-neutrino charged-current interactions can in principle produce two spatially separated cascades—one from the initial interaction and another from the tau decay—the short tau lifetime makes such “double-bang” morphologies difficult to resolve below \SI{100}{\TeV}. Illustrations of the two event morphologies are shown in \cref{fig:event_display}. Other common categorizations subdivide track and cascade events according to their degree of containment within the instrumented detector volume.

\begin{figure}[h!]
    \centering
   \includegraphics[width=\textwidth]{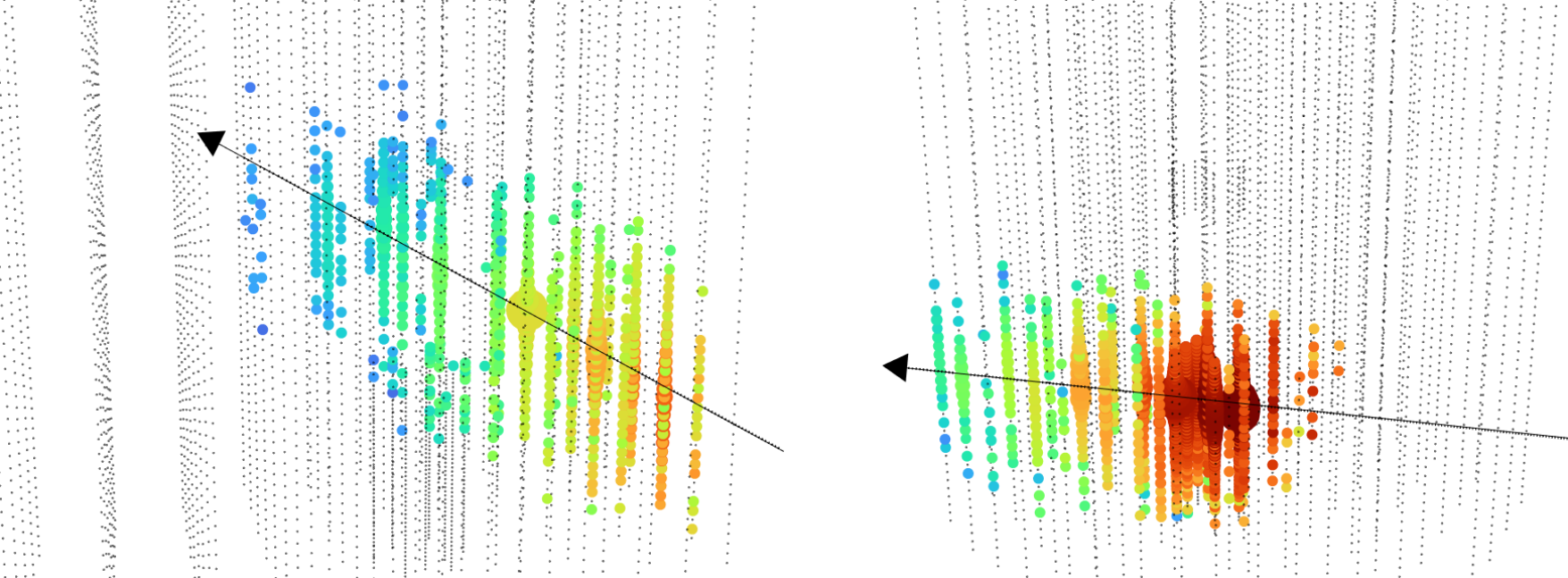}
    \caption{Illustrations of starting- and stopping tracks from our datasets. Left) A \SI{80}{TeV} stopping track. Right) A \SI{44}{TeV} starting track.}
    \label{fig:starting_stopping_display}
\end{figure}

Events are termed \textit{contained} if the neutrino interaction both begins and terminates inside the volume, and \textit{uncontained} otherwise. Uncontained events may be further classified based on which part of the interaction lies outside the detector. For example, a \textit{stopping} event occurs when a neutrino interaction begins outside the volume but propagates into it, terminating inside and leaving its initial signature undetected. Conversely, a \textit{starting} event begins inside the volume but extends beyond the detector boundaries. These categorizations are often used to identify relevant neutrino samples and inform reconstruction methods. In~\cref{fig:starting_stopping_display}, a \SI{80}{TeV} stopping track can be seen in the illustration on the left. On the right in ~\cref{fig:starting_stopping_display}, a \SI{44}{TeV} starting track is shown. 

\subsection{Neutrino Event Reconstruction}
\label{sec:tasks}
Reconstruction of neutrino events involves solving a series of inverse problems that are central to the operation of neutrino telescopes. These problems bear conceptual resemblance not only to reconstruction in other PMT-based Cherenkov detectors~\cite{sno_reco,hyperk_reco}, but also to \textit{jet tagging} at the LHC~\cite{jet_tagging_review_lhc} and event reconstruction in imaging atmospheric Cherenkov telescopes such as CTA~\cite{cta_reco}. In neutrino telescopes, \textit{neutrino event reconstruction} seeks to recover key physical properties of the incident neutrino from the pattern of observed pulses produced by its interaction. While reconstruction needs may vary between samples and telescopes, five key neutrino attributes are generally sought across experiments: energy, direction, inelasticity, interaction vertex, and event morphology (track or cascade). The different reconstructed attributes each play distinct roles in physics analyses, and the difficulty of reconstructing them depends on both the neutrino energy and the event’s containment and morphology. The role of each attribute, along with the challenges of its reconstruction, is briefly outlined below. 

\textbf{Energy} --- Estimation of the energy of incident neutrinos is an important reconstruction task that several key areas of neutrino analyses rely on. Examples include the detection of high-energy cosmic neutrino sources~\cite{icecube_TXS_0506_056_prior, icecube_ngc1068, icecube_galactic_plane} and characterization of the cosmic neutrino flux~\cite{icecube_astrophysical_flux}. At lower energies, the neutrino energy is used to measure neutrino oscillations~\cite{icecube_osc_golden}. To leading order, the energy of the incident neutrino is proportional to the amount of detected Cherenkov radiation, but several factors make energy estimation a non-trivial inference task. For example, $\nu^{NC}_{e, \mu, \tau}$ and $\nu^{CC}_{e,\mu}$ interactions each represent general modalities in energy reconstruction, as their physically different interactions lead to distinctively different relationships between detectable Cherenkov light and neutrino energy. In the case of $\nu^{NC}_{e, \mu, \tau}$ interactions, the outgoing neutrino escapes with part of the incident energy. In contrast, $\nu^{CC}_{e,\mu}$ interactions may deposit the entire neutrino energy within the detector volume, a fraction of which as Cherenkov light. The degree to which an event is contained within the detector volume further complicates the relationship between observed pulses and neutrino energy. For example, stopping $\nu_\mu^{\text{CC}}$ events may produce light patterns similar to, for example, a 1~TeV event but be induced by a neutrino of much higher initial energy.  While the primary goal is often to estimate the incident neutrino energy, a common technique for mitigating the multimodality of the task is to regress proxy labels, such as the  deposited energy within the detector volume, which are more strongly correlated with the observed pulses. The proxy label can subsequently be statistically related to the incident neutrino energy, as is done in analyses such as~\cite{icecube_ngc1068, icecube_galactic_plane}.

\textbf{Direction} --- The direction of the incident neutrino is central to both astrophysical and oscillation studies~\cite{icecube_ngc1068,icecube_galactic_plane,icecube_osc_golden}. It is the directional capability that defines the telescopic function of large neutrino detectors, with the angular resolution setting the discovery potential for identifying astrophysical point sources. For oscillation studies, the reconstructed zenith angle serves as a proxy for the distance travelled by atmospheric neutrinos through the Earth, which directly enters atmospheric neutrino oscillation measurements~\cite{barger2012physics, icecube_osc_golden}. Because the rate of observed atmospheric muons depends strongly on inclination, the zenith angle may also be used to produce neutrino samples with low levels of atmospheric muon contamination. The difficulty of direction reconstruction depends strongly on event morphology, as the elongation seen in sufficiently energetic track events generally correlates well with the direction of the incident neutrino, a feature cascade events do not have.

\textbf{Inelasticity} --- The inelasticity \(y = \frac{E_X}{E_\nu}\), where \(E_X\) and \(E_\nu\) denote the energies of the hadronic system and incident neutrino respectively, is the fraction of the neutrino’s energy transferred to hadronic products and is of high interest for studies of astrophysical and atmospheric neutrinos~\cite{icecube_inelasticity_2018, inelasticity_in_nmo}. While neutrino telescopes may not distinguish neutrinos from anti-neutrinos on an event-by-event basis, differences in the inelasticity distributions between \(\nu\) and \(\bar{\nu}\) events provide a statistical handle for separating the two~\cite{icecube_inelasticity_2025}. Reconstructing inelasticity is generally very challenging and is practically only feasible for CC interactions, where one can equivalently write \(y = \frac{E_X}{E_\ell + E_X}\) with \(E_\ell\) the energy of the outgoing lepton. Because disentangling the hadronic and leptonic components requires clearly separated energy deposits within the detector, measurements of inelasticity are often limited to sufficiently energetic starting track events. In practice, this means that inelasticity measurements are almost exclusively performed with sufficiently energetic starting tracks. Similarly to energy reconstruction, inelasticity reconstruction can be simplified by introducing proxy observables for the energy components, as in~\cite{icecube_inelasticity_2018, icecube_inelasticity_2025}. We adopt such an approach in this work, where the inelasticity is defined using the \textit{visible} energy, i.e. the fraction of the incident neutrino energy that is deposited as Cherenkov photon emission within the detector. 

\textbf{Vertex} --- 
The interaction vertex of the incident neutrino is the point at which the neutrino interacts with matter in or around the detector. While the vertex is not often used directly as an analysis observable, it is widely applied as a selection variable for defining neutrino samples~\cite{flercnn_osc, km3net_orca_osc, icecube_hese, icecube_mese}. For example, the reconstructed vertex may be used to classify events as starting or entering, or as contained or uncontained, which is essential for rejecting atmospheric muon backgrounds and for isolating clean neutrino samples. The precision of vertex reconstruction depends strongly on the distance between the interaction vertex and the nearest optical module, since greater separation reduces the number and timing accuracy of detected photons.

\textbf{Track/Cascade Classification} --- 
Categorization of events into the two primary event morphologies — track and cascade — is used extensively as an analysis observable in atmospheric neutrino oscillation measurements~\cite{icecube_osc_golden,flercnn_osc,km3net_orca_osc}, and as a key tool for constructing neutrino samples in astrophysical studies. For example, because sufficiently energetic track events provide superior angular resolution, searches for astrophysical sources such as~\cite{icecube_ngc1068,icecube_hese,icecube_mese} seek to identify pure track samples. Conversely, cascade samples are exploited in studies such as~\cite{icecube_galactic_plane}, where their reduced contamination from atmospheric muons and improved calorimetric energy reconstruction are advantageous.

\subsection{Reconstruction Algorithms}
%\vspace{1cm}

\noindent With the five commonly reconstructed attributes of the neutrino defined, we now turn to methods for estimating these quantities. A \textit{reconstruction algorithm} is a procedure that infers one or more physical attributes from the observed pulses, i.e.\ a mapping of the form
\[
f: \mathbb{R}^{n \times j} \longrightarrow \mathbb{R}^k ,
\]
where $n$ is the number of recorded pulses, $j$ the number of features associated with each pulse, and $k$ the number of reconstructed attributes. In practice, the pulse features typically include quantities such as photon arrival time, measured charge, and the position of the optical module that detected the pulse. The pulse features available on our datasets are described in~\cref{pulse_level_info}.
\vspace{0.25cm}

\noindent The landscape of reconstruction algorithms can generally be subdivided into likelihood-based methods and machine-learning-based methods. While the overall goal of both categories remains the same, significant differences exist between these two approaches to neutrino event reconstruction. In the following, these differences are outlined.

\subsection{Likelihood-Based Reconstruction}
\label{sec:likelihood_reconstructions}

Maximum likelihood estimation is a standard technique for parameter inference~\cite{maximum_likelihood_history} and has historically been the primary approach for event reconstruction in neutrino telescopes~\cite{SHIOZAWA1999240,AMANDA:2003vtt,SNO:2011hxa}. A central requirement for MLE is the existence of a likelihood function
\begin{equation}
    \mathcal{L}(x|\theta): \mathbb{R}^{n \times j} \times \mathbb{R}^k \longrightarrow \mathbb{R},
    \label{eq:reco_map}
\end{equation}
which models the probability of observing a set of pulses $x \in \mathbb{R}^{n \times j}$ given an event hypothesis $\theta \in \mathbb{R}^k$. The best-fitting event hypothesis is then identified by maximizing $\mathcal{L}(x|\theta)$. The initial challenge in applying MLE for event reconstruction is defining a suitable likelihood function. Often, the assumption that observations on individual OMs are independent is utilized to define likelihoods of the form
\begin{equation}
    \mathcal{L}(x|\theta) = \prod_{i=1}^d  A_i(h|\theta)\cdot \left(\prod_{j=1}^h p_i(t_j|\theta) \right),
\end{equation}
where $d$ denotes the total number of OMs and $h$ is the number of pulses observed on the $i$th OM. The factor $A_i(h|\theta)$ gives the probability of observing $h$ pulses on the $i$th OM (often modeled as a Poisson distribution), while $p_i(t_j|\theta)$ gives the probability density for observing the $j$th pulse at time $t_j$ on the same OM. The detector response functions $A_i(h|\theta)$ and $p_i(t_j|\theta)$ depend strongly on the optical properties of the detection medium, the characteristics of the optical instrumentation, and other detector systematics, making their accurate modeling a central challenge in applying MLE for event reconstruction. While the response functions may be approximated using extensive forward simulation~\cite{icecube_direct_reco}, such approaches are often computationally prohibitive due to the complexity of accurate event simulation in neutrino telescopes.

To mitigate the challenges in obtaining the full reconstruction likelihood, MLE-based techniques often rely on approximations that balance inference speed and accuracy. Such approximations may involve neglecting certain detector effects (e.g., stochastic noise), assuming particular event morphologies and energy range, or reducing the number of free parameters by reconstructing only a subset of the neutrino event properties, each of which may significantly simplify the likelihood. For example, by assuming events are fully contained cascades in an idealized detector, energy reconstruction reduces to finding the proportionality constant between the number of detected photons and the neutrino energy. Similarly, in the case of track-like events, direction reconstruction can be simplified by assuming that the neutrino direction is identical to the muon direction and that the muon propagates along a straight path with continuous energy loss. Such assumptions reduce the likelihood to a comparison between the observed light pattern and that expected from an idealized line source (LineFit)~\cite{AMANDA:2003vtt}. While such approaches are useful as fast first-guess methods, MLE-based reconstructions of analysis observables are often produced using more sophisticated likelihood approximations to improve accuracy. For example, methods such as~\cite{icecube_retro, icecube_splinempe} approximate $A_i(h|\theta)$ and $p_i(t_j|\theta)$ by querying a vast set of precomputed forward simulations, which significantly increases reconstruction quality at speeds far greater than running full simulations.

Within IceCube, the state-of-the-art direction reconstruction algorithm for track events remains likelihood-based, but the remaining properties of interest are predominantly reconstructed using ML-based alternatives.

\subsection{Machine-Learning-Based Reconstructions}
\label{sec:ml_reconstruction}
In recent years, reconstruction methods based on ML, and in particular on deep learning, have seen rapid adoption in neutrino telescopes. By relying on vast labeled datasets, deep learning–based approaches typically formulate the reconstruction of the properties of the incident neutrino as supervised learning problems. The common approach is to introduce an overparameterized model architecture $g(x)$ that approximates the mapping seen in~\cref{eq:reco_map}. The model parameters are adjusted to minimize a loss function that quantifies the error of the predictions, which is typically simpler to evaluate than a reconstruction likelihood. The model parameters are often optimized using variations of Stochastic Gradient Descent (SGD)~\cite{stochastic_approximation_old}, which optimizes the model predictions \textit{on average} across samples of events, as opposed to on an event-by-event basis as in MLE reconstructions. A central challenge in applying deep learning–based reconstruction methods is identifying sufficiently expressive model architectures, descriptive data representations, and suitable loss functions. The following provides a brief overview of the progression of model architectures and their application to neutrino event reconstruction.

The advance of deep learning-based model architectures for neutrino telescope event reconstruction has seen a similar progression as related fields such as jet reconstruction in particle accelerator experiments.
In jet reconstruction, methods have moved from convolutional neural networks (CNNs) ~\cite{jet_cnn_2014} (2014) to graph neural networks (GNNs)~\cite{jet_gnn_2019} (2019), then transformer-based architectures~\cite{jet_transformer_2022} (2022), and recently multi-modal foundation models~\cite{jet_foundation_model_2024} (2024).
Along similar lines, CNNs in IceCube~\cite{mirco_cnn} (2021) have been used for energy reconstruction in recent astronomy results, such as observations of neutrino emissions from the galactic plane~\cite{icecube_galactic_plane} and from the Seyfert galaxy NGC1068~\cite{icecube_ngc1068}.
CNNs have also been used to provide both classifications and reconstructions for measurements of atmospheric neutrino oscillations in IceCube~\cite{flercnn_osc}.
While the benefits of GNNs for event classification were demonstrated in IceCube in 2018~\cite{icecube_gnn_old}, it was first around 2022~\cite{icecube_dynedge} that GNNs began seeing widespread use in the field.
In IceCube, GNNs have been used to produce various reconstructions and classifications, including noise removal, for projecting the sensitivity of a coming detector extension of IceCube to atmospheric neutrino oscillations, mass ordering, and tau appearance~\cite{icecube_queso}.
GNNs have also been applied to infer the mass composition of cosmic rays in IceCube~\cite{icecube_gnn_cosmic_ray_composition}.
The potential of transformer-based architectures was demonstrated by many participants in the public Kaggle Competition "IceCube - Neutrinos in Deep Ice"~\cite{icecube_kaggle} (2024), but their performance on non-public data is still being studied.
Recently, a foundation model along with a novel pre-training task was proposed~\cite{polarbert} (2024).
Other recent techniques include Single Image Super Resolution for neutrino telescopes~\cite{SISR_neutrino_telescope} (2024) and neutrino event representation learning~\cite{dom2vec} (2024). In parallel, hybrid methods that rely on both deep learning and maximum likelihood techniques have been proposed. For example,~\cite{icecube_event_generator} (2021) utilizes neural networks to parameterize likelihood terms, and the method was used to provide direction reconstructions for the galactic plane study in IceCube~\cite{icecube_galactic_plane}.
Similarly, a technical paper~\cite{icecube_normalizing_flow} (2023) from IceCube has demonstrated that normalizing flows conditioned on latent representations of neutrino events may be used to learn asymmetric conditional posterior distributions of neutrino events, allowing production of contours and point predictions through maximum likelihood estimation. 

\vspace{0.5cm}

\noindent The primary appeal of ML-based reconstruction algorithms is their independence from complex likelihood functions and their ability to produce predictions with competitive accuracy at speeds often orders of magnitude faster than MLE-based alternatives. Additionally, since the model architectures are universal function approximators~\cite{universal_function_approximation} and the low-level observations across detectors are highly similar, architectures found to be expressive in one neutrino telescope are likely to be expressive in others, providing a solid foundation for cross-experimental collaboration.

\section{The NuBench Datasets}
\label{sec:datasets}
The Neutrino event reconstruction Benchmark (NuBench) datasets are a collection of seven datasets with nearly 130 million simulated $\nu_\mu^\text{CC}$ and $\nu_\mu^\text{NC}$ interactions in the energy range 10 \si{GeV}–100 \si{TeV}. The events were simulated in six different detector geometries that we refer to as \flowers{}, \flowerl{}, \flowerxl{}, \Triangle{}, \cluster{}, and \hexagon{}, named after their geometric characteristics. 
\setlength{\tabcolsep}{6pt}
\begin{table}[h!]
  \centering
  \caption{Overview of datasets processed for this work. A total of over 129.7 million events were distributed across 7 datasets with geometries similar to existing or proposed neutrino telescopes. Datasets marked with * are simulated in ice, whereas the remainder is simulated in water.}
  \label{table:datasets}
    \begin{tabular}{lccccccc} 
    \hline
    \textbf{Dataset}  & \textbf{Events} & \textbf{Inspiration} & \textbf{$\nu_\mu^\text{CC}/\nu_\mu^\text{NC}$} & \textbf{Strings/DOMs} &  \textbf{Energy Range}\\ 
     & (millions) & & (\%) & & (\si{GeV}) \\
    \hline
    \hline
    \Triangle & 23.1  & P-ONE   & 35/65  & 3/60 & $10$ - $10^5$  \\
    \cluster           & 22.9  & GVD     & 49/51 &  8/288 & $10$ - $10^5$ \\
    \flowers & 20.5  & ORCA & 40/60 &  150/3300 & $10$ - $10^3$\\
    \flowerl& 24.0  & ARCA & 35/65 & 115/2070 & $10$ - $10^5$ \\ 
    \flowerxl& 10.1  &  TRIDENT & 88/12 &  1211/24220 & $10$ - $10^5$\\
    \hexagon& 20.5  & IceCube & 48/52 & 86/5160 & $10$ - $10^5$\\
    $\hexagonice^*$& 8.6  & IceCube  & 57/43 & 86/5160 & $10$ - $10^3$\\
    \hline
    \hline
    \textbf{Total}: &  \texttt{129.7} & & & & & 
    \end{tabular}
\end{table}

The geometries were chosen because of their resemblance to existing or proposed detector designs, representing a range of approaches to detector layouts in the field. The purpose of the datasets is to provide a common resource, not tied to any specific detector geometry or reconstruction task, to facilitate cross-experimental collaboration in the development and application of ML-based reconstruction algorithms.
A brief overview of the dataset catalogue can be seen in~\cref{table:datasets}, which contains essential details such as sample size, morphology ratio ($\nu_\mu^\text{CC}/\nu_\mu^\text{NC}$), detector array details, and energy range.
In the following sections, we provide an overview of the datasets, their contents, and their production methodology, with further details provided in~\cref{sec:app_datasets}. The datasets can be downloaded \href{https://github.com/graphnet-team/NuBench}{here}.

\subsection{Detector Geometries}
\begin{figure}[h!]
    \centering
    \includegraphics[width=1\textwidth]{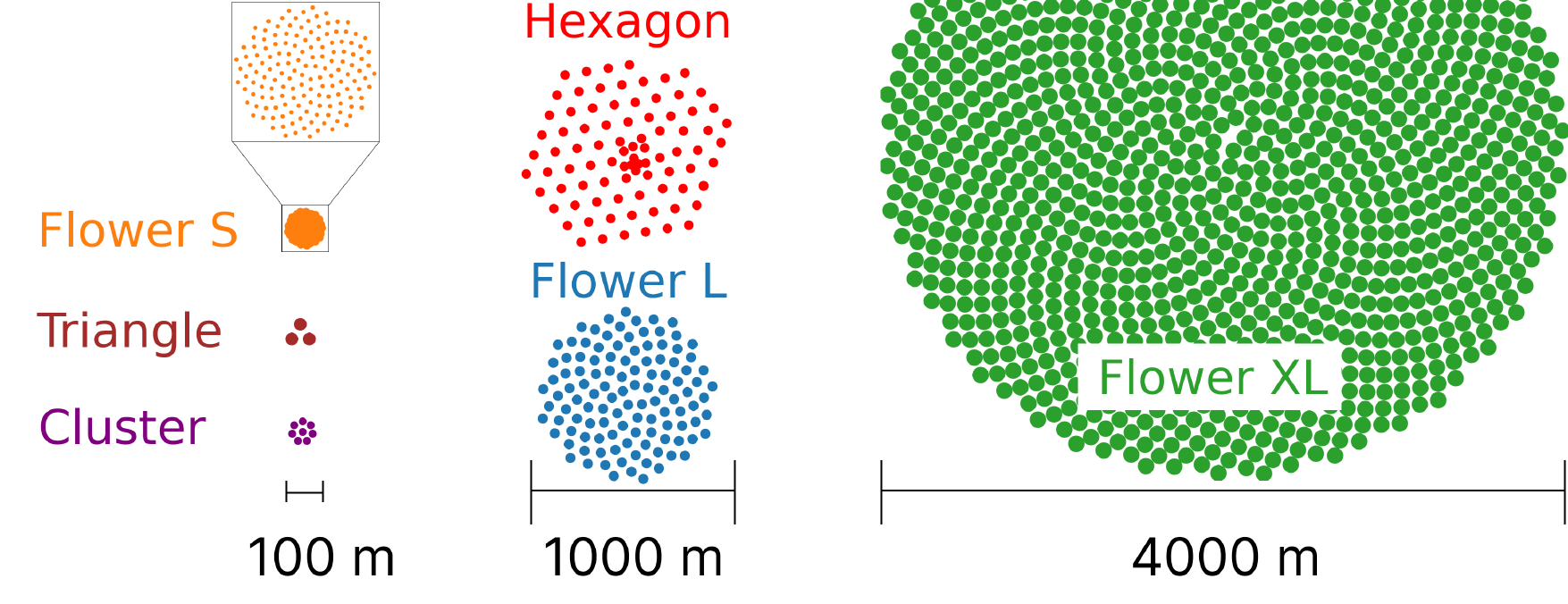}
    \caption{Top-down view of the 6 different detector geometries: \flowers{}, \flowerl{}, \flowerxl{}, \Triangle{}, \cluster{}, and \hexagon{}.
    Approximate length scales are annotated for comparison.}
    \label{fig:xy_geometry}
\end{figure}

\begin{figure}[h!]
    \centering
    \includegraphics[width=1\textwidth]{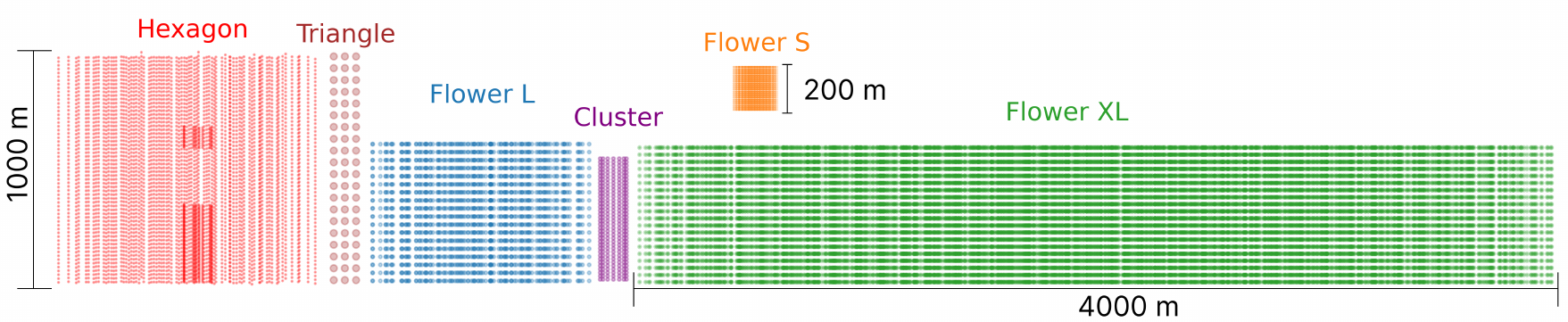}
    \caption{Side view of the 6 different detector geometries: \flowers{}, \flowerl{}, \flowerxl{}, \Triangle{}, \cluster{}, and \hexagon{}.
    Approximate length scales are annotated for comparison.}
    \label{fig:depth_geometry}
\end{figure}

The six geometries are inspired by, but not strictly identical to, the following existing or proposed neutrino telescopes: KM3NeT-ORCA~\cite{km3net_orca_osc}, KM3NeT-ARCA~\cite{arca_astronomy}, TRIDENT~\cite{TRIDENT:2022hql}, P-ONE~\cite{pone}, Baikal-GVD~\cite{baikal_gvd}, and IceCube~\cite{icecube86}, respectively. This diversity of geometries allows us to benchmark reconstruction algorithms across detectors of varying size and density of optical instrumentation, thereby providing insight into their generalizability beyond a single experimental setup.
 The geometries span a wide range of volumes and layouts, with their differences illustrated in~\cref{fig:xy_geometry} (top-down view) and~\cref{fig:depth_geometry} (side view).

As seen in~\cref{fig:xy_geometry}, three of the detectors, namely \flowers{}, \flowerl{}, and \flowerxl{}, have strings arranged in a sunflower-style geometry designed to minimize long corridors where charged particles could otherwise propagate undetected. These detectors consist of 150, 115, and 1211 strings with inter-string spacings of approximately \SI{12}{\m}, \SI{72}{\m}, and \SI{90}{\m}, respectively.

By comparison, \Triangle{} and \cluster{} are small clusters of three and eight strings with inter-string spacings of approximately \SI{100}{\m} and \SI{52}{\m}, respectively. Lastly, \hexagon{} consists of 78 strings arranged in a near-hexagonal main array with an inter-string spacing of \SI{125}{\m}, together with eight additional strings embedded within the array at a smaller spacing to reduce the detector’s energy threshold. This additional infill represents the DeepCore subvolume in IceCube~\cite{deepcore}.

As shown in~\cref{fig:depth_geometry}, the vertical spacing of OMs along detector strings varies greatly across the geometries. In the \Triangle{} geometry, OMs are spaced by about \SI{52}{\m}, the largest among the geometries, followed by about \SI{30}{\m} in \flowerl{} and \flowerxl{}, and \SI{15}{\m} in \cluster{}. The smallest vertical spacing is found in \flowers{}, at just \SI{9}{\m}. In \hexagon{}, the vertical distance between OMs varies between the main array and in the infill, introducing additional irregularity into the geometry. Together, the variations in both horizontal layout and vertical spacing highlight the diversity of detector designs considered in the field.

\subsection{Simulation}
In official simulations from neutrino telescope collaborations, neutrino events are produced in a two-step procedure, neither of which is typically publicly available. First, physical interactions are simulated and Cherenkov photons are traced to the optical instrumentation. In the second step, Cherenkov photons are subject to experiment-specific processing that emulates the hardware response to the Cherenkov photons. This second processing step often includes emulation of PMT response, PMT noise, environmental backgrounds such as bioluminescence~\cite{km3net_noise} and radioactive decays~\cite{icecube_pmt_calibration}, as well as DAQ effects such as dead-time, event triggers, and filters~\cite{icecube_daq, k3mnet_daq}. Accurate modeling of detector response and systematic uncertainties is essential for agreement between simulated and observed neutrino interactions, and remains an active area of research~\cite{km3net_bio_patterns, icecube_pocam, killian_phd}. 

The simulation presented here was performed in a similar two-step procedure. First, neutrino interactions were simulated using \prometheus{}~\cite{prometheus}, a recently published open-source neutrino telescope simulation tool capable of simulating events in arbitrary detector geometries, in both water and ice. Second, the output of \prometheus{}, which consists of raw Cherenkov photons arriving at the OMs, was processed to emulate a simplified detector response, resulting in triggered neutrino events comparable to those produced in official neutrino telescope simulations, but with significant differences. Further details on the physics simulation using \prometheus{} can be found in~\cref{sec:app_datasets}. In the following, we describe the main components of the simplified detector response, including the treatment of triggering, noise, and features at both the pulse and event level.

\subsection*{Detector Response \& Triggering}
\label{sec:detector_response_sim}
The output of \prometheus{} consists of Cherenkov photons at individual OMs without information on their arrival direction. The photons originate from the simulated neutrino interaction, and no stochastic noise or atmospheric muon contamination is included. The OMs are modeled as perfect spheres with a quantum efficiency of 20\% and a radius of \SI{30}{cm}. Differences in photon propagation between water and ice lead to differences in the assumed angular acceptance of the OMs. For water-based simulations, which account for most of the datasets in \cref{table:datasets}, the angular acceptance is uniform across the sphere. In ice, corrections are applied to emulate the angular acceptance measured in IceCube, where photons arriving from above have a reduced probability of detection. Because the \prometheus{} output does not include directional information for individual photons, the subsequent detector response emulation cannot directly simulate multi-PMT optical modules. Instead, each OM is treated as a single effective PMT. In the water-based simulations, this corresponds to a simplified spherical PMT with uniform angular acceptance, while in the ice-based simulations, the effective PMT is modeled with reduced sensitivity to down-going photons.
 Additional details regarding the physics simulation can be found in~\cref{sec:app_datasets}. 
\begin{figure}[h!]
    \centering
    \includegraphics[width=1\textwidth]{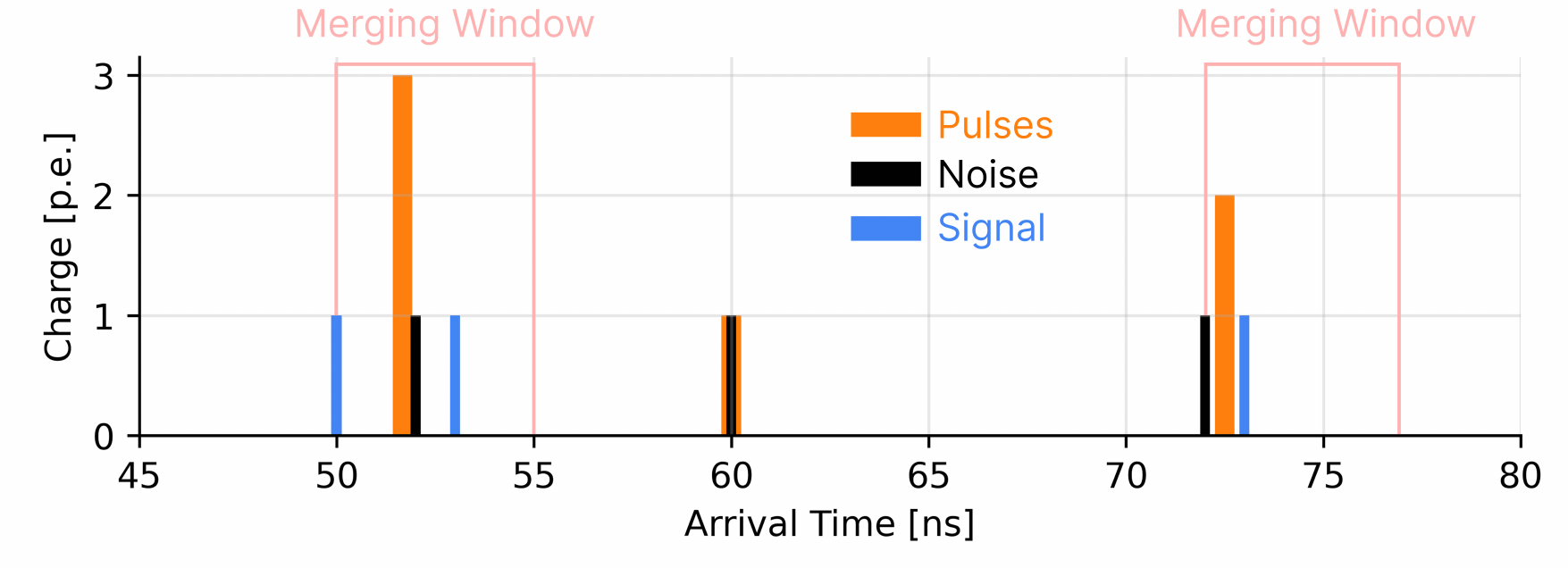}
    \caption{Illustration of the photon merging procedure applied to transform photons from \prometheus{} simulation into pulses of Cherenkov radiation. The merging window is applied starting from the first photon and is set to the TTS assigned to each dataset. The associated charge of the pulses is set to the number of photons found within each merging window.}
    \label{fig:merging_procedure}
\end{figure}

Our detector response emulation begins at the event level by linearly shifting the arrival times of individual photons into a trigger window of at least \SI{5}{\mu s}, centered around the mean photon arrival time. Because photon incidence direction and impact point on the OM are unavailable, we do not apply angle-dependent acceptance corrections; instead, we rescale the overall OM efficiency to reflect designs with reduced angular coverage (see Table~\ref{table:datasets_processing}) to a maximum of the given 20\% through subsampling.  Subsequently, the arrival times of stochastic noise photons are sampled uniformly within the trigger window, using an expected number of noise photons derived from the values in \cref{table:datasets_processing} and according to \cref{eq:data_augmentations_noise}. The OM at which the noise is observed is randomly chosen from the geometry. Following the noise injection, photons at each OM, originating from either neutrino interactions or stochastic noise, are then subject to a merging procedure that combines coincident photons into a single pulse. Starting from the first photon, other photons are considered coincident if they occur within the merging window, and the collection of photons is replaced with a pulse. The average arrival time of the photons within the merging window becomes the arrival time of the pulse, and the induced charge of the pulse is set to the number of photons found within the merging window. The merging window is set to the assigned Transit Time Spread (TTS) of the dataset, chosen, when possible, based on publicly available information from the respective experiments. The TTS represents the intrinsic timing resolution of the PMT. An overview of the assigned TTS values for each dataset can be seen in~\cref{table:datasets_processing}, and the merging procedure is illustrated in~\cref{fig:merging_procedure}. Following the merging procedure, the arrival time and charge of pulses are smeared using a perturbation $\epsilon$ drawn from a normal distribution with standard deviations of \SI{1}{ns} and \SI{0.25}{p.e.}, respectively.
After the smearing, events are filtered based on the number of pulses. Events with fewer than four or exceeding one million signal pulses are removed.
 
While the procedure outlined in this section produces triggered neutrino events resembling those from official simulations, several simplifications should be noted. Each OM is modeled as a single PMT with simplified angular acceptance. Because photon incidence direction and impact point on the OM are unavailable, we do not apply angle-dependent acceptance corrections; instead, we rescale the overall OM efficiency to reflect designs with reduced angular coverage (see Table~\ref{table:datasets_processing}). Additionally, PMT effects such as charge saturation are not modeled, which leaves the total observed charge strongly correlated with neutrino energy. These choices yield an experiment-agnostic approximation suitable for developing and benchmarking reconstruction algorithms across detector geometries.

\newpage
\subsection{Content of Datasets}

Each of the seven NuBench datasets contains both the true properties of the incident neutrino, referred to as event-level information, and the corresponding detector response, referred to as pulse-level information. Each dataset in \cref{table:datasets} is split into a training sample and a test sample, with the test sample containing approximately $3\times 10^6$ events.

Because several parts of the event creation procedure are stochastic, events in the training sample have only been subject to efficiency adjustments and the merging procedure, leaving the remaining processing to be applied as real-time data augmentations during training. Events in the test partitions, on the other hand, have been subject to the full event creation procedure described in ~\cref{sec:detector_response_sim} and therefore represent a single realization of the stochastic processes. To limit dataset size, noise-induced pulses have been removed from events in the test partitions. A separate but compatible extension of the NuBench catalogue is planned, which will include noise pulses on a subset of the test partition, enabling direct comparison of noise-cleaning techniques \cite{icecube_queso}. 

A detailed description of the required real-time data augmentations for the training partitions is provided in Sec.~\cref{sec:data_augmentation}. Further details on the pulse- and event-level information are provided below.

\subsection*{Pulse-level information} 
\label{pulse_level_info}
The pulse-level information available as input to reconstruction algorithms for the datasets in \cref{table:datasets} includes DOM position, pulse arrival time, and pulse charge. An overview is provided in \cref{table:input_features}. 
\setlength{\tabcolsep}{4pt}
\begin{table}[h!]
  \centering
  \caption{Overview of available input data for reconstruction algorithms on the datasets.}
  \label{table:input_features}
    \begin{tabular}{lccc} 
    \hline
    \textbf{Variable}  & \textbf{Description} & \textbf{Dimensionality}\\
    \hline
    \hline
     \texttt{sensor\_pos\_{xyz}} & Position of DOM in meters & $\mathbb{R}^3$ \\
     \texttt{t} & Arrival time of pulse in ns & $\mathbb{R}$ \\
     \texttt{charge} & Charge of pulse in p.e. & $\mathbb{R}$ \\
     \texttt{string\_id} & Integer ID of detector string & $\mathbb{Z}$ \\
     \texttt{is\_signal} & Fraction of signal pulses in merging window & $\mathbb{R}$ \\
    \hline
    \hline
    \end{tabular}
\end{table}
Each event can be represented as an $[n,d]$-dimensional geometric time series, where $n$ denotes the number of observed pulses and $d$ the dimensions listed in \cref{table:input_features}. The number of observed pulses $n$ depends strongly on the neutrino energy, the detector geometry, and in particular the density of optical instrumentation. 

\begin{figure}[h!]
    \centering
    \includegraphics[width=1\textwidth]{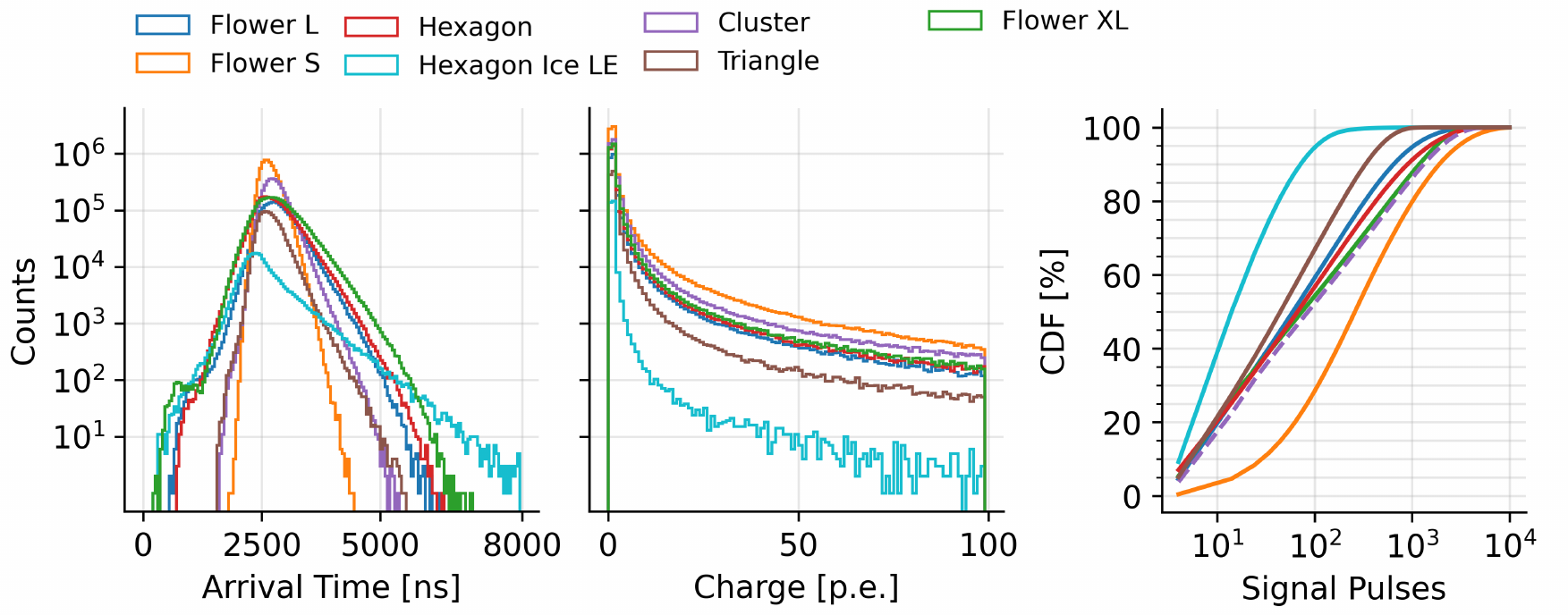}
    \caption{Quantification of pulse-level information in the test set of each dataset in \cref{table:datasets}. Left: Distribution of pulse arrival time. Middle: Distribution of pulse charge. Right: The cumulative percentage of events with respect to the number of signal pulses.}
    \label{fig:n_pulses}
\end{figure}

In~\cref{fig:n_pulses}, the arrival time (left) and charge distribution (middle) are shown for a subsample of the assigned test partition on each of the seven NuBench datasets. On the right of \cref{fig:n_pulses}, the accumulated percentage of events is shown w.r.t. the number of signal pulses for each of the datasets separately.  As seen from~\cref{fig:n_pulses}, large differences between datasets in the number of observed pulses in individual events exist. These differences are primarily driven by differences in the density of optical instrumentation, but are also affected by energy range.

For example, 80\% of events in the \Triangle{} dataset have 200 or fewer signal pulses, whereas 80\% of events in the \flowers{} dataset have 1000 or fewer. These differences are expected: the \Triangle{} detector has only 60 DOMs distributed across three strings with horizontal spacing of 100 meters, while the \flowers{} detector has 3300 DOMs distributed among 150 strings with horizontal spacing of about 12 meters. All datasets also contain events that exceed the 1000-pulse threshold in \cref{fig:n_pulses} by large margins.
Similarly, it can be seen from~\cref{fig:n_pulses} that the charge is predominantly around one but with large tails, and that the arrival times primarily range between 0 and 5~$\mu$s. 

\subsection*{Event-level information}\label{event_level_info}
Each dataset in \cref{table:datasets} contains detailed information about the incident neutrino, along with auxiliary event-level labels. Several of these labels serve as reconstruction targets, and the full set is listed in \cref{table:truth}. The datasets span neutrino energies from a few GeV to 100~TeV, with most events above 1~TeV. Neutrino flavor is restricted to $\nu_\mu$. The zenith and azimuthal angles describe the direction of travel of the incident neutrino, and for CC interactions, the outgoing muon direction is also provided. 
\setlength{\tabcolsep}{4pt}
\begin{table}[h!]
  \centering
  \caption{Overview of available event-level labels in each dataset. Zenith and azimuthal angles describe the direction of travel from the perspective of the respective particle. The interaction channel is encoded as 1 for CC and 2 for NC.}
  \label{table:truth}
    \begin{tabular}{lccc} 
    \hline
    \textbf{Variable}  & \textbf{Description} & \textbf{Dimensionality}\\
    \hline
    \hline
     \texttt{initial\_state\_energy} & Energy of incident neutrino in GeV & $\mathbb{R}$\\
     \texttt{initial\_state\_xyz} & Point of interaction of incident neutrino in meters & $\mathbb{R}^3$\\
     \texttt{initial\_state\_zenith} & Zenith angle of incident neutrino & $\mathbb{R}$\\
     \texttt{initial\_state\_azimuth} & Azimuthal angle of incident neutrino & $\mathbb{R}$\\
     \texttt{initial\_state\_type} & PDG encoding of neutrino flavor~\cite{pdg_encoding} & $\mathbb{Z}$ \\
     \texttt{interaction} & Neutrino interaction channel & $\mathbb{Z}$\\
     \texttt{bjorken\_xy} & Lorentz-invariant kinematic parameters & $\mathbb{R}^2$ \\
     \texttt{visible\_inelasticity} & See~\cref{sec:tasks} & $\mathbb{R}$\\
     \texttt{muon\_(zenith, azimuth)} & Angles of outgoing muon in CC interactions & $\mathbb{R}^2$\\
    \hline
    \hline
    \end{tabular}
\end{table}
Neutrino interaction channels are represented with integers 1 and 2 for CC and NC interactions, respectively. The Bjorken kinematic parameters are provided alongside a proxy label for inelasticity, which describes the ratio of visible hadronic energy to total visible energy and is described in greater detail in ~\cref{sec:tasks}.

\begin{figure}[h!]
    \centering
    \includegraphics[width=1\textwidth]{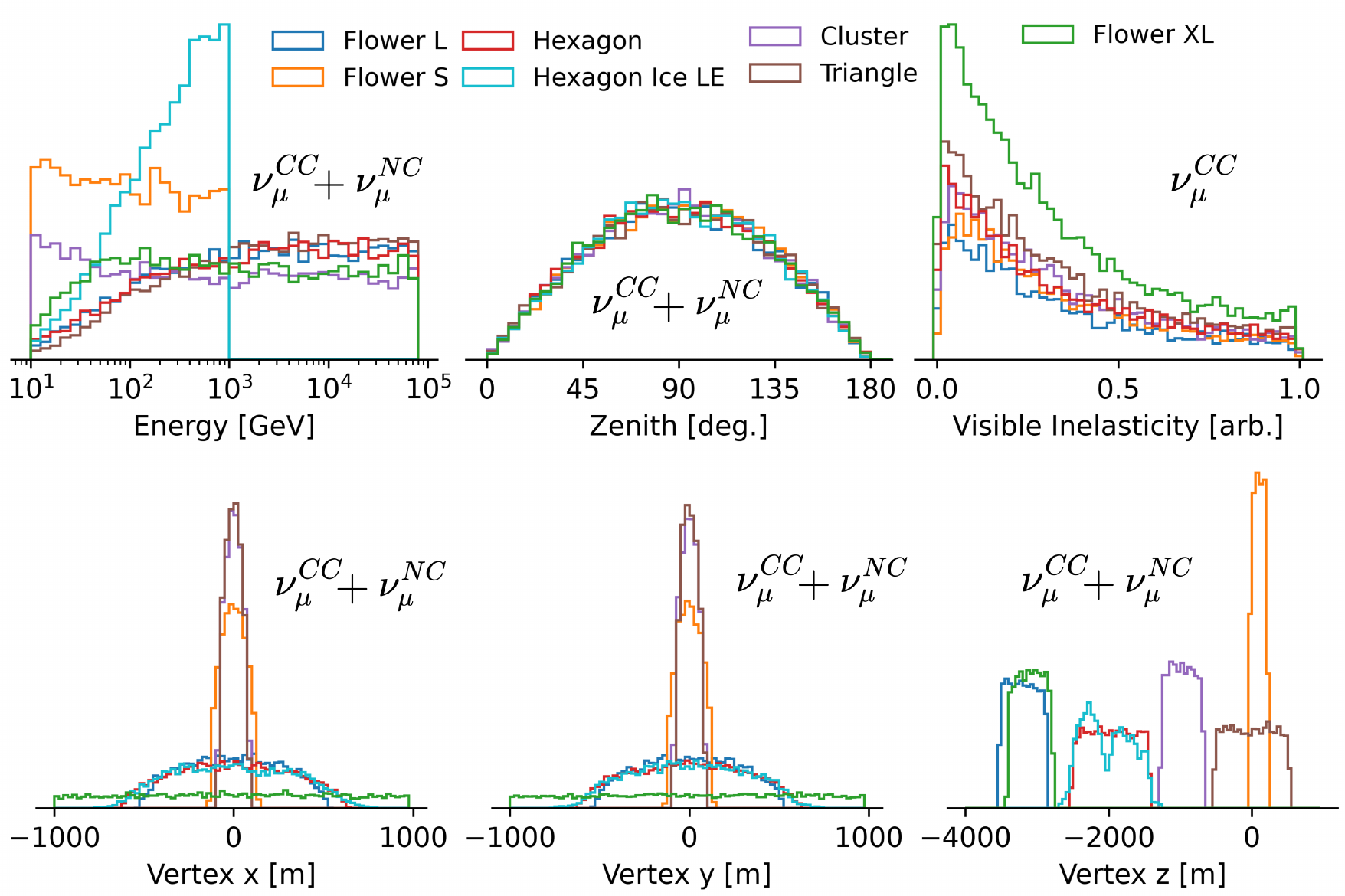}
    \caption{Distributions of neutrino energy, zenith, visible inelasticity, and interaction vertex in the seven NuBench datasets. Note that \hexagonice{} is omitted from the visible inelasticity distribution as that dataset does not cover the full numerical range.}
    \label{fig:truth_variables}
\end{figure}

The distributions of neutrino energy, zenith angle, visible inelasticity, and interaction vertex are shown in~\cref{fig:truth_variables} for each of the seven NuBench datasets. The distributions show both $\nu_\mu^\text{CC}$ and $\nu_\mu^\text{NC}$ events, except for visible inelasticity, which is shown for $\nu_\mu^\text{CC}$ only. While the energy distributions are influenced by the common trigger condition requiring at least three signal pulses, with energy thresholds that depend on the density of optical instrumentation in each geometry, the primary differences arise from variations in the injected energy ranges and assumed spectral indices. The differences in the z-coordinate of the interaction vertex are caused by arbitrary offsets in the geometry coordinates.

Together, these event-level labels provide a wide range of reconstruction targets of common interest, making the datasets broadly applicable for benchmarking reconstruction algorithms across diverse tasks and geometries.

\section{Results \& Comparison}
\label{sec:compare}
In the following, we compare four reconstruction algorithms on the five neutrino attributes introduced in~\cref{sec:tasks}. Two of these algorithms, \particlenet{} and \dynedge{}, are in active use within the IceCube and KM3NeT collaborations, respectively, and both rely on GNNs. These models have been used to reconstruct the full set of attributes on each of the seven NuBench datasets. In addition, one of the winning solutions from the open-data challenge "IceCube -- Neutrinos in Deep Ice," named \deepice{}, is included for direction reconstruction and employs a transformer-encoder architecture.
\begin{table}[h!]
  \label{table:model_overview}
    \begin{tabular}{lcccc} 
    \hline
    \textbf{Model}  & \textbf{Parameters} & \textbf{Paradigm} & 
    \textbf{Data Representation}\\
    & (millions) & & \\
    \hline
    \hline
     \particlenet{} (\cref{subsec:particlenet}) & 0.3  & GNN & Graph\\
     \dynedge{} (\cref{subsec:dynedge}) & 1.3  & GNN & Graph\\
     \grit{} (\cref{subsec:grit}) & 8.8  & GNN+Transformer Hybrid & Graph\\
     \deepice{} (\cref{subsec:deepice}) & 114  & Transformer & Sequence\\
    \hline
    \hline
    \end{tabular}
    \caption{An overview of the models chosen for comparison on the NuBench datasets. Further technical details of each model, their training procedure, data representations and task-specific modifications can be found in~\cref{sec:models}.}
\end{table}

Additionally, we include \grit{}, a new algorithm that combines graph representations with attention mechanisms, bridging GNN- and transformer-based methods. For the comparisons, each model is trained to reconstruct a single attribute on each dataset, yielding multiple instances of the models. Due to the computational complexity of \grit{}, the method is not trained on the \flowers{} dataset, as the high density of optical instrumentation of that geometry leads to increased computational cost.  In addition to differences in model architecture, differences in loss functions and training procedures affect the results. In~\cref{sec:models}, technical details regarding each reconstruction model, its training procedure, loss functions, and data representations are provided. The results shown in each of the following sections are computed on the test partitions of the NuBench datasets. Each comparison includes relevant discussion on evaluation metrics and is summarized using tables with selected performance scores, and the statistical uncertainty of scores is provided when relevant. The statistical error of each score is computed using bootstrapping.
The uncertainties do not contain effects stemming from stochasticity in the training procedure of each model, as that would require several repetitions of the whole training procedure, which is computationally expensive. The best-performing model is marked in bold. In cases where two models are statistically compatible, both are marked. 

The predictions and model artifacts can be downloaded \href{https://github.com/graphnet-team/NuBench}{here}.

\subsection{Energy}
This section presents the results of neutrino energy reconstruction performed by \particlenet{} (grey), \dynedge{} (purple), and \grit{} (green) on the seven NuBench datasets. All three models were trained using the \textsc{LogCosh} loss function defined in \cref{eq:logcosh}, with further details provided in~\cref{sec:models}. 

\begin{figure}[h!]
    \centering
    \includegraphics[width=1.1\textwidth]{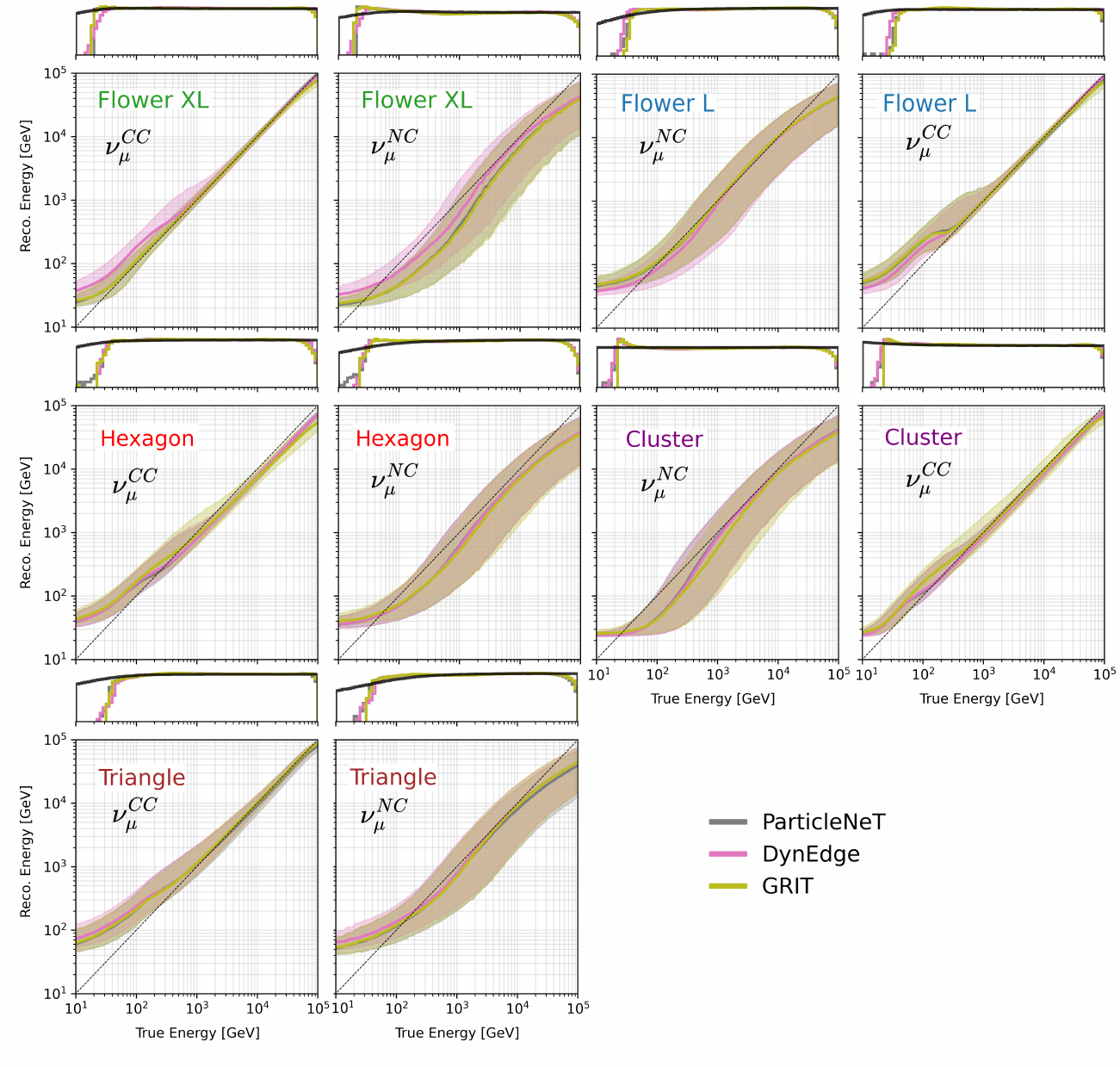}
    \caption{Energy reconstruction performance on the high-energy NuBench datasets using band plots. Shown are results for \particlenet{} (grey), \dynedge{} (purple), and \grit{} (green), compared to the true distributions (black). The diagonal denotes ideal reconstruction; shaded regions show the 16th–84th percentile spread.}
    \label{fig:energy_performance_he}
\end{figure}

Since neutrino energies span several orders of magnitude -- from \SI{10}{GeV} up to $10^5$~GeV in our datasets -- energy reconstruction methods often employ a logarithmic residual such as
\begin{equation}
    R_E = \log_{10}(E_\text{Reco}) - \log_{10}(E_\text{True})
    = \log_{10}\left(\frac{E_\text{Reco}}{E_\text{True}}\right),
\end{equation}
where $E_\text{True}$ and $E_\text{Reco}$ denote the true and reconstructed neutrino energies. Models are therefore often trained to minimize the fractional error in energy reconstruction, as opposed to the absolute error, which is the case in our study. In some studies, such as~\cite{icecube_dynedge}, the percentage error $\text{R}_E = \frac{E_\text{True} - E_\text{Reco}}{E_\text{True}} \times 100$ binned according to true energy is used as a performance metric. Here, the median is used as a measure of bias and the distribution width as a measure of resolution. Another common visualization is the band plot, which shows reconstructed energy versus truth together with percentile bands, simultaneously capturing both bias and variance~\cite{icecube_retro, km3net_orca_osc}. In this work, we adopt band plots for their concise visualization of model performance, as shown in \cref{fig:energy_performance_he}.

\begin{figure}[h!]
    \centering
    \includegraphics[width=1.1\textwidth]{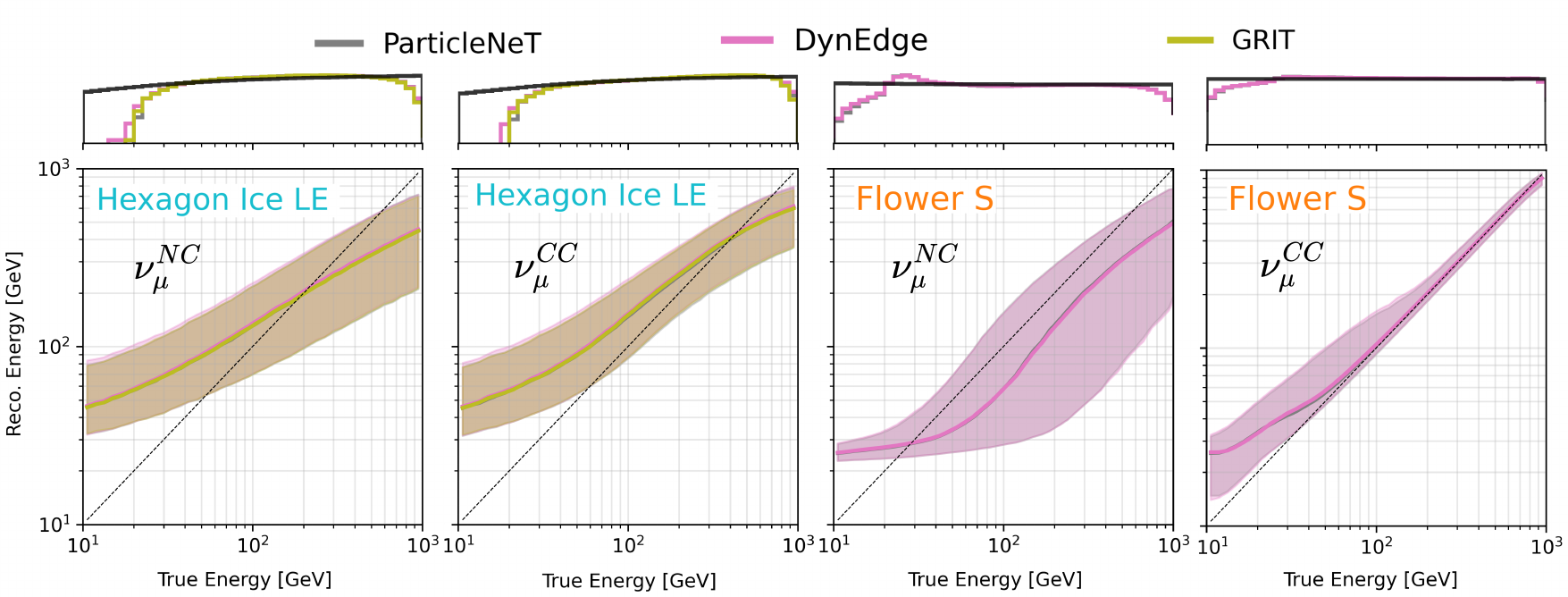}
    \caption{Energy reconstruction performance on the low-energy NuBench datasets using band plots. Shown are results for \particlenet{} (grey), \dynedge{} (purple), and \grit{} (green), compared to the true distributions (black). The diagonal denotes ideal reconstruction; shaded regions show the 16th–84th percentile spread.}
    \label{fig:energy_performance_le}
\end{figure}

In~\cref{fig:energy_performance_he} band plots are shown on the five NuBench datasets that span the full energy range of 10 GeV to $10^5$ GeV (\flowerxl{}, \flowerl{}, \hexagon{} and \cluster{}, \Triangle{}). In \cref{fig:energy_performance_le}, band plots are shown for the remainder of the datasets.  As briefly discussed in~\cref{sec:tasks}, $\nu_\mu^\text{NC}$ and $\nu_\mu^\text{CC}$ events have distinctly different relationships between observed pulses and neutrino energy, and thus represent different modalities in the sample. We therefore report the performance of the models on these two morphologies independently in \cref{fig:energy_performance_he} and \cref{fig:energy_performance_le}. The distributions of predictions from \particlenet{} (grey), \dynedge{} (purple), and \grit{} (green) are shown with respect to the true distribution (black) on top of each subfigure in log-scale. The black diagonal line denotes the ideal reconstruction, while the bands represent the width between the 84th and 16th percentiles. Deviations from the idealized black line represent bias, and narrower bands indicate smaller variance. 

It can clearly be seen in \cref{fig:energy_performance_he} and \cref{fig:energy_performance_le} that significantly higher variance in model predictions exists on $\nu_\mu^\text{NC}$ events compared to $\nu_\mu^\text{CC}$ events, and that this behavior is independent of reconstruction method and detector geometry. For example, on the \flowerxl{}
 dataset seen in \cref{fig:energy_performance_he}, the three models follow the ideal black line tightly for $\nu_\mu^\text{CC}$ events from around 1 TeV to 100 TeV. In comparison, the variance seen on $\nu_\mu^\text{NC}$ events covers around an order of magnitude in predicted energy in the same energy range. This significant difference in variance of model predictions is a direct result of the physical difference between NC and CC interactions. However, secondary effects such as the density of optical instrumentation, our simplified detector response emulation, and event containment affect the results. For example, the \Triangle{} geometry, which has the lowest density of optical instrumentation of the six detector geometries, has significantly higher variance in model predictions on $\nu_\mu^\text{CC}$ events at 1 TeV than other geometries, such as \hexagon{}, and in particular \flowers{} in \cref{fig:energy_performance_le}, which has the highest density of optical instrumentation between the six geometries. Additionally, reconstruction of energy on the NuBench datasets is less challenging than official neutrino telescope collaboration simulation due to the absence of PMT saturation, which induces a strong correlation between the total observed charge and incident neutrino energy, and intentional injection bias in the physics simulation using \prometheus{}. 

The injection bias seeks to generate incident neutrinos with a high probability of detection, increasing the simulation efficiency, and yielding a neutrino sample with a high rate of starting events, which are easier to estimate the energy of, as the hadronic component of the CC interaction is detectable in complement to the leptonic component.  

When comparing performance between models on the seven datasets, it can be observed that \particlenet{}, \dynedge{}, and \grit{} generally perform well, exhibiting only marginal differences on most datasets and event morphologies. The most noticeable differences are seen in \flowerxl{}, where \dynedge{} tend to over-estimate the energy of $\nu_\mu^{\text{CC}}$ at and below 1 TeV, which is not observed for \particlenet{}  and \grit{} until energies of 100 GeV and below. For $\nu_\mu^{\text{NC}}$ events, however, all three models appear to be biased towards underestimation of energy, but \dynedge{} less so. Additionally, it can be observed in \cref{fig:energy_performance_he} that \grit{} has a higher variance on \hexagon{} and \cluster{} for $\nu_\mu^\text{CC}$ events than \particlenet{} and \dynedge{}, but for $\nu_\mu^\text{NC}$ events the difference in variance appear marginal. 

To further quantify the differences in performance for \particlenet{}, \dynedge{} and \grit{}, we provide estimates of model bias and energy resolution in the three energy ranges of $E \leq 10^2$ GeV, $10^2 < E \leq 10^3$ GeV and $10^3 < E \leq 10^5$ in~\cref{tab:energy}. In each energy range, bias and resolution are reported using both $\nu_\mu^\text{CC}$ and $\nu_\mu^\text{NC}$ events, and the metrics are defined using the percentile error as described at the beginning of the section. The provided uncertainties describe the statistical fluctuation and are obtained through bootstrapping, but do not contain the stochasticity involved in the training procedure of the models. The model with the best performance is highlighted in bold in each energy range and for both metrics.   

\begin{table}[H]
    \centering
    \begin{tabular}{ccccccc}
    \multicolumn{7}{c}{Energy Reconstruction} \\
    \toprule
    Model 
    & \multicolumn{2}{|c|}{$E \leq 10^2~\text{GeV}$}
    & \multicolumn{2}{|c|}{$10^2 < E \leq 10^3~\text{GeV}$}
    & \multicolumn{2}{|c}{$10^3 < E \leq 10^5~\text{GeV}$} \\
    \midrule
    & \text{Bias} \text{[\%]}& $\sigma$ \text{[\%]}& \text{Bias} \text{[\%]}& $\sigma$ \text{[\%]}& \text{Bias} \text{[\%]} & $\sigma$ \text{[\%]}\\
    \midrule
    \midrule
    \multicolumn{7}{c}{\texttt{Flower XL}} \\
    \midrule
    \particlenet{} & \textbf{-8.92 $\pm$ 0.18} & \textbf{120.16 $\pm$ 0.31} & 18.18 $\pm$ 0.14 & 101.61 $\pm$ 0.18 & \textbf{3.78 $\pm$ 0.05} & \textbf{90.11 $\pm$ 0.18} \\
\dynedge{} & -70.08 $\pm$ 0.37 & 203.08 $\pm$ 0.54 & \textbf{-6.93 $\pm$ 0.18} & 169.47 $\pm$ 0.58 & 4.14 $\pm$ 0.04 & 99.1 $\pm$ 0.33 \\
\grit{} & -9.37 $\pm$ 0.21 & 122.63 $\pm$ 0.38 & 21.15 $\pm$ 0.18 & \textbf{99.74 $\pm$ 0.17} & 9.33 $\pm$ 0.06 & 90.42 $\pm$ 0.17  \\
    
    \midrule
    \multicolumn{7}{c}{\texttt{Flower L}} \\
    \midrule
    \particlenet{} & -159.37 $\pm$ 0.34 & 279.99 $\pm$ 0.62 & -14.87 $\pm$ 0.11 & 215.52 $\pm$ 0.42 & 8.65 $\pm$ 0.03 & \textbf{106.54 $\pm$ 0.21} \\
\dynedge{} & \textbf{-106.46 $\pm$ 0.29} & \textbf{224.34 $\pm$ 0.42} & \textbf{-8.03 $\pm$ 0.1} & \textbf{189.24 $\pm$ 0.31} & \textbf{3.13 $\pm$ 0.02} & 107.41 $\pm$ 0.16 \\
\grit{} & -168.98 $\pm$ 0.29 & 289.28 $\pm$ 0.63 & -13.81 $\pm$ 0.1 & 212.77 $\pm$ 0.43 & 6.77 $\pm$ 0.03 & 108.47 $\pm$ 0.21  \\
    
    \midrule
    \multicolumn{7}{c}{\texttt{Hexagon}} \\
    \midrule
    \particlenet{} & \textbf{-72.71 $\pm$ 0.21} & \textbf{202.89 $\pm$ 0.4} & 15.56 $\pm$ 0.05 & \textbf{141.0 $\pm$ 0.25} & 26.23 $\pm$ 0.02 & \textbf{86.25 $\pm$ 0.11} \\
\dynedge{} & -74.7 $\pm$ 0.17 & 206.51 $\pm$ 0.42 & 14.05 $\pm$ 0.07 & 145.73 $\pm$ 0.28 & \textbf{25.76 $\pm$ 0.01} & 87.93 $\pm$ 0.11 \\
\grit{} & -85.85 $\pm$ 0.18 & 226.39 $\pm$ 0.46 & \textbf{6.04 $\pm$ 0.1} & 156.1 $\pm$ 0.23 & 32.44 $\pm$ 0.03 & 94.11 $\pm$ 0.11  \\
    
    \midrule
    \multicolumn{7}{c}{\texttt{Hexagon Ice LE}} \\
    \midrule
    \particlenet{} & \textbf{-90.01 $\pm$ 0.17} & \textbf{226.98 $\pm$ 0.42} & 13.98 $\pm$ 0.04 & \textbf{98.35 $\pm$ 0.07} & -- & -- \\
\dynedge{} & -95.05 $\pm$ 0.14 & 240.7 $\pm$ 0.42 & \textbf{12.27 $\pm$ 0.04} & 101.0 $\pm$ 0.07 & -- & -- \\
\grit{} & -90.85 $\pm$ 0.18 & 228.06 $\pm$ 0.36 & 14.58 $\pm$ 0.04 & 98.41 $\pm$ 0.06 & -- & --   \\
    
    \midrule
    \multicolumn{7}{c}{\texttt{Flower S}} \\
    \midrule
    \particlenet{} & \textbf{-29.29 $\pm$ 0.08} & \textbf{135.74 $\pm$ 0.12} & \textbf{4.44 $\pm$ 0.02} & \textbf{78.93 $\pm$ 0.09} & -- & -- \\
\dynedge{} & -30.38 $\pm$ 0.07 & 138.61 $\pm$ 0.14 & \textbf{4.44 $\pm$ 0.02} & 79.85 $\pm$ 0.1 & -- & -- \\
\grit{} & -- & -- & -- & -- & -- &  --  \\
    
    \midrule
    \multicolumn{7}{c}{\texttt{Cluster}} \\
    \midrule
    \particlenet{} & -48.44 $\pm$ 0.09 & 145.94 $\pm$ 0.13 & 11.91 $\pm$ 0.05 & 151.16 $\pm$ 0.34 & 15.84 $\pm$ 0.02 & \textbf{90.17 $\pm$ 0.17} \\
\dynedge{} & \textbf{-46.42 $\pm$ 0.09} & \textbf{145.23 $\pm$ 0.15} & 12.57 $\pm$ 0.05 & \textbf{148.77 $\pm$ 0.27} & \textbf{15.66 $\pm$ 0.02} & \textbf{90.3 $\pm$ 0.16} \\
\grit{} & -58.18 $\pm$ 0.12 & 164.38 $\pm$ 0.15 & \textbf{10.27 $\pm$ 0.09} & 163.84 $\pm$ 0.25 & 17.96 $\pm$ 0.06 & 112.7 $\pm$ 0.13  \\
    
    \midrule
    \multicolumn{7}{c}{\texttt{Triangle}} \\
    \midrule
    \particlenet{} & \textbf{-153.37 $\pm$ 0.45} & \textbf{341.05 $\pm$ 1.03} & \textbf{-21.32 $\pm$ 0.16} & 213.14 $\pm$ 0.31 & 14.54 $\pm$ 0.04 & \textbf{110.3 $\pm$ 0.12} \\
\dynedge{} & -196.2 $\pm$ 0.52 & 397.03 $\pm$ 1.07 & -27.62 $\pm$ 0.13 & 219.81 $\pm$ 0.37 & \textbf{5.03 $\pm$ 0.03} & 119.68 $\pm$ 0.13 \\
\grit{} & -160.75 $\pm$ 0.44 & \textbf{341.64 $\pm$ 1.0} & \textbf{-21.24 $\pm$ 0.12} & \textbf{198.59 $\pm$ 0.39} & 5.76 $\pm$ 0.04 & 117.82 $\pm$ 0.11  \\
    \midrule
    
    \end{tabular}
  \caption{Bias and resolutions for energy reconstructions from  \particlenet{}, \dynedge{}, and \grit{} on each of the test partitions in the seven datasets. The results are computed for the three distinct energy ranges of $E \leq 10^2$ GeV, $10^2 < E \leq 10^3$ GeV, and $10^3 < E \leq 10^5$ GeV and include both $\nu_\mu^\text{CC}$ and $\nu_\mu^\text{NC}$ events. Statistical errors represent one standard deviation and are obtained through bootstrapping. Effects stemming from stochasticity in the training procedure are not included.\label{tab:energy}}
\end{table}

\label{sec:energy_comparison}

\newpage
\subsection{Direction}
In this section, we review the performance of \particlenet{} (grey), \dynedge{} (pink), and \grit{} (green) in reconstructing the direction of incident neutrinos on the seven NuBench datasets. For reference, one of the winning solutions from the "IceCube -- Neutrinos in Deep Ice" open data challenge, \deepice{} (black), is included in the comparison. Each of the four models is trained using the von Mises–Fisher loss function (\cref{eq:vmf}), which minimizes the opening angle between the true direction vector $\vec{D}_\text{True}$ and the reconstructed direction $\vec{D}_\text{Reco}$. The opening angle is defined as
\begin{equation}
    \psi = \cos^{-1}\left( \frac{\vec{D}_\text{True} \cdot \vec{D}_\text{Reco}}{|\vec{D}_\text{True}| \, | \vec{D}_\text{Reco}|}\right),
    \label{eq:opening_angle}
\end{equation}
where $|\vec{D}|$ denotes the vector norm. The opening angle is a standard performance metric for evaluating direction reconstruction algorithms~\cite{icecube_retro, icecube_ngc1068, mirco_cnn, icecube_galactic_plane}. 

As discussed in \cref{sec:tasks}, the difficulty of direction reconstruction depends strongly on the neutrino energy, event morphology, and containment. Accordingly, opening angles are typically smaller for $\nu_\mu^\text{CC}$ events than for $\nu_\mu^\text{NC}$ events, and they decrease with increasing energy. Consequently, opening angles can be used in several ways to compare reconstruction algorithms. The choice of metric usually depends on the intended application. For offline studies such as~\cite{icecube_ngc1068, icecube_galactic_plane}, the angular resolution—defined as the median opening angle binned according to neutrino energy—is often used to identify the best-performing algorithms in a given energy range. In complement, the fraction of events reconstructed within a given opening-angle threshold can also be used for comparison. This latter approach is particularly relevant for real-time applications, where neutrino telescopes issue alerts intended for follow-up observations. In such cases, the angular threshold is typically motivated by the field-of-view limitations of external instruments, often below 5 degrees \cite{magic_technical_report, swift_technical_report}.

\begin{figure}[h!]
    \centering
    \includegraphics[width=1.1\textwidth]{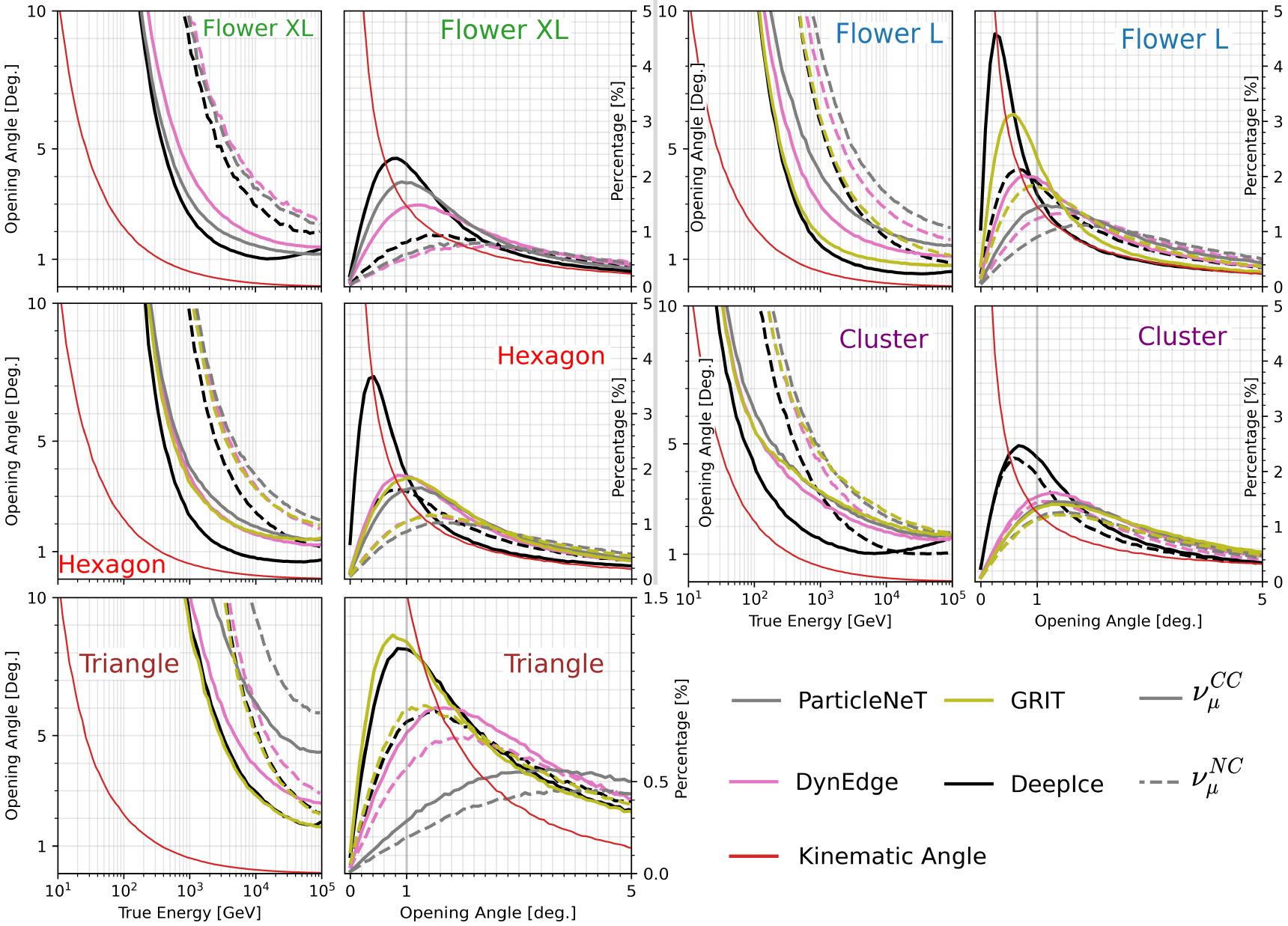}
    \caption{Performance of \particlenet{}, \dynedge{}, \grit{} and \deepice{} for direction reconstruction on the five high-energy NuBench datasets. Figures show both the median opening angle as a function of neutrino energy and the distribution of opening angles below 5 degrees. Percentages are given w.r.t. to each topology in the datasets.}
    \label{fig:direction_panel_he}  
\end{figure}

\begin{figure}[h!]
    \centering
    \includegraphics[width=1.1\textwidth]{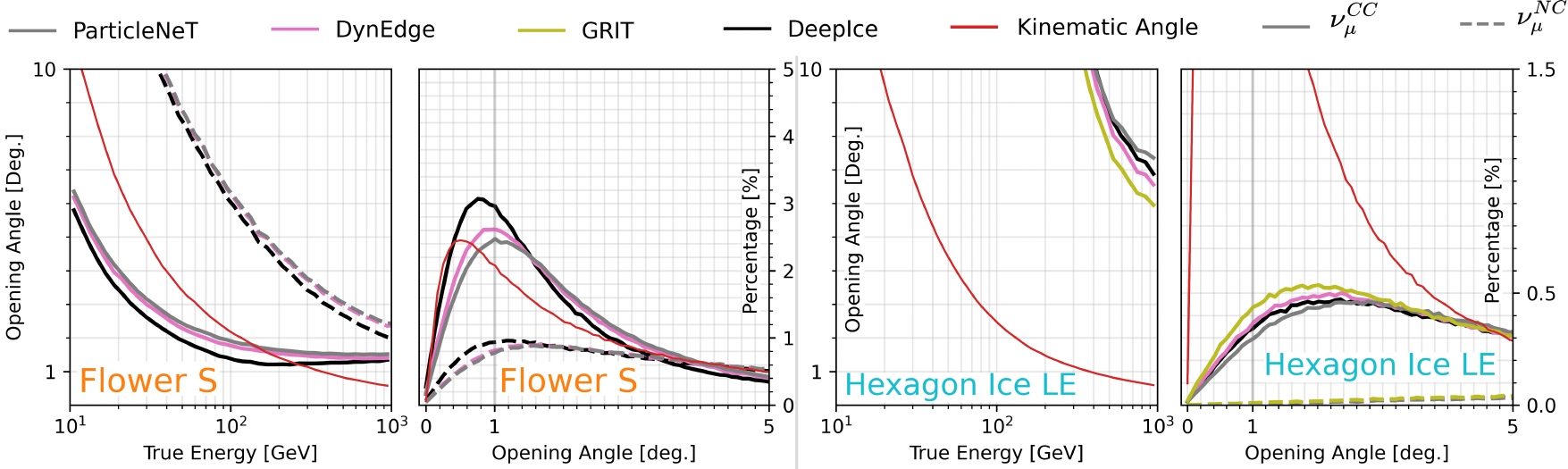}
    \caption{Performance of \particlenet{}, \dynedge{}, \grit{} and \deepice{} for direction reconstruction on the two low-energy NuBench datasets. Figures show both the median opening angle as a function of neutrino energy and the distribution of opening angles below 5 degrees. Percentages are given w.r.t. to each topology in the datasets.}
    \label{fig:direction_panel_le}  
\end{figure}

Figures~\ref{fig:direction_panel_he} and \ref{fig:direction_panel_le} compare the four models using both angular resolution and distributions of opening angles. Results are shown separately for $\nu_\mu^\text{CC}$ (solid) and $\nu_\mu^\text{NC}$ (dashed) events. Percentages in the opening angle distributions are calculated with respect to each topology, and a one-degree threshold is highlighted with a vertical grey line. The x-axis ticks are spaced in 0.2° increments. In addition to the four models, we include the \textit{kinematic angle} (red), defined as the angle between the incident neutrino and the outgoing lepton in CC interactions. For $\nu_\mu^\text{CC}$ events, the kinematic angle represents a practical lower bound on the opening angle for most geometries. However, because it only reflects the leptonic component of the interaction, reconstruction algorithms may surpass this bound in detectors with sufficiently dense optical instrumentation by resolving the hadronic component, which provides additional information that improves neutrino direction estimates. Due to convergence issues, \grit{} has been omitted from the comparison on the \flowerxl{} dataset. 

As expected, \cref{fig:direction_panel_he} and \cref{fig:direction_panel_le} show a significant difference in both angular resolution and opening angle distributions between $\nu_\mu^\text{CC}$ and $\nu_\mu^\text{NC}$ events, independent of reconstruction algorithm and detector geometry. When comparing angular resolution across the seven datasets, general differences emerge that can be attributed to detector layout. For example, when comparing \flowerxl{} and \flowerl{}, several reconstruction methods achieve sub-degree resolutions on \flowerl{} but not on \flowerxl{}. In addition, the gap in angular resolution between $\nu_\mu^\text{CC}$ (solid) and $\nu_\mu^\text{NC}$ (dashed) events appears larger on \flowerxl{} than on \flowerl{}. While \flowerxl{} spans a larger volume, it has a wider inter-string spacing (\SI{90}{m}) than \flowerl{} (\SI{72}{m}). The higher density of instrumentation in \flowerl{} likely explains the substantially larger fraction of $\nu_\mu^\text{NC}$ events reconstructed with opening angles below 1°.

However, due to the significantly larger volume of \flowerxl{}, the geometry would likely yield superior angular resolutions on $\nu_\mu^\text{CC}$ events beyond the energy range available in the NuBench datasets, as resolutions seen on \flowerl{} would eventually plateau as events would be increasingly uncontained, similar to what can be observed for \flowers{} in \cref{fig:direction_panel_le} at around \SI{200}{GeV} and above. The large differences between the achieved angular resolution on \hexagonice{} (\cref{fig:direction_panel_le}) and the achieved angular resolution in the remainder of the datasets are likely caused by the difference in angular acceptance between water- and ice-based simulation in \prometheus{}.   

When comparing the performance of individual algorithms, a clear picture emerges from \cref{fig:direction_panel_he} and \cref{fig:direction_panel_le}. On most datasets, and often with large margins, \deepice{} outperforms \particlenet{}, \dynedge{} and \grit{} both in terms of angular resolution and fraction of events reconstructed with angular errors below 1 degree. In the \Triangle{} dataset, however, the relative differences in performance between models are smaller, and the resolution curves from \deepice{} and \grit{} are nearly identical. The distribution of angular errors on \Triangle{} reveals a slightly increased fraction of events reconstructed with an angular error below 1 degree from \grit{}. The results generally imply that dot-product attention mechanisms -- which are used in \deepice{} and \grit{} to encode information globally as opposed to locally -- appear to be more expressive for direction reconstruction than localized message-passing convolutions on graphs, as employed in \particlenet{}, \dynedge{}. 

To further quantify the performance of each model, we provide performance metrics in \cref{tab:direction}. Here, the median opening angle ($\psi_\text{median}$), along with the percentage of events reconstructed at or below 1 and 5 degrees ($\psi \leq x^\circ$), is shown. These percentages are provided according to the event morphology, similar to \cref{fig:direction_panel_he} and \cref{fig:direction_panel_le}. Due to their dependence on event morphology, the metrics are reported for $\nu_\mu^\text{CC}$ and $\nu_\mu^\text{NC}$ events separately. To account for the energy dependence, the metrics are shown for the two energy ranges $10 < E \leq 10^3$ GeV and $10^3 < E \leq 10^5$ GeV. Statistical errors have been computed using bootstrapping and are $\mathcal{O}(10^{-3})$ for both metrics, but have been omitted from \cref{tab:direction} for brevity.

\newpage

\begin{table}[H]
    \centering
    \begin{tabular}{ccccccc}
    \multicolumn{7}{c}{Direction Reconstruction} \\
     \toprule
    Model & \multicolumn{3}{|c|}{$E \leq 10^3~\text{GeV}$ ($\nu_\mu^\text{CC}$ / $\nu_\mu^\text{NC}$)}
    & \multicolumn{3}{|c}{$10^3 < E \leq 10^5~\text{GeV}$ ($\nu_\mu^\text{CC}$ / $\nu_\mu^\text{NC}$)} \\
    \midrule
    & $\psi_\text{median}$ [deg.] & $\psi \leq 1^\circ$ [\%] & \multicolumn{1}{c|}{$\psi \leq 5^\circ$ [\%]} & \multicolumn{1}{|c}{$\psi_\text{median}$ [deg.]} & $\psi \leq 1^\circ$ [\%] & $\psi \leq 5^\circ$ [\%]\\

    %\midrule

    \midrule
    \multicolumn{7}{c}{\texttt{Flower XL}} \\
    \midrule
    \particlenet{} & 14.33 / \textbf{33.38} & 1.56 / 0.3 & 21.75 / 6.54 & 1.6 / 3.81 & 27.23 / 6.74 & 91.05 / 61.8 \\
\dynedge{} & 16.22 / 33.47 & 1.0 / 0.3 & 16.8 / 5.98 & 2.03 / 4.07 & 19.43 / 5.92 & 85.79 / 59.1 \\
\grit{} & -- & -- & -- & -- & -- & -- \\
\deepice{} & \textbf{14.23} / 33.6 & \textbf{2.12} / \textbf{0.39} & \textbf{24.11} / \textbf{7.01} & \textbf{1.29} / \textbf{3.21} & \textbf{36.65} / \textbf{9.98} & \textbf{93.23} / \textbf{67.15}  \\
    
    \midrule
    \multicolumn{7}{c}{\texttt{Flower L}} \\
    \midrule
    \particlenet{} & 13.21 / 20.66 & 1.26 / 0.6 & 20.45 / 11.54 & 2.12 / 3.5 & 18.36 / 8.27 & 83.66 / 64.89 \\
\dynedge{} & 11.6 / 18.95 & 2.12 / 0.84 & 26.46 / 14.2 & 1.54 / 2.9 & 30.67 / 11.99 & 88.76 / 70.54 \\
\grit{} & 10.03 / 17.54 & 4.51 / 1.25 & 33.09 / 17.87 & 0.99 / 2.14 & 50.41 / 21.11 & 92.12 / 77.03 \\
\deepice{} & \textbf{9.92} / \textbf{17.19} & \textbf{5.81} / \textbf{1.32} & \textbf{33.98} / \textbf{18.47} & \textbf{0.66} / \textbf{1.86} & \textbf{65.13} / \textbf{28.14} & \textbf{92.87} / \textbf{78.44}  \\
    
    \midrule
    \multicolumn{7}{c}{\texttt{Hexagon}} \\
    \midrule
    \particlenet{} & 16.91 / 32.44 & 1.35 / 0.45 & 19.8 / 8.38 & 2.03 / 3.86 & 19.93 / 8.1 & 82.15 / 59.55 \\
\dynedge{} & 16.61 / 31.92 & 1.5 / 0.49 & 20.89 / 8.94 & 1.76 / 3.48 & 25.62 / 10.25 & 83.92 / 62.12 \\
\grit{} & 16.52 / 32.34 & 1.79 / 0.54 & 22.17 / 9.27 & 1.83 / 3.42 & 23.33 / 10.03 & 84.16 / 63.15 \\
\deepice{} & \textbf{15.02} / \textbf{30.73} & \textbf{3.44} / \textbf{0.82} & \textbf{27.15} / \textbf{11.39} & \textbf{0.89} / \textbf{2.53} & \textbf{54.46} / \textbf{20.08} & \textbf{88.23} / \textbf{68.74}  \\
    
    \midrule
    \multicolumn{7}{c}{\texttt{Hexagon Ice LE}} \\
    \midrule
    \particlenet{} & 23.84 / 56.18 & 1.81 / 0.04 & 20.98 / 0.97 &--&--&--\\
\dynedge{} & 23.6 / 56.12 & 2.26 / 0.04 & 22.24 / 1.03 &--&--&--\\
\grit{} & \textbf{21.91} / \textbf{54.87} & \textbf{2.71} / \textbf{0.05} & \textbf{23.94} / \textbf{1.28} &  --  &  --  &  --  \\
\deepice{} & 23.75 / 56.21 & 2.15 / \textbf{0.05} & 21.51 / 1.08 &--&--&-- \\
    
    \midrule
    \multicolumn{7}{c}{\texttt{Flower S}} \\
    \midrule
    \particlenet{} & 2.27 / 6.73 & 18.37 / 5.19 & 77.24 / 40.23 &--&--&--\\
\dynedge{} & 2.13 / 6.66 & 20.67 / 5.48 & 78.4 / 40.7 &--&--&--\\
\grit{} & -- & -- & -- &--&--&--\\
\deepice{} & \textbf{1.82} / \textbf{6.38} & \textbf{26.32} / \textbf{7.11} & \textbf{81.04} / \textbf{42.29} &--&--&-- \\
    
    \midrule
    \multicolumn{7}{c}{\texttt{Cluster}} \\
    \midrule
    \particlenet{} & 8.29 / 14.55 & 2.58 / 1.3 & 32.69 / 19.83 & 2.06 / 2.47 & 17.83 / 14.08 & 86.93 / 78.43 \\
\dynedge{} & 7.66 / 13.82 & 3.25 / 1.7 & 36.03 / 22.12 & 1.88 / 2.12 & 21.11 / 18.67 & 88.02 / 81.07 \\
\grit{} & 7.66 / 13.74 & 2.88 / 1.48 & 35.14 / 21.62 & 2.18 / 2.6 & 17.21 / 13.64 & 84.45 / 77.34 \\
\deepice{} & \textbf{6.25} / \textbf{12.15} & \textbf{7.83} / \textbf{3.0} & \textbf{43.79} / \textbf{27.16} & \textbf{1.22} / \textbf{1.35} & \textbf{40.45} / \textbf{37.53} & \textbf{92.0} / \textbf{86.23}  \\
    
    \midrule
    \multicolumn{7}{c}{\texttt{Triangle}} \\
    \midrule
    \particlenet{} & 30.47 / 41.35 & 0.22 / 0.1 & 4.9 / 2.24 & 6.41 / 9.59 & 2.66 / 1.52 & 38.81 / 25.55 \\
\dynedge{} & 29.25 / 39.94 & 0.48 / 0.16 & 8.16 / 3.28 & 4.03 / 6.22 & 7.93 / 4.9 & 57.53 / 43.18 \\
\grit{} & \textbf{28.36} / \textbf{39.33} & \textbf{0.94} / \textbf{0.21} & \textbf{10.28} / \textbf{3.87} & \textbf{2.96} / \textbf{5.26} & \textbf{18.21} / \textbf{8.39} & \textbf{64.59} / \textbf{48.59} \\
\deepice{} & 28.49 / 39.51 & 0.72 / 0.17 & 9.35 / 3.6 & 3.08 / 5.38 & 16.17 / 7.64 & 64.03 / 47.88  \\
    \midrule
    
    \end{tabular}
  \caption{Selected performance metrics for direction reconstruction. $\psi_{\text{median}}$ represents the median opening angle and $\psi \leq x^\circ$ represents the percentage of events reconstructed with an opening angle at or below a certain threshold. The metrics are provided for the two energy ranges $10^2 < E \leq 10^3$ GeV and $10^3 < E \leq 10^5$ GeV. Additionally, the metrics are reported for each event morphology separately ($\nu_\mu^\text{CC}$ / $\nu_\mu^\text{NC}$).  Statistical errors are small at $\mathcal{O}(10^{-3})$ and have been omitted for brevity.\label{tab:direction}}
\end{table}

\label{sec:direction_comparison}

\newpage
\subsection{$\mathcal{T}/\mathcal{C}$ Classification}
\label{sec:track_cascade_classification_comparison}
In this section, we review the performance of \particlenet (grey), \dynedge (purple), and \grit{} (green) on the $\mathcal{T}/\mathcal{C}$ classification task, which seeks to distinguish between the two canonical event morphologies known as cascades and tracks, which in our datasets represent the $\nu_\mu$ NC and $\nu_\mu$ CC interactions, respectively.
During training, the loss function for all three models is the $\texttt{BCELoss}$ in \cref{eq:bce}.

\begin{figure}[htbp]
\centering
\includegraphics[width=\textwidth]{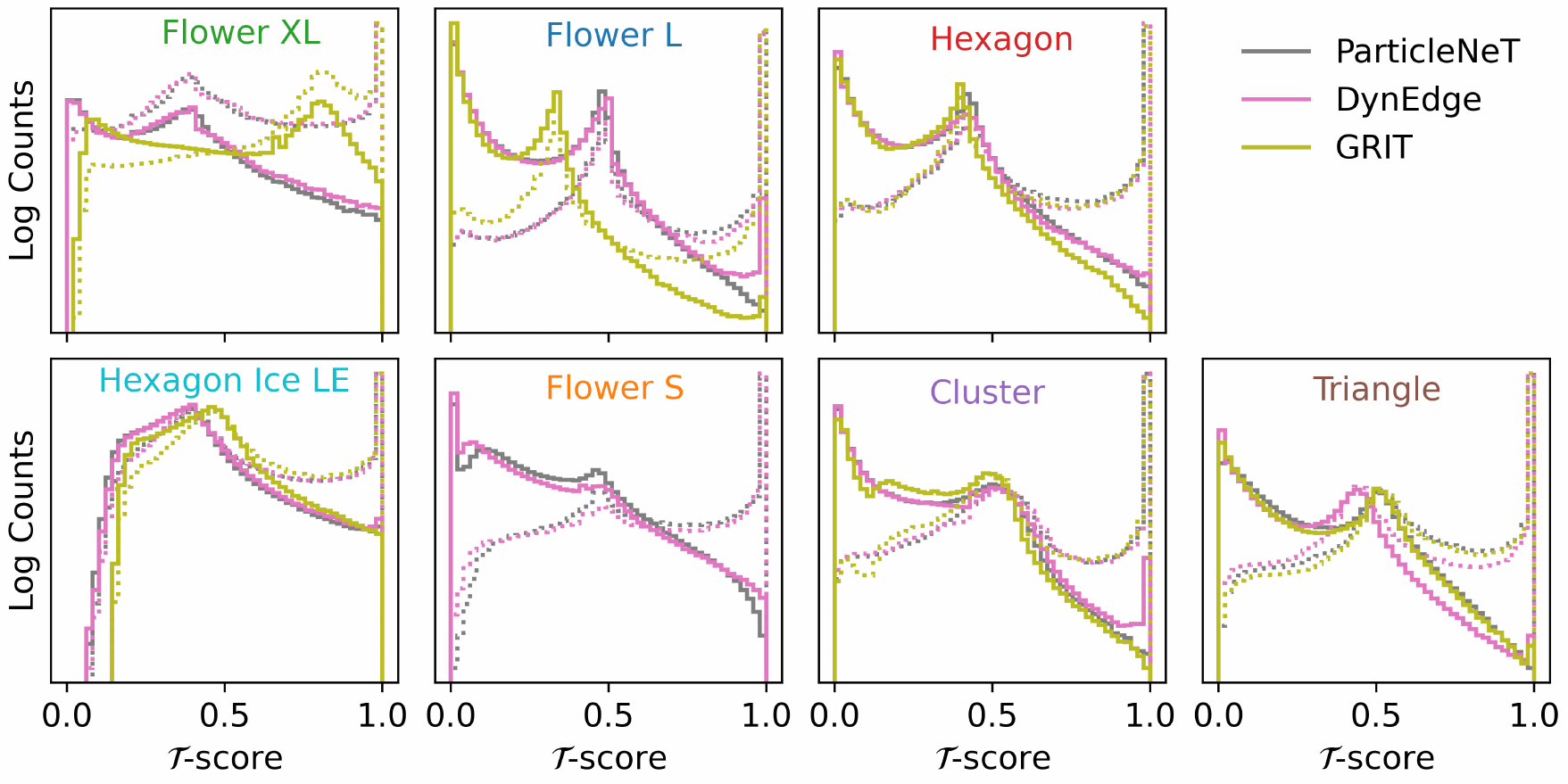}
\caption{Distribution of model predictions on the test partitions of the $\mathcal{T}$ datasets for the $\mathcal{T}/\mathcal{C}$ classification task. Scores close to 1 represent confident placement into the $\mathcal{T}$-category, whereas scores close to 0 indicate confident $\mathcal{C}$-categorization.}
\label{fig:track_score_distributions}
\end{figure}

In \cref{fig:track_score_distributions}, the model predictions across the whole test partition of each of the seven datasets are shown in separate panels.
Each distribution is partitioned into $\mathcal{C}$-events (solid) and $\mathcal{T}$-events (dashed), and predictions from \particlenet{}, \dynedge{} and \grit{} are shown in grey, purple and olive, respectively, and in log10 counts.
From \cref{fig:track_score_distributions} a few key insights about the models, their training procedure, and dataset differences can be seen.
First, most distributions in \cref{fig:track_score_distributions} contain three distinct modalities around 0, 0.5, and 1.
Events concentrated around 0 and 1 represent events that were confidently assigned to the $\mathcal{C}$ and $\mathcal{T}$ categories, respectively.
In contrast, the center modality around 0.5 contains events that the model was unable to assign a confident score to.
This group of events typically contains examples that are challenging to classify due to, for example, the energy of the incident neutrino being too low to induce a distinctive difference in morphology, or because the event is only partly contained within the detector volume.
When comparing the center modalities between models, it can be seen in \cref{fig:track_score_distributions} that in the case of \grit{}, the location is significantly shifted in some datasets. To first order, the location of the middle modality is largely given by the positive-to-negative example ratio in the training dataset.
Because both  \particlenet{} and \dynedge{} were trained on a subsample of the training partitions that was constructed to contain an equal number of $\mathcal{T}$ and $\mathcal{C}$ events, the middle modality is roughly centered around 0.5 for both models in the seven datasets.
Because \grit{} was trained on the entire training partition, the location is driven by the morphology ratio of each dataset.
In the case of \flowerxl{} the ratio of $\mathcal{T}$-to-$\mathcal{C}$ events is 88\%, which is roughly the location of the modality seen in \cref{fig:track_score_distributions}.
Second, when comparing predictions on \hexagon{}, which is simulated in water (10 GeV - 100 TeV), against predictions on \hexagonice{}, which is simulated in ice (10 Gev - 1 TeV), significant differences can be seen in the concentration of confident $\mathcal{C}$-predictions.
These differences are primarily induced by variations in dataset energy ranges and differences in simulation techniques, as water-based simulation in \prometheus{} provides full angular coverage on an OM level.

To produce a binary categorization of neutrino events into the $\mathcal{T} / \mathcal{C}$ categories, a threshold in the continuous scores seen in \cref{fig:track_score_distributions} has to be defined. Events below this choice in threshold are defined as $\mathcal{C}$ events, whereas events at or above are defined as $\mathcal{T}$ events.
As seen in \cref{fig:track_score_distributions}, different choices in the threshold yield a different number of true and false positives, and the optimal choice is therefore  problem-dependent.

\begin{figure}[htbp]
    \centering
    \includegraphics[width=\textwidth]{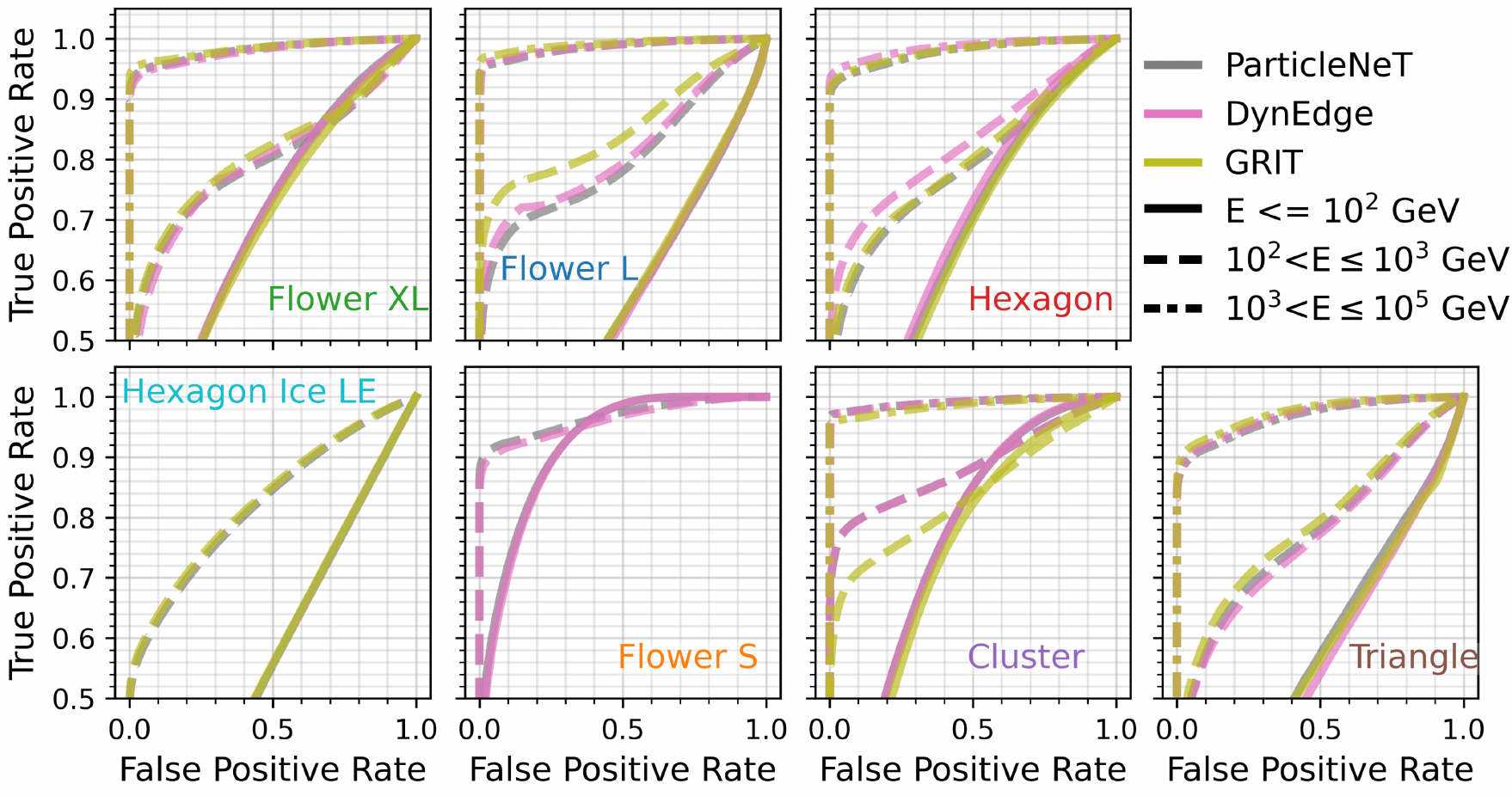}
    \caption{Receiver-Operator-Characteristic (ROC) curves for $\particlenet$, $\dynedge$ and $\grit$ on the seven datasets for the $\mathcal{T}/\mathcal{C}$ classification task.
    Curves are provided for three distinct energy ranges of $E \leq 10^2$ GeV, $10^2 < E \leq 10^3$ GeV, and $10^3 < E \leq 10^5$ GeV.
    Note that \grit{} was not trained on \flowers{} due to the computational demand.
    }
    \label{fig:roc_panel}
\end{figure}

\noindent A standard procedure to assess the performance of a binary classifier across the threshold range is to record the false positive rate (FPR) and true positive rate (TPR) at each threshold, yielding the receiver operating characteristic (ROC) curves, allowing for direct comparison of classifiers across the threshold range.
In our case, false positives represent cases where $\mathcal{C}$ events are misclassified as $\mathcal{T}$ events, and true positives denote correctly classified $\mathcal{T}$ events.
The area under the ROC curve (AUC) is a standard metric that summarizes the discrimination performance visualized by ROC curves to a single number.
A random classifier yields an AUC score of 0.5, whereas perfect separation yields a score of 1.0 \cite{roc}.
Because the difficulty of the $\mathcal{T}/\mathcal{C}$ classification task depends on the energy of the incident neutrino, we provide the ROC curves (\cref{fig:roc_panel}) and AUC scores  (\cref{tab:track_cascade}) for each model in the energy ranges $E \leq 10^2$ GeV, $10^2 < E \leq 10^3$ GeV and $10^3 < E \leq 10^5$ GeV. The datasets \flowers{} and \hexagonice{} do not contain events in the highest energy range; therefore, model performance is quantified in only the first two ranges for those datasets. 
In \cref{fig:roc_panel}, the ROC curves for the three energy ranges are shown for \particlenet{}, \dynedge{}, and \grit{} on the test partitions of the seven datasets.
The three energy ranges are depicted with solid ($E \leq 10^2$ GeV), dashed ($10^2 < E \leq 10^3$ GeV) and dash-dot-dash ($10^3 < E \leq 10^5$ GeV).
Note that \grit{} is not available on the \flowers{} dataset due to computational demand.
As seen in \cref{fig:roc_panel}, a significantly higher TPR is achieved at lower FPR on events in the higher energy range compared to events in the lower energy range.
This finding is consistent with our expectations, as the morphological differences between the $\mathcal{T}$ and $\mathcal{C}$ interactions increase with the energy of the incident neutrino.
The energy regime required to measure distinctively different morphologies largely depends on the detector geometry and, especially, the density of optical instrumentation within the detector volume.
For example, in the \flowers{} detector, which has the highest density of optical instrumentation in our study, both \particlenet{} and \dynedge{} can correctly categorize 90\% of $\mathcal{T}$ events in the  $E\leq 100$ GeV energy range at a false positive rate of less than 30\%.
Compared with other detector geometries in our study, it is clear that \flowers{} has a significantly enhanced sensitivity to events in this energy range.
When comparing the performance of the three models, it is clear from \cref{fig:roc_panel} that each model performs well for the $\mathcal{T}/\mathcal{C}$ classification task, and on the datasets \flowerxl{}, \hexagonice{}, and \flowers{} only minor deviations in performance can be seen.
In the remaining datasets, larger differences in performance are observed between the models. For example, in the \flowerl{} dataset, performance on events at or below 100 GeV is virtually identical, but \grit{} appears significantly better in the $10^2 < E \leq 10^3$ GeV range.
The opposite can be observed in the $10^2 < E \leq 10^3$ GeV range of \cluster{}, where \particlenet{} and \dynedge{} perform significantly better than \grit{}.

\cref{tab:track_cascade} shows the AUC scores for each model on the seven datasets. The first column represents the AUC score computed across all events in each test partition of the datasets.
In the remaining columns, the AUC score is computed on neutrino events with energies in the $E \leq 10^2$ GeV, $10^2 < E \leq 10^3$ GeV, and $10^3 < E \leq 10^5$ GeV ranges.
\cref{tab:track_cascade} shows that \grit{} has the best overall AUC score on four of the six datasets it was applied to.
In comparison, \particlenet{} and \dynedge{} hold the best overall AUC scores on two and one of the seven datasets, respectively.
The largest lead in overall AUC score for \grit{} as compared to the two other models is seen on \flowerxl{} and \flowerl{}, which are also the datasets with the highest class imbalance between $\mathcal{T}$ and $\mathcal{C}$ events.
As a result, the loss of training examples from the balancing scheme employed in the training procedures of both \particlenet{} and \dynedge{}, which aims to obtain an equal number of $\mathcal{T}$ and $\mathcal{C}$ training examples through subsampling of the training partition, is highest on these datasets.
Around 7.2 million ($\approx30\%$) and 7.6 million ($\approx76\%$) examples have been omitted from training for \flowerl{} and \flowerxl{} by the subsampling procedure, respectively, which is likely to partly explain the significantly higher AUC score for \grit{} on those datasets.

\begin{table}[H]
    \centering
    \begin{tabular}{ccccc}
    \multicolumn{5}{c}{$\mathcal{T}/\mathcal{C}$ Classification} \\
    \toprule
    Model & AUC & $\text{AUC}_{E \leq 10^2~\text{GeV}}$ & $\text{AUC}_{ 10^2<E\leq 10^3~\text{GeV}}$ & $\text{AUC}_{ 10^3<E\leq 10^5~\text{GeV}}$ \\
    \midrule
    
    \midrule
    \multicolumn{5}{c}{\texttt{Flower XL}} \\
    \midrule
    ParticleNeT & 0.8954 $\pm$ 0.0004 & \textbf{0.6797 $\pm$ 0.0013} & 0.7969 $\pm$ 0.0009 & 0.9816 $\pm$ 0.0003 \\
DynEdge & 0.889 $\pm$ 0.0004 & \textbf{0.678 $\pm$ 0.0014} & 0.797 $\pm$ 0.0012 & 0.9796 $\pm$ 0.0002 \\
GRIT & \textbf{0.9005 $\pm$ 0.0003} & 0.6702 $\pm$ 0.0013 & \textbf{0.8108 $\pm$ 0.0011} & \textbf{0.984 $\pm$ 0.0002}  \\
    
    \midrule
    \multicolumn{5}{c}{\texttt{Flower L}} \\
    \midrule
    ParticleNeT & 0.9308 $\pm$ 0.0002 & 0.5371 $\pm$ 0.0012 & 0.8012 $\pm$ 0.0007 & 0.986 $\pm$ 0.0001 \\
DynEdge & 0.9318 $\pm$ 0.0002 & 0.5327 $\pm$ 0.001 & 0.8136 $\pm$ 0.0007 & 0.9859 $\pm$ 0.0001 \\
GRIT & \textbf{0.9462 $\pm$ 0.0001} & \textbf{0.5462 $\pm$ 0.0013} & \textbf{0.8512 $\pm$ 0.0006} & \textbf{0.9909 $\pm$ 0.0001}  \\
    
    \midrule
    \multicolumn{5}{c}{\texttt{Hexagon}} \\
    \midrule
    ParticleNeT & 0.9177 $\pm$ 0.0002 & 0.656 $\pm$ 0.001 & 0.7898 $\pm$ 0.0005 & 0.9786 $\pm$ 0.0001 \\
DynEdge & \textbf{0.934 $\pm$ 0.0001} & \textbf{0.6719 $\pm$ 0.001} & \textbf{0.8273 $\pm$ 0.0005} & \textbf{0.9853 $\pm$ 0.0001} \\
GRIT & 0.9219 $\pm$ 0.0002 & 0.6444 $\pm$ 0.001 & 0.798 $\pm$ 0.0006 & 0.9801 $\pm$ 0.0001  \\
    
    \midrule
    \multicolumn{5}{c}{\texttt{Hexagon Ice LE}} \\
    \midrule
    ParticleNeT & 0.7739 $\pm$ 0.0003 & 0.5473 $\pm$ 0.0009 & 0.8241 $\pm$ 0.0003 & --\\
DynEdge & 0.7714 $\pm$ 0.0003 & 0.5465 $\pm$ 0.0006 & 0.8215 $\pm$ 0.0003 & --\\
GRIT & \textbf{0.7793 $\pm$ 0.0003} & \textbf{0.5505 $\pm$ 0.0007} & \textbf{0.8295 $\pm$ 0.0003} & -- \\
    
    \midrule
    \multicolumn{5}{c}{\texttt{Flower S}} \\
    \midrule
    ParticleNeT & \textbf{0.9398 $\pm$ 0.0001} & \textbf{0.9198 $\pm$ 0.0002} & \textbf{0.9669 $\pm$ 0.0002} & -- \\
DynEdge & 0.9339 $\pm$ 0.0001 & 0.9168 $\pm$ 0.0002 & 0.9619 $\pm$ 0.0002 & -- \\
GRIT & -- & -- & -- & -- \\
    
    \midrule
    \multicolumn{5}{c}{\texttt{Cluster}} \\
    \midrule
    ParticleNeT & \textbf{0.9242 $\pm$ 0.0002} & \textbf{0.7536 $\pm$ 0.0007} & \textbf{0.8849 $\pm$ 0.0005} & \textbf{0.9908 $\pm$ 0.0001} \\
DynEdge & 0.9223 $\pm$ 0.0001 & \textbf{0.754 $\pm$ 0.0005} & \textbf{0.8852 $\pm$ 0.0004} & 0.9891 $\pm$ 0.0001 \\
GRIT & 0.9014 $\pm$ 0.0002 & 0.7257 $\pm$ 0.0005 & 0.836 $\pm$ 0.0005 & 0.9856 $\pm$ 0.0001  \\
    
    \midrule
    \multicolumn{5}{c}{\texttt{Triangle}} \\
    \midrule
    ParticleNeT & 0.9205 $\pm$ 0.0002 & \textbf{0.5561 $\pm$ 0.0016} & 0.7748 $\pm$ 0.0005 & 0.9667 $\pm$ 0.0002 \\
DynEdge & 0.9207 $\pm$ 0.0001 & 0.5322 $\pm$ 0.0013 & 0.7681 $\pm$ 0.0006 & 0.97 $\pm$ 0.0001 \\
GRIT & \textbf{0.9292 $\pm$ 0.0001} & 0.5479 $\pm$ 0.0011 & \textbf{0.7902 $\pm$ 0.0005} & \textbf{0.9744 $\pm$ 0.0001}  \\
    \midrule
    
    \end{tabular}
    \caption{AUC scores for \particlenet{}, \dynedge{}, and \grit{} on each of the test partitions in the seven datasets. The first column represents the overall AUC score of each model, whereas the remaining AUC scores are computed for the three distinct energy ranges of $E \leq 10^2$ GeV, $10^2 < E \leq 10^3$ GeV, and $10^3 < E \leq 10^5$ GeV. Statistical errors represent one standard deviation and are obtained through bootstrapping. Effects stemming from stochasticity in the training procedure are not included. Entries marked in bold represent the best score, and cases where two scores are statistically compatible, both are marked in bold.\label{tab:track_cascade}}
\end{table}
Interestingly, the \Triangle{} dataset has a similar class imbalance, resulting in an omission of 6.9 million ($\approx 30\%$) training examples by the subsampling procedure, but results in a significantly smaller lead in AUC score for \grit{}. 
This difference could be explained by the significantly smaller detector volume of the \Triangle{} detector, which holds just three lines, thereby significantly constraining the possible event morphologies and benefiting less from increased training data.  

%In comparison, the subsampling procedure on datasets \texttt{Cluster}, \texttt{Hexagon}, and \texttt{Hexagon Ice LE} omits around $2\%$, $4\%$, and $14\%$, respectively.    

\newpage
\subsection{Interaction Vertex}
\label{sec:vertex_comparison}

In this section, we review the performance of \particlenet{}, \dynedge{}, and \grit{} on the interaction vertex reconstruction task. Reconstructing the interaction vertex is a challenging problem in neutrino telescopes, as it depends strongly on detector geometry, optical module density, and event containment. For this task, the three models were trained with different loss functions: \texttt{GaussianNegativeLogLikelihood} (\cref{eq:gnll}), \texttt{EuclideanDistance} (\cref{eq:euclidean_loss}), and \texttt{LogCosh} (\cref{eq:logcosh}), respectively. Each loss penalizes model predictions in a different way, leading to distinctively different optimization problems. For example, \texttt{GaussianNegativeLogLikelihood} requires the model to predict both the vertex position and an associated uncertainty under Gaussian assumptions, similar in spirit to the von Mises–Fisher approach in \cref{eq:vmf}. In contrast, \texttt{EuclideanDistance} penalizes only the straight-line distance between the predicted and true vertex positions in $\mathbb{R}^3$. Further details on the loss functions, models and their training procedures can be found in \cref{sec:models}.

\begin{figure}[htbp]
    \centering
    \includegraphics[width=\textwidth]{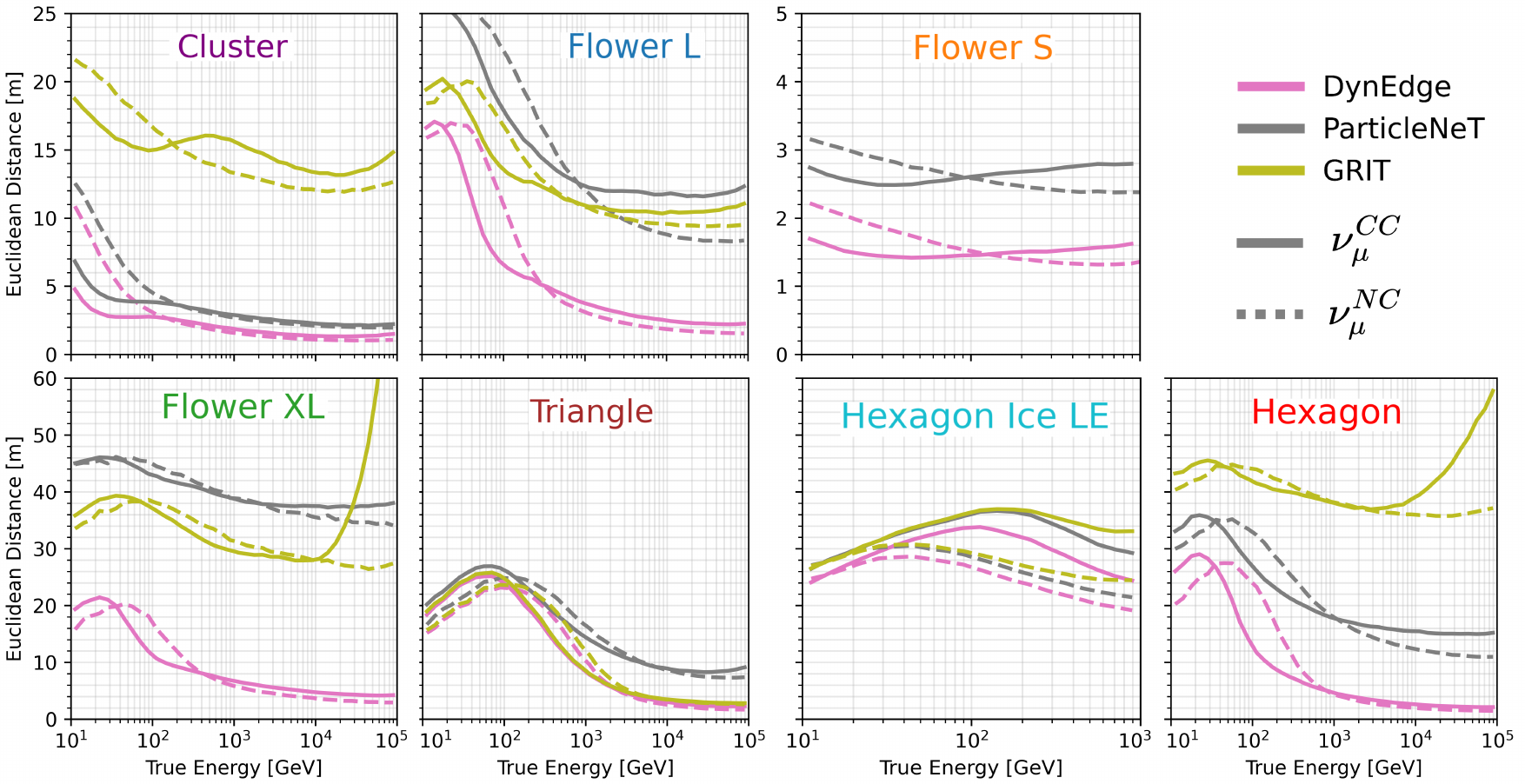}
    \caption{Median Euclidean distance between the true and reconstructed vertex as a function of neutrino energy for the three models across the seven NuBench datasets. Smaller values correspond to more accurate vertex reconstructions. Results are shown separately for $\nu_\mu^{\text{CC}}$ (solid) and $\nu_\mu^{\text{NC}}$ (dotted).}
    \label{fig:median_vertex}
\end{figure}

For evaluating the performance of the models on the vertex reconstruction task, we first use the Euclidean distance between the true and reconstructed vertices, defined as
\begin{equation}    
    D_{xyz} = \sqrt{(x_\mathrm{true} - x_\mathrm{reco})^2 + (y_\mathrm{true} - y_\mathrm{reco})^2 + (z_\mathrm{true} - z_\mathrm{reco})^2}
    \label{eq:euclidean_vertex}
\end{equation}
which is a commonly applied metric to quantify spatial distance in $\mathbb{R}^3$ \cite{icecube_dynedge}. 
In \cref{fig:median_vertex}, the Euclidean distance defined in \cref{eq:euclidean_vertex} is shown as a function of neutrino energy for each dataset, illustrating its energy dependence. The performance of \particlenet{} (grey), \dynedge{} (purple), and \grit{} (green) is displayed, with solid lines representing $\nu_\mu^{\text{CC}}$ events and dotted lines representing $\nu_\mu^{\text{NC}}$ events. 

By comparing the overall scale of the Euclidean distances in \cref{fig:median_vertex}, a clear relationship emerges between the density of optical instrumentation and the difficulty of vertex reconstruction. Detectors with closer OM spacing, such as \flowers{} and \cluster{}, tend to achieve lower errors than those with larger inter-string distances. In particular, the \flowers{} detector shows consistently strong performance across the entire energy range, reflecting the advantages of its high instrumentation density. While a clear difference in the median Euclidean distance can be seen between $\nu_\mu^\text{CC}$ and $\nu_\mu^\text{CC}$ events, the difference is relatively smaller than what can be observed for energy and direction reconstruction in \cref{sec:energy_comparison} and \cref{sec:direction_comparison}.

When comparing performance between models, a clear picture emerges in \cref{fig:median_vertex}. The \dynedge{} model consistently achieves the best vertex reconstruction, with its median distance curve lying below those of the other two models across nearly all datasets. The main exception is the \Triangle{} dataset, where \grit{} performs nearly identically for $\nu_\mu^{\text{CC}}$ events and only slightly worse for $\nu_\mu^{\text{NC}}$ events. The largest separation between models appears in the \flowerxl{} dataset, a geometry with 1211 strings and inter-string spacings of around 90~m horizontally and 30~m vertically. Here, the median Euclidean distance of \dynedge{} is roughly a factor of two smaller than that of \grit{}, the next-best model. Given that \particlenet{} and \dynedge{} share similar architectures and employ the same graph convolution, this pronounced performance gap is unexpected. The gap was initially attributed to differences in loss function, since \dynedge{} is trained directly on the Euclidean distance shown in \cref{fig:median_vertex}, whereas \particlenet{} uses \cref{eq:gnll}, which requires the model to produce realistic uncertainties in addition to vertex predictions. Controlled tests were carried out in which \particlenet{} was trained using the Euclidean distance. These experiments showed that this modification did not significantly improve its vertex reconstruction performance. This finding suggests that architectural differences between \particlenet{} and \dynedge{} are the primary cause of the performance gap observed in \cref{fig:median_vertex}.

In the sub-figures of \cref{fig:median_vertex} that represent the \flowerxl{} and \hexagon{} datasets, it can be seen that for the $\nu_\mu^{\text{CC}}$ events at around 10~TeV and above, the \grit{} model provides significantly less accurate vertex reconstructions. This behavior is unique to \grit{} and limited to these two datasets. Since the training procedure of \grit{} does not rebalance the fraction of $\nu_\mu^{\text{CC}}$ and $\nu_\mu^{\text{NC}}$ events—as is done in \particlenet{} and \dynedge{}—one might expect a bias toward $\nu_\mu^{\text{CC}}$ in \flowerxl{}, which contains about 88\% $\nu_\mu^{\text{CC}}$ events. Yet, contrary to this expectation, the median Euclidean distance increases with energy for $\nu_\mu^{\text{CC}}$ events, while $\nu_\mu^{\text{NC}}$ events are unaffected. The same trend is observed in \hexagon{}, where the $\nu_\mu^{\text{CC}}$ to $\nu_\mu^{\text{NC}}$ ratio is roughly equal, indicating that morphology imbalance is not the cause. A more plausible explanation is the choice of loss function: \texttt{LogCosh} is effectively linear for large residuals, in contrast to quadratic losses such as \cref{eq:gnll} or MSE that penalize large deviations more heavily. High-energy, uncontained tracks, where the true vertex lies far from the first observed pulse, could therefore be under-penalized by \grit{}. However, since \flowerxl{} and \hexagon{} do not contain a substantially larger fraction of such events compared to the other datasets, the degradation is more likely attributed to the architecture or its training procedure.
%The behavior is likely a result of the differences in the ratio of $\nu_\mu^{\text{CC}}$ to $\nu_\mu^{\text{NC}}$ events between the datasets, which was also discussed in \cref{sec:track_cascade_classification_comparison}.  
%As seen from \cref{table:datasets}, just 88\%  and 48\% of events are $\nu_\mu^{\text{CC}}$ in \flowerxl{} and \hexagon{}, respectively. The training strategies of \particlenet{} and \dynedge{} includes a procedure that seeks to balance the number of $\nu_\mu^{\text{CC}}$ and $\nu_\mu^{\text{NC}}$ events in the training dataset through subsampling, yielding a smaller but balanced training dataset. Such a procedure is not employed in the training of \grit{}, which could explain the bias seen Finally, \grit{} model exhibit a increase worsening when increasing in energy---as it is seen in \flowerxl{} and \hexagon{}---which does not happen at other models and that can be explained by its training strategy, where an unbalanced sample with a large number of tracks is considered, hence, worsening the reconstruction.\\

\begin{figure}[htbp]
    \centering
    \includegraphics[width=\textwidth]{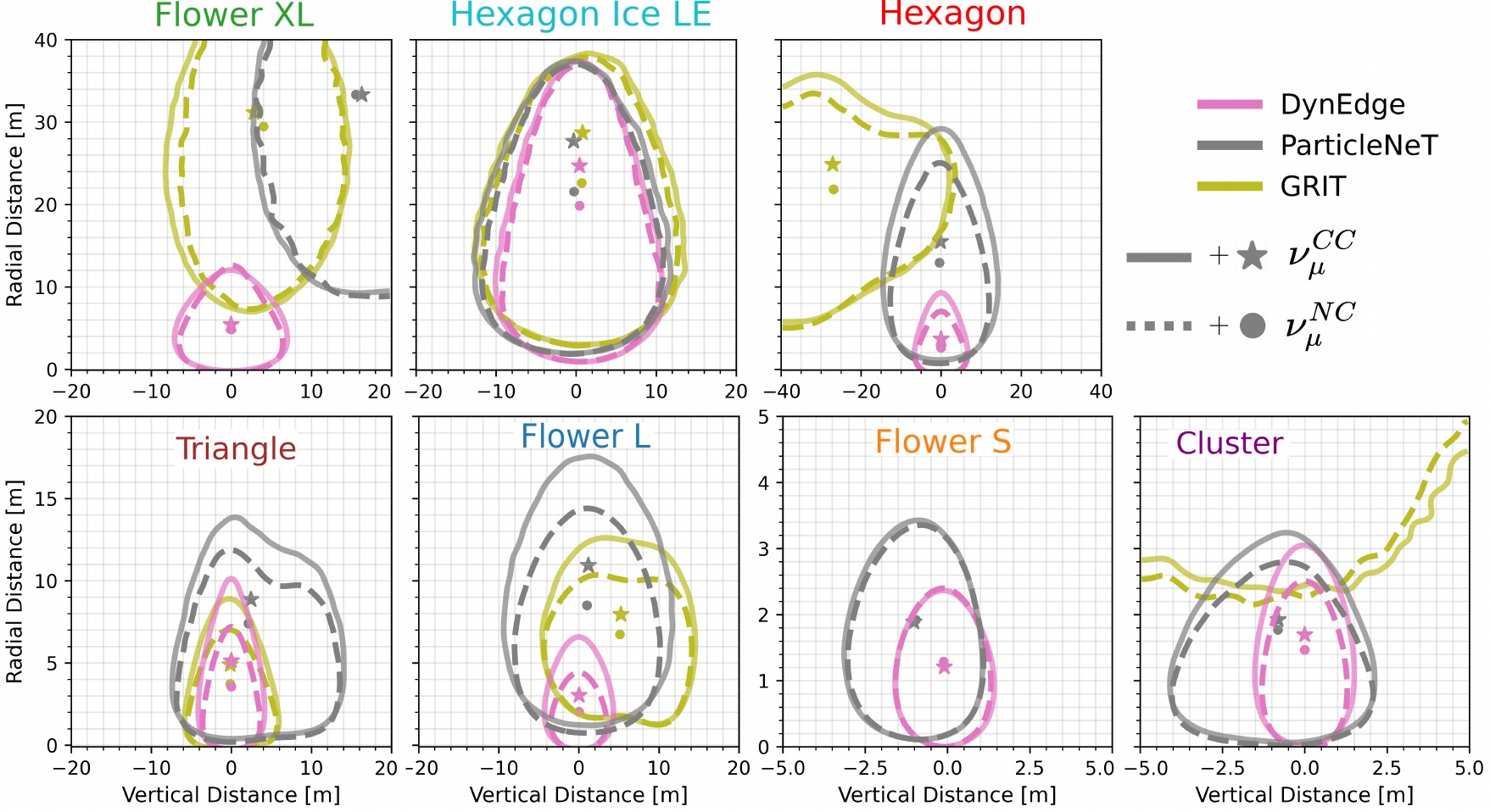}
    \caption{Error contours for the vertex reconstruction task. The markers indicate the median error, and the contours show the 68\% quantile distributions. Smaller contour areas correspond to lower variance, while centers close to zero indicate reduced bias. Results are shown separately for $\nu_\mu^\text{CC}$ (solid lines, "$\star$") and $\nu_\mu^\text{NC}$ (dashed lines, "$\medbullet$").}
    \label{fig:residuals_vertex}
\end{figure}

In addition to considering the full Euclidean distance, which provides a dimension-agnostic measure of the error, we also decompose the Euclidean distance into horizontal ($D_\text{xy}$) and vertical ($D_\text{z}$)  components defined by $D_{xy} = \sqrt{(x_\mathrm{true} - x_\mathrm{reco})^2 + (y_\mathrm{true} - y_\mathrm{reco})^2}$ and $D_{z} = (z_\mathrm{true} - z_\mathrm{reco})$. The horizontal component ($D_\text{xy}$) corresponds to the horizontal contribution to the overall Euclidean distance, as shown in \cref{fig:residuals_vertex}, while the vertical component ($D_\text{z}$)  quantifies the error along the vertical dimension of the vertex. Unlike  $D_\text{xy}$, the vertical component may take on negative values, which represent overestimation, whereas positive values indicate underestimation. By showing the horizontal error against vertical error, contours are obtained with an intuitive interpretation: the contour area reflects the variance of predictions, and the contour center indicates bias. In \cref{fig:residuals_vertex}, we show the median error and 68\% contours for $\nu_\mu^\text{CC}$ ($\nu_\mu^\text{NC}$) events separately, using "$\star$" ("$\medbullet$") and solid (dashed) lines.

By decomposing the full Euclidean distance into horizontal and vertical components, several nuances in the model and geometry comparisons emerge, as shown in \cref{fig:residuals_vertex}. Overall, $\nu_\mu^\text{NC}$ events tend to exhibit slightly narrower contours than $\nu_\mu^\text{CC}$ events, though the two morphologies remain strongly correlated in area and shape. In the \Triangle{} dataset, \dynedge{} and \grit{} produce contours that are narrow in the vertical direction but elongated horizontally. In contrast, \particlenet{} yields an asymmetric, multi-modal contour with large vertical variance. On most datasets, all three models achieve vertical errors close to zero, indicating that estimating vertex depth is generally less challenging than reconstructing the horizontal component, which is expected, as the horizontal distance between OMs is typically larger than their vertical separation. However, notable exceptions exist: for the \hexagon{} dataset, predictions from \grit{} are heavily biased in both components, while for \flowerxl{} a similar bias is observed in \particlenet{}. Across all datasets, \dynedge{} consistently produces the smallest contour areas, indicating the lowest overall variance.  
\begin{table}[H]
    \centering
    \begin{tabular}{ccccccc}
    \multicolumn{7}{c}{Vertex Reconstruction} \\
     \toprule
    Model & \multicolumn{3}{|c|}{$E \leq 10^3~\text{GeV}$}
    & \multicolumn{3}{|c}{$10^3 < E \leq 10^5~\text{GeV}$} \\
    \midrule
    & $D_\text{xyz}$ [m]& $|D_\text{z}|$ [m]& \multicolumn{1}{c|}{$D_\text{xy}$ [m]} & $D_\text{xyz}$ [m]& $|D_\text{z}|$ [m]& \multicolumn{1}{c}{$D_\text{xy}$ [m]} \\

    %\midrule

    \midrule
    \multicolumn{7}{c}{\texttt{Flower XL}} \\
    \midrule
    \particlenet{} & 42.9 & 16.13 & 36.77 & 36.82 & 16.55 & 30.54 \\
\dynedge{} & \textbf{11.98} & \textbf{3.04} & \textbf{10.06} & \textbf{4.31} & \textbf{1.8} & \textbf{3.35} \\
\grit{} & 35.51 & 7.68 & 32.86 & 30.0 & 6.91 & 28.09  \\
    
    \midrule
    \multicolumn{7}{c}{\texttt{Flower L}} \\
    \midrule
    \particlenet{} & 17.47 & 6.15 & 14.09 & 10.2 & 4.36 & 7.84 \\
\dynedge{} & \textbf{6.52} & \textbf{1.96} & \textbf{5.28} & \textbf{2.2} & \textbf{0.88} & \textbf{1.75} \\
\grit{} & 13.59 & 6.5 & 9.69 & 10.08 & 6.01 & 6.46  \\
    
    \midrule
    \multicolumn{7}{c}{\texttt{Hexagon}} \\
    \midrule
    \particlenet{} & 25.46 & 6.96 & 22.0 & 13.99 & 5.08 & 11.45 \\
\dynedge{} & \textbf{12.02} & \textbf{2.09} & \textbf{10.33} & \textbf{2.37} & \textbf{0.87} & \textbf{1.93} \\
\grit{} & 41.3 & 26.43 & 25.72 & 38.2 & 28.14 & 22.06  \\
    
    \midrule
    \multicolumn{7}{c}{\texttt{Hexagon Ice LE}} \\
    \midrule
    \particlenet{} & 28.73 & 8.49 & 24.2 & -- & -- & -- \\
\dynedge{} & \textbf{26.04} & \textbf{6.84} & \textbf{21.94} & -- & -- & -- \\
\grit{} & 30.11 & 9.39 & 25.25 & -- & -- &  -- \\
    
    \midrule
    \multicolumn{7}{c}{\texttt{Flower S}} \\
    \midrule
    \particlenet{} & 2.64 & 1.3 & 1.9 & -- & -- & -- \\
\dynedge{} & \textbf{1.55} & \textbf{0.63} & \textbf{1.25} & -- & -- & -- \\
\grit{} & -- & -- & -- & -- & -- & --  \\
    
    \midrule
    \multicolumn{7}{c}{\texttt{Cluster}} \\
    \midrule
    \particlenet{} & 4.3 & 1.84 & 3.16 & 2.25 & 1.45 & 1.15 \\
\dynedge{} & \textbf{2.97} & \textbf{0.93} & \textbf{2.52} & \textbf{1.29} & \textbf{0.44} & \textbf{1.09} \\
\grit{} & 16.43 & 5.04 & 14.23 & 13.06 & 4.31 & 10.89  \\
    
    \midrule
    \multicolumn{7}{c}{\texttt{Triangle}} \\
    \midrule
    \particlenet{} & 21.74 & 6.67 & 18.27 & 9.27 & 4.56 & 5.88 \\
\dynedge{} & \textbf{18.37} & \textbf{2.97} & \textbf{16.07} & \textbf{2.9} & \textbf{0.81} & 2.54 \\
\grit{} & 19.32 & 3.66 & 16.96 & 3.6 & 1.95 & \textbf{2.34}  \\
    \midrule
    
    \end{tabular}
  \caption{Selected performance metrics for vertex reconstruction. $D_{\text{xyz}}$ denotes the full Euclidean distance, whereas $D_{\text{z}}$ and $D_{\text{xy}}$ represent the depth and radial components of the Euclidean distance, respectively. The metrics are provided for the two energy ranges $ E \leq 10^3$ GeV and $10^3 < E \leq 10^5$ GeV. Statistical errors are small at $\mathcal{O}(10^{-3})$ and have been omitted for brevity.\label{tab:vertex}}
\end{table}

To further quantify the difference between \particlenet{}, \dynedge{} and \grit{} on the vertex reconstruction task, selected metrics are provided in \cref{tab:vertex}. The metrics include the median Euclidean distance ($D_\text{xyz}$), the median, absolute vertical ($D_\text{z}$) distance and the median horizontal ($D_\text{xy}$) distance. The metrics are given for the two energy ranges of $E\leq 10^3$ GeV and $10^3 < E \leq 10^5$ GeV to account for the different energy ranges in the NuBench datasets. Statistical errors are of $\mathcal{O}(1e-3)$ and omitted for brevity.

\newpage
\subsection{Inelasticity}
\label{sec:inelasticity_comparison}
In this section, we evaluate the performance of \particlenet{}, \dynedge{}, and \grit{} on the task of reconstructing the visible inelasticity,
$y_\text{vis} = \frac{E_X^\text{vis}}{E_\ell^\text{vis} + E_X^\text{vis}}$
where $E_X^\text{vis}$ and $E_\ell^\text{vis}$ denote the visible hadronic and leptonic energy components, respectively. 
As discussed in \cref{sec:tasks}, this definition of inelasticity applies only to CC interactions; consequently, the comparison in this section is restricted to $\nu_\mu^\text{CC}$ events. 
Furthermore, the \hexagonice{} dataset is excluded from the comparison, as its distribution of $y_\text{vis}$ does not span the full numerical range $[0,1]$ observed in \cref{fig:truth_variables}.
 
The difficulty of reconstructing the visible inelasticity largely depends on the spatial separation between the hadronic and leptonic components of the visible neutrino energy within the detector volume. This separation, in turn, is primarily determined by the energy of the incident neutrino and the degree of event containment. Accordingly, the reconstruction error, defined in this section as
\begin{equation}
    R_y = |y_\text{reco} - y_\text{vis}|,
\end{equation}
tends to decrease with increasing neutrino energy, since distinguishing the hadronic and leptonic components becomes particularly challenging at lower energies.

\begin{figure}[h!]
    \centering
    \includegraphics[width=1\textwidth]{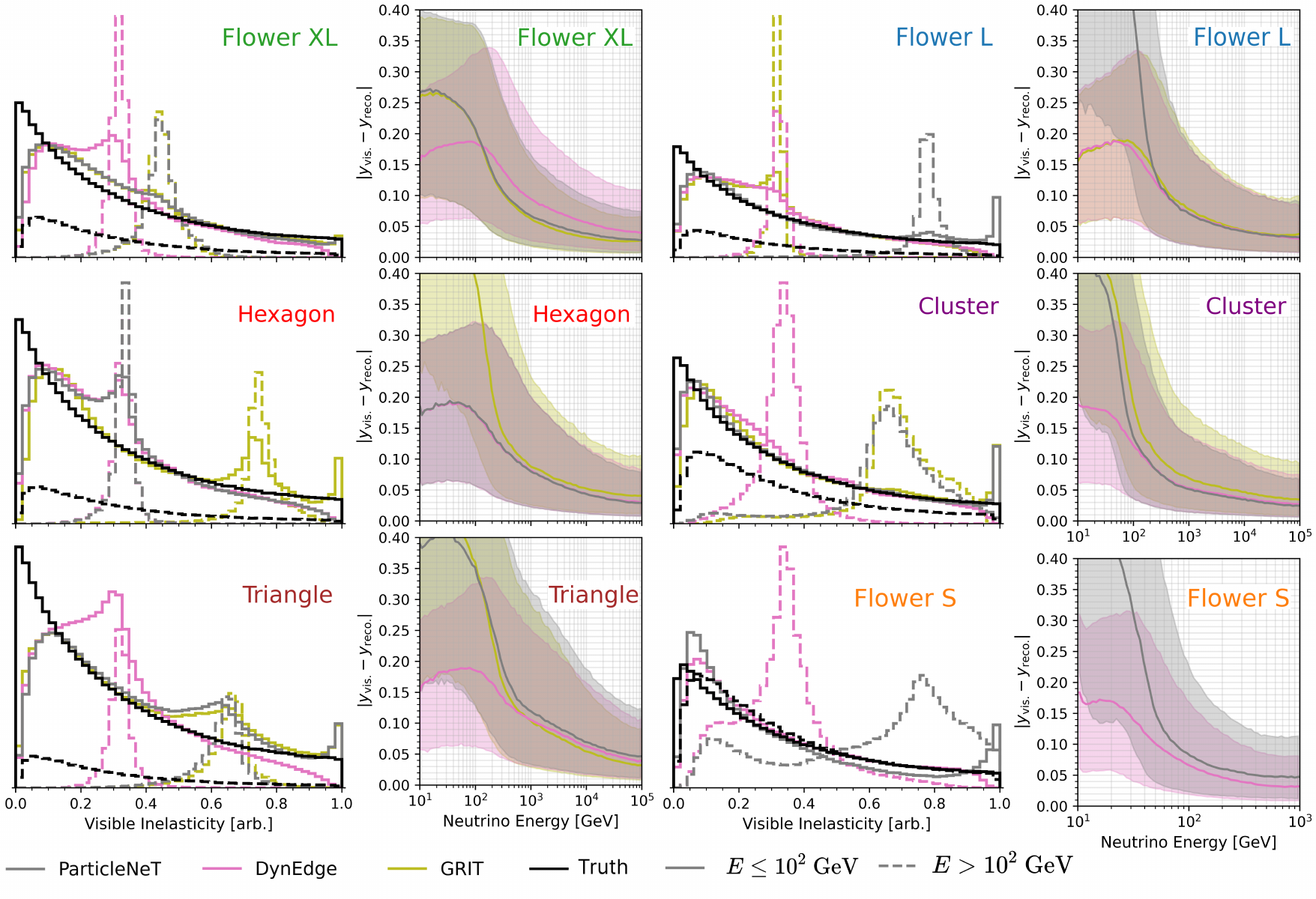}
    \caption{Performance of \particlenet{}, \dynedge{}, \grit{}, and \deepice{} on the reconstruction of visible inelasticity for the six water-based NuBench datasets. 
    The resolution figures show the median absolute error ($|y_\text{reco} - y_\text{vis}|$) as a function of neutrino energy, with shaded bands indicating the 16th–84th percentile range. 
    The distributions show reconstructed $y_\text{reco}$ compared to the true $y_\text{vis}$ for two energy ranges: $E \leq 10^2$~GeV (solid) and $E > 10^2$~GeV (dashed).}
    \label{fig:inelasticity_panel}  
\end{figure}

To compare model performance, we report the median $R_y$ (solid lines) as a function of neutrino energy, along with the 16th–84th percentile range (shaded bands) in \cref{fig:inelasticity_panel}, which indicate model bias and variance, respectively. 
Additionally, the distributions of model predictions are shown alongside the true $y_\text{vis}$ distributions for the two energy ranges $E \leq 10^2$~GeV and $E > 10^2$~GeV.

For this task, the models were trained with different loss functions: \particlenet{} uses the mean squared error loss (\cref{eq:MSE}), while both \dynedge{} and \grit{} employ the \texttt{LogCosh} loss (\cref{eq:logcosh}) combined with a sigmoid activation on the model output. 

When comparing the distributions of reconstructed and true inelasticity in \cref{fig:inelasticity_panel}, it can be seen that all models tend to produce multimodal predictions, similar to the morphology classification results shown in \cref{fig:track_score_distributions}. 
For instance, in the \hexagon{} dataset, predictions in the $E > 10^2$~GeV range (solid) exhibit a pronounced excess around $y_\text{vis} \approx 0.3$ for both \particlenet{} and \dynedge{}, while \grit{} shows an excess near $y_\text{vis} \approx 0.7$. 
Notably, these excesses coincide with the dominant modes of the lower-energy ($E \leq 10^2$~GeV) distributions, suggesting that the multimodal behavior is primarily driven by low-energy events for which inelasticity reconstruction is most challenging. 
In these cases, the models appear to favor narrow, locally optimal prediction ranges that minimize the overall loss across such events. The fact that the modal positions overlap between models for some datasets but diverge for others indicates that the mode location is not primarily driven by the mean target value, as was the case in the modalities observed in \cref{fig:track_score_distributions}. 
Instead, inspection of the energy-dependent reconstruction error suggests that the mode location reflects model-specific optimization strategies at different neutrino energies. 
This effect is particularly visible in the \flowerxl{} dataset, where both \particlenet{} and \grit{} exhibit higher bias and variance than \dynedge{} between $10^1$ and $10^2$~GeV, but the inverse trend is observed at higher energies.

Geometry-dependent effects are also apparent in the resolution curves. 
For example, the median error between $10^2$ and $10^3$~GeV is substantially lower in the \flowers{} dataset—whose high optical module density enhances spatial resolution—than in detectors with sparser instrumentation such as \flowerxl{} or \Triangle{}. 
This trend is expected, as resolving the spatially distinct hadronic and leptonic components of the visible energy requires higher optical density, especially for low-energy events.

To summarize the comparison between models, performance metrics showing the median $R_y$ and the 84th–16th percentile width of $R_y$ ($\sigma$) are provided in \cref{tab:inelasticity}. 
To account for the clear energy dependence of these metrics, the quantities are reported for three energy ranges: $10^1 \leq E \leq 10^2$~GeV, $10^2 < E \leq 10^3$~GeV, and $10^3 < E \leq 10^5$~GeV. 
The statistical uncertainties on the reported metrics are small, on the order of $\mathcal{O}(10^{-4})$, and are omitted for brevity.

From \cref{tab:inelasticity}, it can be seen that models achieving the lowest median $R_y$ also tend to exhibit the smallest $\sigma$. 
In the lowest energy range ($10^1 \leq E \leq 10^2$~GeV), \dynedge{} generally yields the most accurate reconstructions across all six datasets, with the sole exception of \flowerl{}, where \grit{} achieves a slightly smaller $\sigma$. 
At intermediate energies ($10^2 < E \leq 10^3$~GeV), \particlenet{} and \grit{} perform best on \cluster{} and \flowerxl{}, respectively, while \dynedge{} outperforms the other models on the remaining datasets. 
In the highest energy range ($10^3 < E \leq 10^5$~GeV), \grit{} and \particlenet{} tend to provide the most accurate predictions overall.

\begin{table}[H]
    \centering
    \begin{tabular}{ccccccc}
    \multicolumn{7}{c}{Inelasticity Reconstruction} \\
     \toprule
    Model & \multicolumn{2}{|c|}{$10^1 \leq E \leq 10^2~\text{GeV}$}
    & \multicolumn{2}{|c}{$10^2 < E \leq 10^3~\text{GeV}$}  & \multicolumn{2}{|c}{$10^3 < E \leq 10^5~\text{GeV}$}\\
    \midrule
    & Median  $R_y$ & $\sigma$ & \multicolumn{1}{|c}{Median $R_y$} & $\sigma$ & \multicolumn{1}{|c}{Median $R_y$}  & \multicolumn{1}{c}{$\sigma$} \\

    %\midrule

    \midrule
    \multicolumn{7}{c}{\texttt{Flower XL}} \\
    \midrule
    \particlenet{} & 0.249 & 0.2981 & 0.1124 & 0.1904 & 0.0398 & 0.0958 \\
\dynedge{} & \textbf{0.181} & \textbf{0.237} & 0.1399 & 0.2423 & 0.0577 & 0.1348 \\
\grit{} & 0.2467 & 0.2932 & \textbf{0.1088} & \textbf{0.1843} & \textbf{0.0345} & \textbf{0.0836}  \\
    
    \midrule
    \multicolumn{7}{c}{\texttt{Flower L}} \\
    \midrule
    \particlenet{} & 0.5185 & 0.4921 & 0.127 & 0.2239 & \textbf{0.0436} & \textbf{0.1053} \\
\dynedge{} & \textbf{0.1812} & 0.2405 & \textbf{0.1107} & \textbf{0.2065} & \textbf{0.0435} & 0.1071 \\
\grit{} & 0.1821 & \textbf{0.2352} & 0.1231 & 0.2197 & 0.0464 & 0.1124  \\
    
    \midrule
    \multicolumn{7}{c}{\texttt{Hexagon}} \\
    \midrule
    \particlenet{} & 0.1861 & 0.2441 & 0.1193 & 0.2178 & \textbf{0.0446} & \textbf{0.1145} \\
\dynedge{} & \textbf{0.1851} & \textbf{0.2438} & \textbf{0.1182} & \textbf{0.2171} & 0.0454 & 0.1167 \\
\grit{} & 0.4912 & 0.484 & 0.1554 & 0.2798 & 0.0545 & 0.131  \\
    
    \midrule
    \multicolumn{7}{c}{\texttt{Flower S}} \\
    \midrule
    \particlenet{} & 0.2653 & 0.5384 & 0.0554 & 0.0811 & -- & -- \\
\dynedge{} & \textbf{0.1276} & \textbf{0.2453} & \textbf{0.0405} & \textbf{0.0713} & -- &  --\\
\grit{} & -- & -- & -- & -- &  -- &  -- \\
    
    \midrule
    \multicolumn{7}{c}{\texttt{Cluster}} \\
    \midrule
    \particlenet{} & 0.3591 & 0.4713 & \textbf{0.077} & \textbf{0.1193} & \textbf{0.0345} & \textbf{0.0858} \\
\dynedge{} & \textbf{0.1712} & \textbf{0.2568} & 0.0775 & 0.1468 & 0.0369 & 0.0908 \\
\grit{} & 0.3962 & 0.4659 & 0.1034 & 0.1512 & 0.0469 & 0.1123  \\
    
    \midrule
    \multicolumn{7}{c}{\texttt{Triangle}} \\
    \midrule
    \particlenet{} & 0.3858 & 0.4044 & 0.1987 & 0.2984 & 0.0705 & 0.1796 \\
\dynedge{} & \textbf{0.1846} & \textbf{0.2401} & \textbf{0.141} & \textbf{0.2411} & 0.0629 & 0.1554 \\
\grit{} & 0.4061 & 0.4158 & 0.1769 & 0.2818 & \textbf{0.055} & \textbf{0.1524}  \\
    \midrule
    
    \end{tabular}
  \caption{Selected performance metrics for inelasticity reconstruction. The median $R_y$ along with the 84th-16th percentile width of $R_y$ ($\sigma$) are provided for the three energy ranges $ 10^1 \leq E \leq 10^2$ GeV, $10^2 < E \leq 10^3$ GeV and $10^3 < E \leq 10^5$. Statistical errors are small at $\mathcal{O}(10^{-4})$ and have been omitted for brevity.\label{tab:inelasticity}}
\end{table}

\section{Conclusion}
\label{sec:conclusion}

This work expands the open-data corpus available to the neutrino telescope community by introducing the \textbf{NuBench} datasets, a collection of seven datasets containing nearly 130 million simulated neutrino events across six detector geometries inspired by existing or proposed neutrino telescopes. 
The datasets are designed for the development and comparison of reconstruction algorithms across different detector layouts and for several key reconstruction tasks of common interest within neutrino telescope collaborations. 
They span neutrino energies from 10~GeV to 100~TeV and include both $\nu_\mu^{\text{CC}}$ and $\nu_\mu^{\text{NC}}$ interactions simulated in water and ice. 

Using the NuBench datasets, we compared up to four reconstruction algorithms across five reconstruction targets: neutrino energy, direction, interaction vertex, inelasticity, and event morphology. 
Two of these algorithms, \particlenet{} and \dynedge{}, are currently used within the IceCube and KM3NeT collaborations, respectively. 
Both are based on graph neural networks and were applied to the complete NuBench catalogue. 
For direction reconstruction, we additionally included \deepice{}, one of the winning transformer-based solutions from the open-data challenge \textit{IceCube -- Neutrinos in Deep Ice}, and \grit{}, a hybrid GNN–attention model evaluated on most datasets and reconstruction tasks. 

In our comparisons, we recover rules of thumb that have governed detector geometry design and its relationship to certain reconstruction tasks for the past decades: in tasks where spatial resolution is critical, such as vertex and inelasticity reconstruction, detectors with high optical module densities perform markedly better than sparser, high-volume geometries. 
Conversely, low-density but large-volume detectors achieve better results for the reconstruction of high-energy track directions. 
When comparing reconstruction methods directly, no single architecture was consistently superior across all tasks or energy ranges. 
Instead, the best method appears to depend on both task and energy range. 
For direction reconstruction, \deepice{} achieved the most accurate results on nearly all datasets, closely followed by \grit{}, which we attribute to their use of dot-product attention---a global operation in contrast to the localized graph convolutions employed by \particlenet{} and \dynedge{}. 
In other tasks where global attention might be expected to provide an advantage, such as track–cascade classification, the performance gap between \particlenet{}, \dynedge{}, and \grit{} was smaller, with no architecture consistently outperforming the others. 
For energy reconstruction, performance among \grit{}, \particlenet{}, and \dynedge{} was strongly correlated, with the best results alternating between \particlenet{} and \dynedge{}, suggesting that global attention mechanisms provide limited benefit for this task. 
For vertex reconstruction, \dynedge{} produced the most accurate results across datasets, outperforming both \particlenet{} and \grit{}. 
This finding is particularly notable given the architectural similarity between \dynedge{} and \particlenet{}, suggesting that even small architectural differences can have significant effects on optimization and reconstruction performance.
While direct comparisons with state-of-the-art likelihood-based reconstruction techniques, such as \cite{icecube_splinempe}, could add further nuance to this study, their computational demands make such comparisons very challenging on the full NuBench catalogue.

Overall, our results demonstrate that deep-learning-based architectures, which are expressive on one detector geometry for a given task, tend to remain effective across other geometries and reconstruction targets. 
This reinforces the importance of cross-experimental collaboration in the development of future reconstruction techniques and highlights the role of open, reproducible benchmarks such as NuBench in advancing neutrino telescope research. Using such benchmarks, future work may seek to expand the comparisons to new model architectures, existing likelihood techniques, and measure generalization between open benchmarks and official collaboration simulations.

\acknowledgments
The NuBench datasets are hosted on the Electronic Research Data Archive (ERDA) provided by the University of Copenhagen (UCPH). We thank UCPH and the ERDA team for supporting open data access through their service and long-term commitments to storing the datasets. Additionally, we'd like to thank Dr. Philipp Eller (TUM) and Professor Troels C. Petersen (UCPH) for their valuable discussions, feedback, and support. The authors also acknowledge Dr. Antonin Vacheret, Directeur de Recherche CNRS at the Laboratoire de Physique Corpusculaire (LPC), who granted access to the local HGX 8x A100 GPU server for parts of the model training shown in this work. This work has been supported by the Deutsche Forschungsgemeinschaft (DFG, German Research Foundation) under the SFB 1258 – 283604770 ‘Neutrinos and Dark Matter in Astro- and Particle Physics’, and the PUNCH4NFDI consortium fund “NFDI 39/1”. The authors also acknowledge the support of MICIU for PRE2022-104211,  PID2021-124591NB-C41 and PID2024-156285NB-C4,1 funded by MICIU/AEI/10.13039/501100011033 and by FEDER, EU, and Generalitat Valenciana for CIPROM/2023/51 Spain. The authors gratefully acknowledge the computer resources at Artemisa and the technical support provided by the Instituto de Fisica Corpuscular, IFIC(CSIC-UV). Artemisa is co-funded by the European Union through the 2014-2020 ERDF Operative Programme of Comunitat Valenciana, project IDIFEDER/2018/048.

\appendix

\bibliographystyle{JHEP}
\bibliography{biblio.bib}

\section{Simulation and Processing}
\label{sec:app_datasets}

\subsection{Particle Physics Simulation}

The first step of the simulation chain is event injection (also called generation). This is a process by which the initial properties of the neutrino and of the interaction are selected. Software addressing this problem has a long history, see, e.g.~\cite{osti_5884484,Hill:1996hzh,Gazizov:2004va,Yoshida:2003js,Bailey:2002,DeYoung:865626,IceCube:2020tcq,KM3NeT:2020tvi}.

While these tools have their differences, e.g., injection medium and the language written in, the most important difference for our case is the injection point. Specifically, if the neutrinos are first injected at the surface of the Earth and then propagated to the detector, or injected at the detector and Earth effects are included after the fact via an \textit{a posteriori} weight. Here, it is important to choose a software that uses the latter approach, as using the former might cause the ML-based reconstruction to learn about a mismodeling of the Earth rather than local details of the detector.
% This was not a practical choice until software that could efficiently compute this propagation, such as, e.g. \texttt{NuSQuIDS}~\cite{Arguelles:2021twb}.
For this reason, the \prometheus{} used in this work relies on \texttt{LeptonInjector}~\cite{IceCube:2020tcq,LeptonInjectorRepository}, a tool that injects neutrinos directly into the detector's simulation volume (which is typically larger than the detector itself).

After the injection, the produced secondaries ($\mu^\pm$, $e^\pm$, $\tau^\pm$, and hadronics) are propagated through the detector medium. Of particular interest are the $\mu$ and $\tau$ due to their long propagation distances. While there are a few codes that have been developed for the propagation of muons (e.g.~\cite{Lipari:1991ut, Lohmann:1985qg, Desiati:2002xg, Chirkin:2004hz}), \prometheus{} uses \texttt{PROPOSAL}~\cite{Koehne:2013gpa, dunsch_2018_proposal_improvements, dunsch_2020_1484180}, which is currently used by the neutrino observatories in their simulation chains and still actively maintained. \texttt{PROPOSAL} is capapble of propagating both $mu$ and $\tau$, but at the energies considered in this work ($\leq 100$ TeV), the $\tau$ propagation can also mostly be neglected.

On the other hand at the scales relevant here, $e$ and hadronics almost instantly produce a "cascade". These are nearly pointlike energy depositions with an energy-dependent longitudinal extension with length $\mathcal{O}(\SI{5}{\m}--\SI{10}{\m})$.

\subsection{Treatment of photons}

After propagation of the final states, \prometheus{} converts the energy deposited in and around the detector into photons. To do this, it uses two different software packages depending on whether the detector is modeled in water or ice. For water-based detectors, \fennel{}~\cite{fennel2022@github} is used, a new package developed specifically for \prometheus{}. For ice-based detectors, it uses a standalone version of \ppc{}~\cite{Charles2013}. Both packages rely on parameterizations~\cite{Radel:2012ij} derived from dedicated \geant{}~\cite{GEANT4:2002zbu} simulations over a range of energies. These parametrizations deviate less than 20\% from the underlying \geant{} simulations~\cite{prometheus}. When simulating neutrino detectors in water, \ppc{} and \fennel{} both calculate the light yield and emission angles from various energy losses along a particle track and from hadronic showers. 

After the light production, the photons need to be propagated to the detection modules. Depending on the medium, a different propagation code is used.

\hyperion{} is used to propagate photons in water, employing a Monte Carlo approach by default.
Photons are represented as photon states, which carry key details such as the photon’s current position, direction, time (or distance) since emission, and wavelength. To initialize the photons, their starting states are drawn from distributions that model the emission spectrum of the simulated source type.
The propagation loop consists of three main steps.
First, \hyperion{} samples the distance to the next scattering event from an exponential distribution.
Then it checks for intersections to determine whether the photon’s path (based on its current position, the sampled distance, and its direction) crosses a detector module.
If an intersection occurs, the photon is stopped, and the intersection point is recorded in the photon state.
If there is no intersection, the photon is advanced to the following scattering site, and a new direction is sampled based on the scattering angle distribution.

In addition to the Monte Carlo mode, \hyperion{} supports using a normalizing flow that computes the transmission probability of a photon to an OM, and uses accept-reject sampling to determine if the photon arrives at the optical module.
These transmission probabilities are computed using the Monte Carlo method to simulate a large number of photons, and then a normalizing flow is fitted to the results.
Due to water's uniformity, this depends only on the distance to the OM and the angle between the photon's momentum vector and the vector connecting the photon emission point and OM position.

In Ice, \ppc{} carries out the photon propagation via Monte Carlo methods very similar to those \hyperion{} uses in water.
This takes into account the non-uniform nature of the Antarctic ice sheet, including the depth dependence of the scattering and absorption lengths, the tilt of the ice layers, and the birefringence of the ice.
Due to these non-uniformities, the normalizing-flow-based method employed by \hyperion{} is not tractable as the flow would depend on the depth, horizontal position, and photon momentum. This would drastically increase the dimensionality and thus require a much larger simulation set.

\section{Models}
\label{sec:models}

\subsection{Shared Techniques}
\label{subsec:common_techniques}
Several of the models presented in this work rely on similar techniques in either model architecture, loss functions, or training procedure.
These overlapping techniques are elaborated upon here.

\subsubsection*{Real-Time Data Augmentations}
\label{sec:data_augmentation}
Events on the training partitions should be subject to the stochastic event creation steps mentioned in \cref{sec:detector_response_sim} in order to represent events in the test partitions well.
Specifically, the following transformations should be employed:

\begin{align}
\label{eq:data_augmentations}
\texttt{t} & \longrightarrow |\texttt{t} + \epsilon_\texttt{t}| && \text{where}~ \epsilon_\texttt{t} \in \mathcal{N}(\mu = 0, \sigma = 1~\text{ns})  \\
\texttt{charge} & \longrightarrow  |\texttt{charge} + \epsilon_\texttt{charge}| && \text{where}~ \epsilon_\texttt{charge} \in \mathcal{N}(\mu = 0, \sigma = 0.25 ~\text{p.e.})
\end{align}
where $||$ represents the absolute values \footnote{As the probability of obtaining negative values from the transformations in \cref{eq:data_augmentations} is very low, the inclusion of || is a formality.} and $\epsilon$ represents a randomly drawn perturbation from the normal distribution $\mathcal{N}$ centered around the original value of the pulse attribute.
The transformations seen in \cref{eq:data_augmentations} should be applied independently to each pulse in the training partition by drawing an $\epsilon$ for each pulse separately. 

The following procedure should be followed in advance of the transformations in \cref{eq:data_augmentations} to sample stochastic noise pulses.
First, the total number of noise pulses is sampled from a Poisson distribution $P(\lambda = N_{\text{total}})$ where the expectation value $N_{\text{total}}$ is given by

\begin{align} \label{eq:data_augmentations_noise}
N_{\text{total}} & = \left( \frac{R_{\text{OM}} \cdot t_{\text{window}}}{10^6} \right) \cdot N_{\text{OMs}},
\end{align}
where $ R_{\text{OM}} = N_{\text{PMT}}\cdot R_{\text{PMT}}$ represent the OM-level noise rate and $t_{\text{window}}$ is the trigger window of 5~$\mu$s.
The OM-level noise rate $R_{OM}$ is computed using the quantities shown in \cref{table:datasets_processing}, where $ N_{\text{PMT}}$ represents the number of PMTs in an OM and $R_{\text{PMT}}$ is the noise rate of a single PMT.
For example, $ N_{\text{total}}$ for the \texttt{Cluster} dataset is given by $ N_{\text{total}} = (24 \cdot 7500 \cdot 5/10^6) \cdot 60$ = 54.
Thus, in this case 54 noise hits are expected on average within the trigger window.
Second, $N_{\text{total}}$ random arrival times are sampled from a uniform distribution starting from 0 ns to $t_{\text{max}}$ ns, where $t_{\text{max}}$ is given by

\[
t_{\text{max}} = \begin{cases} 
      t_{\text{window}} & \text{if } \mathrm{max}(t) < t_{\text{window}} \\
      \mathrm{max}(t) & \text{otherwise}
   \end{cases},
\]
where the quantity $t$ represents the collection of arrival times of the signal pulses in the event subject to the augmentation.
Lastly, $N_{\text{total}}$ OM positions are randomly sampled from the detector geometry, combined with the sampled arrival times, and the associated charge is set to 1 p.e.
\setlength{\tabcolsep}{2pt}
\begin{table}[b!]
  \centering
  \caption{Overview of datasets processed for this work. A total of over 129.7 million events were distributed across 7 datasets with geometries similar to known neutrino telescopes. Transit Time Spread (TTS) provided are typical values reported by the manufacturer at room temperature when available. Noise rates are given per PMT. Datasets marked with * are simulated in ice, whereas the remainder is simulated in water.}
  \label{table:datasets_processing}
    \begin{tabular}{lcccccccc} 
    \hline
    \textbf{Dataset}  & \textbf{Events} & Inspiration & $\nu_\mu^\text{CC} / \nu_\mu^\text{NC}$ & Strings/DOMs/PMTs & TTS/Noise & Optical Eff.\\ 
     & (millions) & & (\%) & & (ns / kHz) & (\%)\\
    \hline
    \hline
    \Triangle & 23.1  & P-ONE \cite{pone}   & 35/65  & 3/60/24 & 1.5/7.5\cite{km3net_noise} & 20 \\
    \cluster           & 22.9  & GVD \cite{baikal_gvd}     & 49/51 &  8/288/1 & 3.4/60 \cite{gvd_noise, gvd_pmt_hamamatsu} & 17.5\\
    \flowers & 20.5  & ORCA \cite{km3net_orca_osc}  & 40/60 &  150/3300/31 & 4.5/7.5 \cite{km3net_pmt_hamamatsu,km3net_noise} & 20\\
    \flowerl& 24.0  & ARCA \cite{arca_astronomy}  & 35/65 & 115/2070/31 & 4.5/7.5 \cite{km3net_pmt_hamamatsu,km3net_noise} & 20\\ 
    \flowerxl& 10.1  &  TRIDENT \cite{TRIDENT:2022hql}  & 88/12 &  1211/24220/31 & 3.7/7.5 \cite{trident_juno_pmt, km3net_noise} & 20\\
    \hexagon& 20.5  & IceCube \cite{icecube86}  & 48/52 & 86/5160/1 & 3.2/7.5 \cite{icecube_pmt_hamamatsu, km3net_noise} & 15\\
    $\hexagonice^*$& 8.6  & IceCube \cite{icecube86}  & 57/43 & 86/5160/1 & 3.2/0.3 \cite{icecube_pmt_hamamatsu, icecube_pmt_calibration} & 15\\
    \hline
    \hline
    \textbf{Total}: &  \texttt{129.7} & & & & & 
    \end{tabular}
\end{table}
The perturbation of arrival time or the sampling of stochastic noise may produce pulses closer in time than what can be observed in the test partitions.
While this discrepancy can be mitigated by merging pulses that fall within the TTS shown in \cref{table:datasets_processing} and adjusting the arrival time of the merged pulses to be the charge-weighted average, such corrections were found to have little impact on generalizability as the probability of occurrence is low.
Because the merging procedure introduces increased computational cost with no noticeable impact on generalizability, the procedure has been omitted from this work.  

\subsubsection*{Standardization of Input Features}
\label{sec:standardization}
The input variables shown in \cref{table:input_features} are given in different units and generally cover large numerical ranges that are unsuitable for deep learning methods.
For example, the OM positions $\texttt{sensor\_pos\_xyz}$ may span several kilometers for the largest detector geometries and the arrival time $\texttt{t}$ range from hundreds to thousands of nanoseconds.  As a prerequisite step, these quantities have been subject to the following transformations 

\begin{align} \label{eq:standardization}
\texttt{sensor\_pos\_xy} & = \texttt{sensor\_pos\_xy} / 100 \\
\texttt{sensor\_pos\_z} & = \texttt{sensor\_pos\_z} / 1000 \\
\texttt{t} & = \texttt{t} / 10^5 \\
\texttt{charge} & = \log_{10}\left(\texttt{charge} + 1\right)
\end{align}
which brings the numerical ranges of the input variables in \cref{table:input_features} within roughly $\pm 10$. The $+1$ in the charge standardization function is added for numerical stability.

\subsubsection*{Data Representation with Percentile Aggregations}
\label{sec:percentile_clusters}
Graph neural networks relies on graph representations of data. Formally, a graph $G = \{N(G),E(G)\}$ is an abstract mathematical object comprised of nodes $n_i \in N(G)$, where $N(G)$ denotes the set of nodes in G, and edges $e_i \in E(G)$, where E(G) represents the set of edges in $G$. The nodes often represent data, whereas edges are used to imply a relationship between the data. Due to the abstract nature of graphs, neutrino events may be represented as graphs in multiple ways, and the choice in graph representation effectively serves as a hyperparamter of the overall GNN model in question.  The large variance in the number of observed pulses in individual neutrino events, as seen in \cref{fig:n_pulses}, leads to challenges in model design and specifically, in the choice of data representation of neutrino events. The computational complexity of reconstruction methods depends primarily on the number of observed pulses, and often aggregation schemes are applied to the neutrino events to standardize the data dimensions and curtail the computation complexity.\\

Two GNNs applied in this work, \particlenet{} (\cref{subsec:particlenet}) and \dynedge{} (\cref{subsec:dynedge}) was originally proposed on datasets where the energy of neutrino events were sufficiently low to employ graph representations where nodes represented individual pulses, removing the need for aggregations completely. However, scaling such graph representations to higher energies comes at a greatly increased computational cost \cite{ComprehensiveStudyLargeScaleGraph}. For this work, an alternative graph representation was chosen for the two models that preserves the geometry but curtails the variance in the time domain by applying statistical aggregations using percentiles. \\

In this work, each events forms a $[n,d]$-dimensional geometric time series, where $n$ denotes the number of observed pulses in the event, varying from event to event, and and $d$ represents the dimensions equal to the input variables from \cref{table:input_features}. In Percentiles Clusters, each unique optical module (OM) that registers light pulses is represented as a node, reducing the number of nodes from the total number of pulses $n$ to the number of unique OMs $n_u$. For each OM, we aggregate the time $\texttt{t}$ and charge $\texttt{charge}$ features of all pulses it recorded by computing a set of percentiles. These percentiles capture the distribution of values and provide a compact statistical summary. In addition, to account for cluster’s statistic, the number of pulses registered at each unique OM is added too. Since each OM has fixed spatial coordinates ($\texttt{sensor\_pos\_{xyz}}$), these are also included as part of the feature set. As a result, the node feature dimensionality $d$ becomes:
    \begin{equation}
        d \mapsto d' = 3+(d’ \times p)+1
        \label{eq:d_redef}
    \end{equation}
where the 3 comes from the OM's spatial coordinates ($\texttt{sensor\_pos\_{xyz}}$), $d'=2$ corresponds to the two features (time $\texttt{t}$ and charge $\texttt{charge}$) used to compute the percentiles, $p$ is the number of percentiles computed for each feature, and the final 1 accounts for the number of pulses observed at that OM.    For this representation, we chose to perform clustering in the following percentiles: 
\begin{equation}
    [0, 10, 50, 90, 100]
    \label{eq:percentiles}
\end{equation}
where not only the median (50\%) is considered to represent the central tendency of each OM, but also percentiles near the extremes (0\%-10\% and 90\%-100\%) are included to provide insight into early pulses in time, which usually makes the basis for event reconstruction, and late pulses—potentially caused by light scattering or afterpulses, if present. \\

Therefore, the input data is a [n$_{u}$-14]-dimensional geometric time series. Even though, the augmentation of $d$-dimension from 5 to 14, choosing the unique OM representation, $n_u$ will be restricted as most up to the number of total OMs in each detector from \cref{table:datasets_processing}, leading to a speed up factor of $\sim 90\%$ using $\sim 50\%$ less memory is achieved while keeping more or less the same performance.

\subsubsection*{Dynamic Edge Convolution}
\label{sec:edgeconv}
Dynamical edge convolution, a convolutional operator for graph-structured data, was originally presented for segmentation of 3D point clouds \cite{edgeconv}. Given a graph $\mathcal{G}$ with $n$ nodes and edges $e$, the EdgeConv operator acting on a target node $i$, is defined as:

\begin{equation}
    \Tilde{x}_i = \texttt{Aggr}\left(\mathcal{H}_{\theta}(x_i, x_i - x_j) ~\forall ~j \in e_i\right)
    \label{eq:edge_conv}
\end{equation}
where $x_i$ and $e_i$ denote the node features and neighborhood of the ith node, respectively. $\mathcal{H}_{\theta}$ represents a learned function that receives in addition to $x_i$ the pair-wise difference between $x_i$ and the node features $x_j$  of the jth member of the neighbourhood of $x_i$, denoted with $e_i$. Each of these outputs of $\mathcal{H}_{\theta}$ is aggregated to form the convoluted node features $\Tilde{x}_i$. Using the node features to specify the neighborhoods for each node, e.g., via distance metrics, \cref{eq:edge_conv} can be interpreted as a translation operation, and a new distance calculation can be executed to obtain an updated neighborhood given the translation. 
\begin{figure}[h!]
    \centering
    \includegraphics[width = \textwidth]{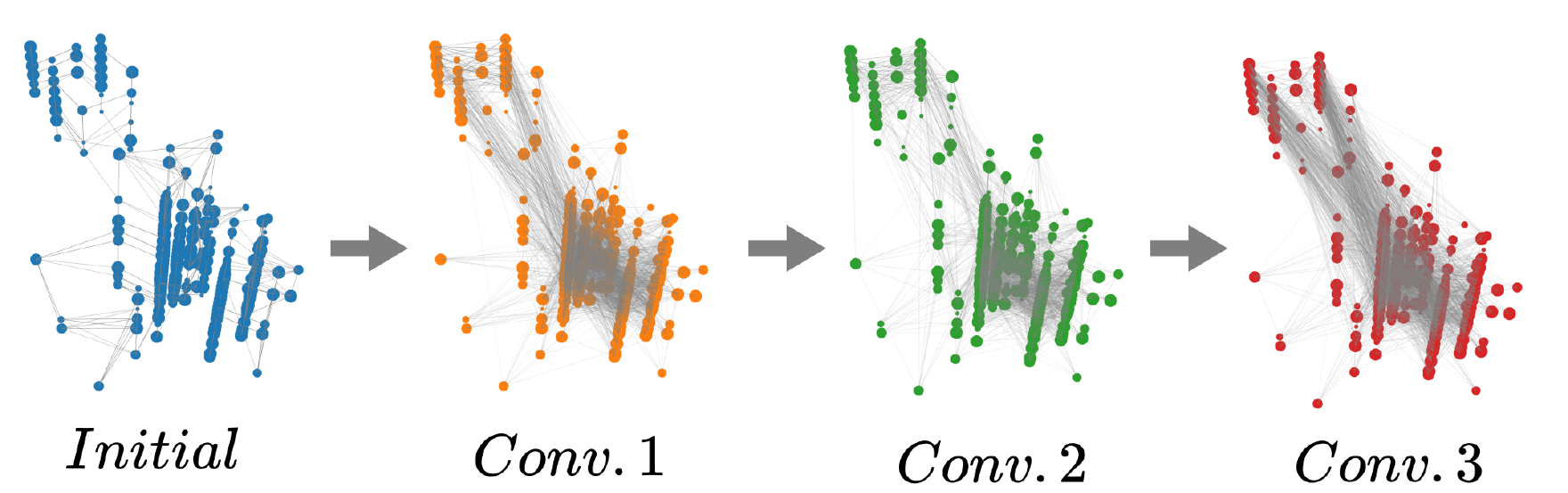}
    \caption{Illustration of dynamic edge convolution.}
    \label{fig:dynamic_edge_conv_illustration}
\end{figure}
Therefore, by applying multiple iterations of  convolution followed by neighborhood updates, the GCN can learn a graph representation given node and edge definitions. A visualization of dynamically learned edges from consecutive graph convolutions and neighbourhood updates can be seen in \cref{fig:dynamic_edge_conv_illustration}.

\subsubsection*{Loss Functions}
For regression of continuous variables, such as neutrino energy, the following loss function 
\begin{equation}
    \texttt{LogCosh}(R) = \ln\left(\cosh(R)\right)
    \label{eq:logcosh}
\end{equation}
may be defined, where $R$ represents the residual, quantifying the deviation between the true ($y$) and predicted quantity ($\Tilde{y}$) for a given regression task. For example, a common definition of residual for regression of neutrino energy is  

\begin{equation}
    R_{E} = \log_{10}(\Tilde{y}) - \log_{10}(y) =  \log_{10}\left( \frac{\Tilde{y}}{y}\right)
    \label{eq:energy_residual}
\end{equation}

 which quantifies the per-event error on a logarithmic scale to curtail the many orders of magnitude that the neutrino energy spans. The main benefit of \texttt{LogCosh} is that it contains a blend of the characteristics of two popular loss functions, namely Mean Square Error (MSE), which is quadratic in $R$, and Mean Absolute Error, which is linear in $R$. \cref{eq:logcosh} is linear in large residuals, parabolic around zero, and, similarly to MSE and MAE, penalizes over- and underestimation equally.   %which for energy reconstruction is computed in logarithmic scale to curtail the large numerical range, and is defined as $R_{E} = \log_{10}(f(\Tilde{x}) - \log_{10}(y) =  \log_{10}\left( \frac{f(\Tilde{x})}{y}\right)$. \cref{eq:logcosh} is linear in large residuals, parabolic around 0 and penalizes over- and underestimation equally.

For direction reconstruction, the neutrino direction can be represented as a direction vector $\Vec{y}$, and the loss quantified using the von Mises-Fisher (vMF) distribution for 3D vectors. By taking the negative natural logarithm, the loss function becomes
\begin{equation}
    \texttt{vMF}_{\texttt{3D}}(\Tilde{y}, \Vec{y}) = - \kappa \cdot \cos\left(\Delta \theta(\Tilde{y}, \Vec{y}) \right) - \ln\left(C_3(\kappa)\right)
    \label{eq:vmf}
\end{equation}
where $\Tilde{y} \in \mathbb{R}^3$ represents the predicted direction vector and $\Delta \theta(\Tilde{y}, \Vec{y})$ denotes the opening angle between the estimated and true direction vector.
The normalization factor $C_3(\kappa)$ depends on modified Bessel-functions \cite{von-mises-fisher-loss}. \cref{eq:vmf} bears resemblance to conventional choices in loss function for directional reconstruction, such as $1 - \cos(\Delta\theta)$, while allowing for uncertainty estimation through the concentration parameter $\kappa$ under Gaussian assumptions, which is an additional input to the loss function from the model. 

For  $\mathcal{T}/\mathcal{C}$ classification, which can be phrased as a binary classification problem, a common choice of loss function is the binary cross-entropy (BCE) given by

\begin{equation}
    \texttt{BCELoss} = -\left(y\ln{\Tilde{y}} + (1 - y)\ln{(1 - \Tilde{y})} \right)
    \label{eq:bce}
\end{equation}

where $y$ represents the binary truth label of either 0 ($\mathcal{C}$) and 1 ($\mathcal{T}$).
\texttt{BCELoss} provides a probability-like interpretation of the model predictions $\Tilde{y}$, where scores close to 1 and 0 are interpreted as likely examples of the $\mathcal{T}$ and $\mathcal{C}$ categories, respectively.

\subsection{ParticleNet}
\label{subsec:particlenet}
\particlenet{} is a deep learning architecture designed initially for jet tagging, where jets are collimated clouds of particles produced in proton-proton collision events at the LHC.
This model also leverages the structure of a Dynamic Graph Convolutional Neural Network (DGCNN), which captures both local and global features without requiring an arbitrary ordering of particles~\cite{Qu_2020}.
This architecture was initially developed within the KM3NeT collaboration for the reconstruction of atmospheric muon bundles for the study of cosmic rays compositions in KM3NeT/ORCA4, as described in~\cite{stefan_phd}. Additionally, \particlenet{} has been applied for reconstruction of neutrino energy, direction and, classification between $\nu/\mu$ and $\mathcal{T}/\mathcal{C}$ in the KM3NeT/ORCA6 detector in studies of neutrino oscillations ~\cite{guderian_phd}.\\

\begin{figure}[htbp]
    \centering
    \includegraphics[width=\textwidth]{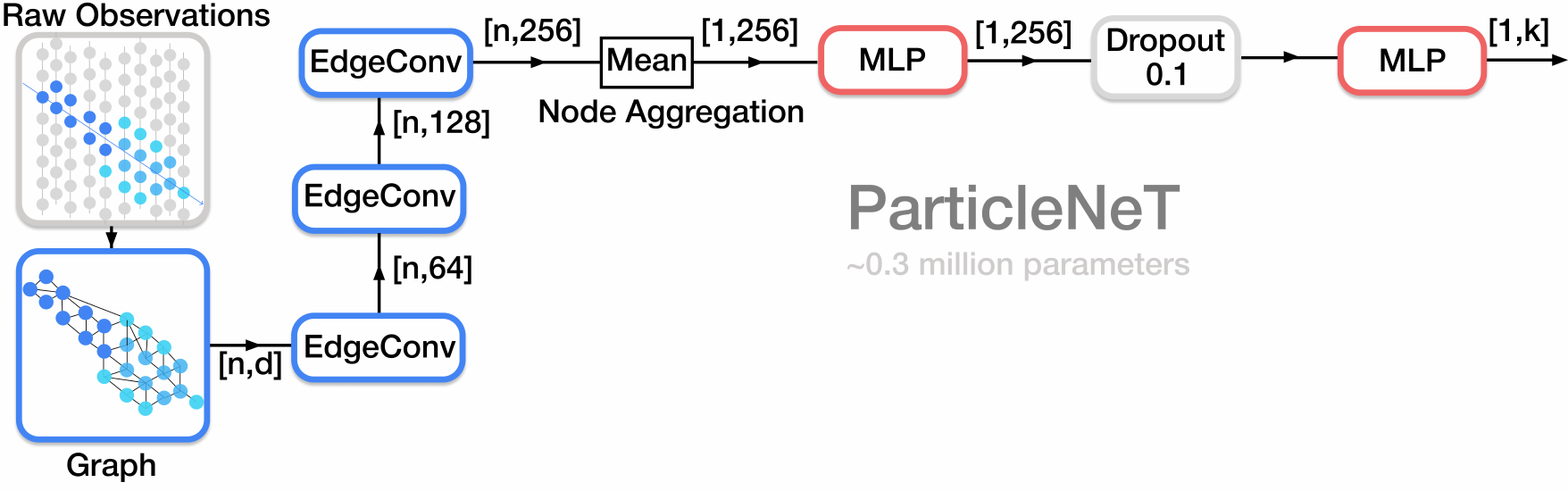}
    \caption{Illustration of the \particlenet{}~\cite{Qu_2020} architecture as commonly configured within the KM3NeT Collaboration, and as used in this work.
    \label{fig:particlenet_structure}}
\end{figure}

In DGCNN-based architectures, the event is represented as a graph.
In this adaptation of \particlenet{}, which shifts the focus from jet studies to neutrino detection in Cherenkov water/ice-based telescopes, the graph nodes correspond to the percentile aggregations mentioned in \cref{sec:edgeconv}, with edges established between each node and its eight nearest neighbors in the latent feature space. The choice of eight neighbors was motivated by a balance between computational efficiency and maintaining model performance. The architecture of the \particlenet{} model is fully depicted in \cref{fig:particlenet_structure}. Besides the change in the number of neighbors and the addition of a dropout to mitigate overfitting, the choice of the parameters constituting the model is identical to those presented in studies for ORCA~\cite{stefan_phd, guderian_phd}.

\subsubsection*{Task Adaptations}
The model slightly changes depending on the task. For regression of neutrino energy, which spans several orders of magnitude, the final output of the model is in log10 and the error is quantified using the $\texttt{LogCosh}$ loss function in \cref{eq:logcosh} with the energy residual from \cref{eq:energy_residual}. For direction reconstruction, the neutrino direction is represented as a direction vector $\Vec{y}$ and the $\texttt{vMF}_{\texttt{3D}}$ loss function from \cref{eq:vmf} is used. This choice of loss functions was previously studied and presented in different conferences inside the KM3NeT collaboration for both ARCA and ORCA with very promising results~\cite{prado_ai_mad2}.

For the neutrino vertex reconstruction, the chosen loss function is the Gaussian Negative Log Likelihood (GNLL)~\cite{NGLL_ref, NGLL_pytorch}
\begin{equation}
    \texttt{GaussianNegativeLogLikelihood}_\texttt{3D} = \sum_{i=1}^{N} \left( \dfrac{(\mathbf{y}_i - \boldsymbol{\mu}_i)^2}{\mathbf{\sigma}_i^2}+ \log (\mathbf{\sigma}_i) \right)
    \label{eq:gnll}
\end{equation}

where $\mathbf{y}_i \in \mathbb{R}^3$ represents the true vertex position, and $\boldsymbol{\mu}_i$ is the predicted mean vertex position.
The covariance matrix $\mathbf{\sigma}_i$ is predicted by the model and encodes the positional uncertainty. The loss function \eqref{eq:gnll} is derived from the negative log-likelihood of a multivariate Gaussian distribution in three dimensions, which allows for both position estimation and uncertainty quantification.
The first term ensures that predictions are penalized based on their Mahalanobis distance from the ground truth, while the second term regularizes the uncertainty by discouraging overly large variances ~\cite{guderian_phd}. 

For the visible inelasticity reconstruction, the MSE function is used
\begin{equation}
    \texttt{MeanSquaredError}(x_i, y_i) = (x_i - y_i)^2
    \label{eq:MSE}
\end{equation}
where $x_i$ are the individual inelasticity predictions and $y_i$ stands for the true ones. The MSE loss penalizes the squared differences between the predicted and true inelasticity values, encouraging the model to minimize large deviations.
This loss function assumes that errors are symmetrically distributed and does not explicitly model predictive uncertainty.

\subsubsection*{Training Procedure}
The \particlenet{} model has been trained separately for each task in \cref{sec:tasks} using the datasets listed in \cref{table:datasets}. First, a preliminary model is trained on a balanced subsample, \( X_{\mathcal{T}/\mathcal{C}} \), consisting of 1 million tracks and 1 million cascades, without applying any additional event-by-event reweighting.
The Adam optimizer~\cite{adam} is used along with the \texttt{ReduceLROnPlateau} learning rate scheduler, starting with an initial learning rate of \( 10^{-3} \), a scheduler patience of 2 epochs, and an early stopping patience of 8 epochs.  

In the second phase, the training dataset is extended to include the full sample, \( X_{\mathcal{T}/\mathcal{C},\text{full}} \), with a balanced number of tracks and cascades.
This time, the dataset size is limited by the smaller of the two categories (i.e., the number of tracks or cascades, whichever is fewer).
The model with the lowest validation loss from the first phase is then used as the starting point for this second training stage. Training on \( X_{\mathcal{T}/\mathcal{C},\text{full}} \) follows the same procedure as before, using the Adam optimizer and the \texttt{ReduceLROnPlateau} scheduler with the same hyperparameter values for the initial learning rate, patience, and early stopping.

This two-stage training procedure was motivated by the long training time required when training a model from scratch on \( X_{\mathcal{T}/\mathcal{C},\text{full}} \). With this approach, the model converges in approximately a factor of 4 fewer training epochs.

\subsection{DynEdge}
\label{subsec:dynedge}
\dynedge{} (Dynamical Edge) is a graph convolutional neural network (GCNN) published by the IceCube collaboration in 2022~\cite{icecube_dynedge}, and was initially presented as a deep-learning alternative to existing maximum likelihood methods on a simulated neutrino sample used primarily for the study of atmospheric neutrino oscillations in IceCube.
Since then, \dynedge{} has been applied to a wide range of tasks both within and outside the IceCube collaboration~\cite{graphnet_2}.
Within IceCube, \dynedge{} has been used to project the expected sensitivities of IceCube Upgrade to neutrino mass ordering and the $\theta_{23}$, $\Delta m^2_{23}$ oscillation parameters~\cite{icecube_queso}. 

\begin{figure}[htbp]
\centering
\includegraphics[width=\textwidth]{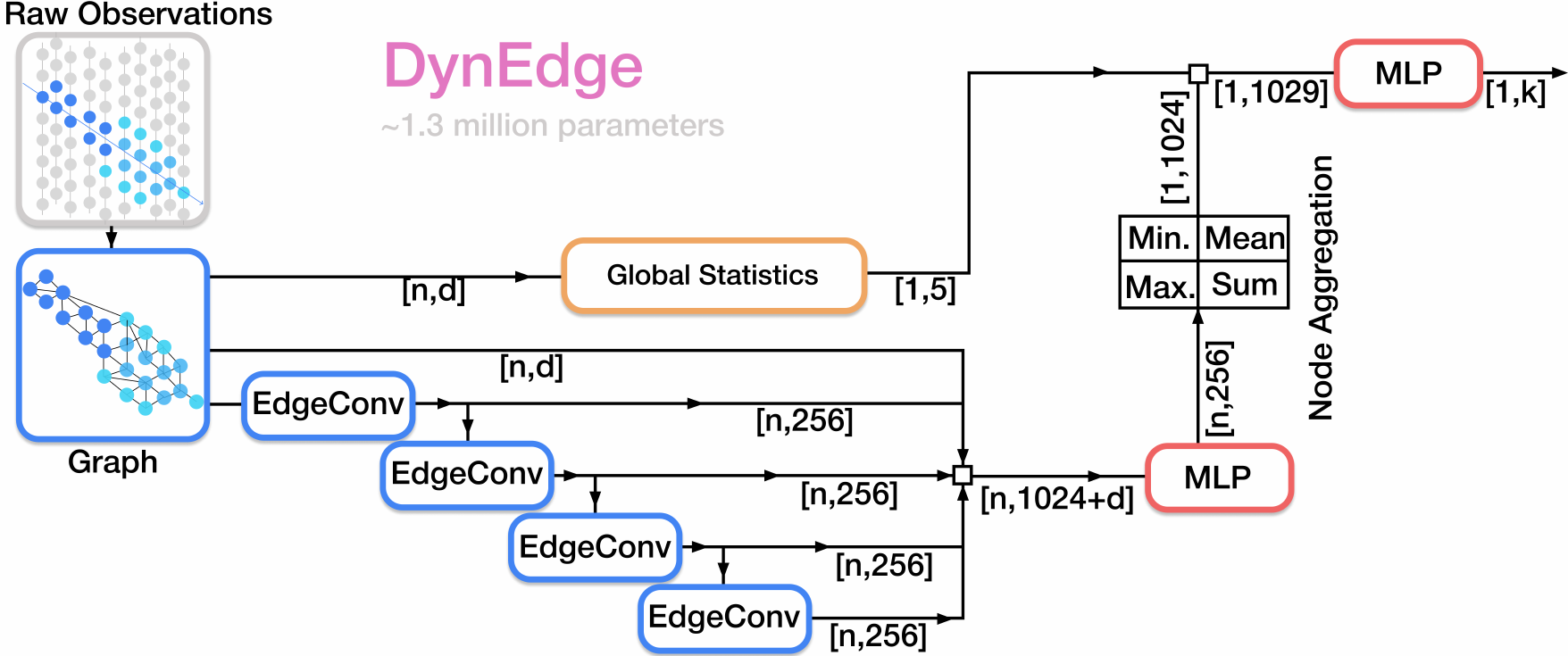}
\caption{ Illustration of the \dynedge{}~\cite{icecube_dynedge} architecture as configured for this work.
Parameters $d$ and $n$ represent the number of node features and the number of nodes, respectively.
\label{fig:dynedge_diagram}}
\end{figure}

\dynedge{} is a domain adaptation of the dynamical edge convolution described in \cref{sec:edgeconv} but with skip connections, global graph heuristics, and other minor modifications for processing events in neutrino telescopes as point cloud graphs.
An illustration of the \dynedge{} model architecture can be found in \cref{fig:dynedge_diagram}.
By comparing the model diagrams of \particlenet{} (\cref{fig:particlenet_structure}) and \dynedge{} (\cref{fig:dynedge_diagram}), considerable similarities can be seen as the original \particlenet{} architecture in~\cite{jet_gnn_2019} was a basis for the development.
The increase in learnable parameters (0.3 million vs. 1.3 million), skip connections, global graph heuristics, and the four-fold statistical node aggregation are the main differences between the two models.
The version of \dynedge{} used in this work is similar to its original implementation, and no modifications to the model architecture have been made to optimize for highly energetic events.
Instead, the graph representation has been altered from the original graph representation used in~\cite{icecube_dynedge}, where every node represents a single pulse of Cherenkov radiation, to the percentile aggregations mentioned in \cref{sec:percentile_clusters}.
Edges are drawn to each node's eight nearest neighbors.
The change in graph representation decreased training time by upwards of 90\% and approximately halved the memory requirements.
Further technical details can be found in publications~\cite{icecube_kaggle, icecube_queso, icecube_dynedge}

\subsubsection*{Training Procedure}
A separate instance of \dynedge{} has been trained for each task in \cref{sec:tasks} on each dataset in \cref{table:datasets}, yielding 35 models.
As a preliminary step, a subsample $X_{\mathcal{T}/\mathcal{C}}$ is created from the pre-defined training partition with an equal number of track and cascade examples to minimize bias towards any of the two topologies.
Each instance has undergone the following training procedure, which consists of two stages.
First, a preliminary instance of the model is trained on a 1 million event subsample of $X_{\mathcal{T}/\mathcal{C}}$ using the Adam~\cite{adam} optimizer and the \texttt{ReduceLROnPlateau}\footnote{PyTorch implementation available \href{https://pytorch.org/docs/stable/generated/torch.optim.lr_scheduler.ReduceLROnPlateau.html\#reducelronplateau}{here}.} learning rate scheduler.
The initial learning rate is set to $10^{-3}$ with a scheduler patience of 2 epochs, and early stopping is used with a patience of 10 epochs.
Then, the pre-trained model is trained using the same optimizer on the full subsample $X_{\mathcal{T}/\mathcal{C}}$ for a maximum of 200 epochs with an early stopping patience of 20 with the same learning rate scheduler but with patience set to 6, providing a generous learning rate scan.
Usually, convergence was achieved within 60 epochs. Due to the high variance in the number of pulses per event in the datasets, different batch sizes were used for each dataset to curb memory usage.
Batch sizes chosen for training \dynedge{} ranged from a few hundred to upwards of 2000 events.

\subsubsection*{Task Adaptations}
The model is configured slightly differently depending on the reconstruction task. These minor differences in configuration span choices in loss function, dimensionality of prediction heads, and handling of numerical ranges of target variables, and are detailed in this section.

For regression of neutrino energy, which spans several orders of magnitude, the final output of the model is in log10 and the error is quantified using the $\texttt{LogCosh}$ loss function in \cref{eq:logcosh} with the energy residual from \cref{eq:energy_residual}.

In the neutrino interaction vertex reconstruction, the position vector is not normalized, as opposed to what is presented in \cite{icecube_dynedge}, and the chosen loss function is the Euclidean distance
\begin{equation}    \texttt{EuclideanDistance}_\texttt{3D}(\Tilde{x}, \Vec{y}) = \sqrt{(\Tilde{x}_1 - \Vec{y}_1)^2 + (\Tilde{x}_2 - \Vec{y}_2)^2 + (\Tilde{x}_3 - \Vec{y}_3)^2}
    \label{eq:euclidean_loss}
\end{equation}
where $\Tilde{x}_1, \Tilde{x}_2, \Tilde{x}_3 \in \Tilde{x} \in \mathbb{R}^3$ and $\Vec{y}_1, \Vec{y}_2, \Vec{y}_3 \in \Vec{y} \in \mathbb{R}^3$ represents the estimated and true position vector. 

For direction reconstruction, the neutrino direction is represented as a direction vector $\Vec{y}$ and the $\texttt{vMF}_{\texttt{3D}}$ loss function from \cref{eq:vmf} is used.

The reconstruction of visible inelasticity and $\mathcal{T}/\mathcal{C}$ classification both apply a sigmoid activation to the model output $\Tilde{y} \in \mathbb{R}$ and use the \texttt{LogCosh} (\cref{eq:logcosh}) and \texttt{BinaryCrossEntropy} (\cref{eq:bce}) loss functions, respectively.
For visible inelasticity reconstruction, the residual is defined as $R_{VI} = \Tilde{y} - y$, where $y \in \mathbb{R}$ represents the true, visible inelasticity.

\subsection{GRIT} % Philip's Model
\label{subsec:grit}

In this section, we describe the use of a graph transformer for neutrino reconstruction tasks.
The chosen architecture is derivative of the \grit{} transformer~\cite{ma2023graph}, which uses a modified attention mechanism to combine edge and node information to update both edges and nodes. 
The GRIT graph transformer model attempts to provide stronger inductive bias, without message-passing elements, by including a learned positional encoding and incorporating graph degree information within the transformer layers.

\begin{figure}[htbp]
\centering
\includegraphics[width=\textwidth]{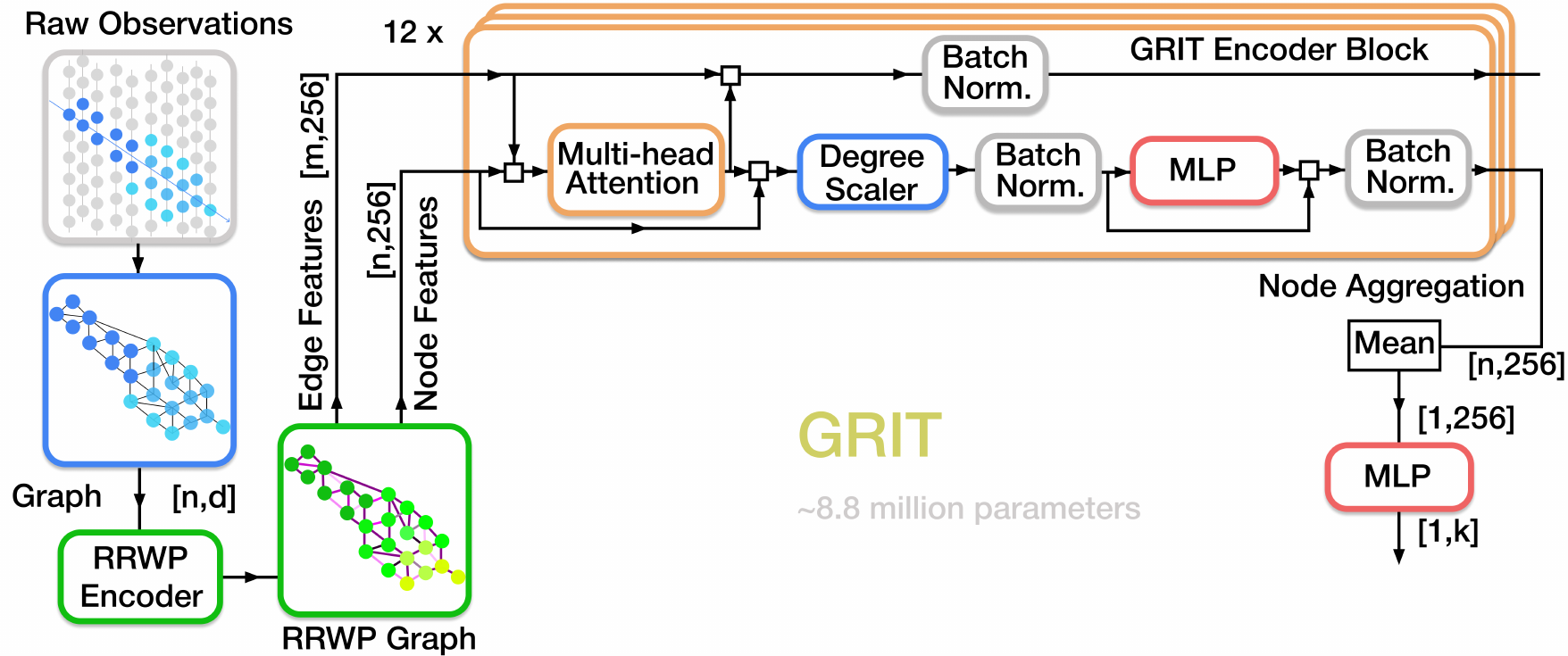}
\caption{ Illustration of the \grit{} architecture as configured for this work.
Parameters $d$ and $n$, $m$ and $k$ represent the number of node features, nodes, edges, and dimension of target label, respectively.
\label{fig:grit_diagram} }
\end{figure}

The \grit{} attention operation starts by computing a relative context score $c_{i,j}$ between the queries ($x_i$), keys ($x_j$), and node-pair/edge $e_{i,j}$:
\begin{equation}
    c_{i,j} = (W_Q x_i + W_K x_j) \odot W_{Ew} e_{i,j},
\end{equation}
where $W_Q$, $W_K$, and $W_{Ew}$ are learnable weights for the queries, keys, and edges respectively. The updated edges, which preserve connectivity, are obtained by combining the context score and edge biases ($W_{Eb}$),
\begin{equation}
    \hat{e}_{i,j} = \textrm{ReLU}(\rho(c_{i,j}) + W_{Eb} e_{i,j}),
\end{equation}
with $\rho = \textrm{sign}(x)\cdot \sqrt{\vert x \vert}$. These modified edges are then multiplied by weight matrix $W_{A}$ and summed over all nodes to get the node pair attention score $\alpha_{i,j}$,
\begin{equation}
    \alpha_{i,j} = \textrm{softmax}_{j\in V}\left( W_{A} \hat{e}_{i,j} \right).
\end{equation}
The attention score is then used to sum over all nodes and edges with value weights $W_V$ and $W_{Ev}$,
\begin{equation}
    \hat{x}_{i} = \sum_{j\in V} \alpha_{i,j} \cdot \left( W_V x_j + W_{Ev} \hat{e}_{i,j} \right).
\end{equation}
The output nodes and edges are obtained by multiplying an output weight matrix and summing over the head dimension $h$,
\begin{equation}
    x_{i}^{out} = \sum_{h=0}^{N_{h}} W_{O}^{h} \hat{x}_{i}^{h}, \quad \quad \quad e_{i,j}^{out} = \sum_{h=0}^{N_{h}} W_{Eo}^{h} \hat{e}_{i,j}^{h}.
\end{equation}

\subsubsection*{Encoding}
Like ordinary transformers, there has been a significant effort to understand the role of position/structural embeddings~\cite{graph_transformer_survey, graph_positional_encodings_benchmark, deep_graph_representation_learning_survey}. 
In the original \grit{} implementation, absolute and relative position encodings are added to the nodes and edges through a random walk process.
Relative random walk probabilities (RRWP) with $k$ steps are obtained by multiplying the one-step probability matrix $M = D^{-1} A$ with itself $k$ times, where $D$ is a diagonal matrix with elements $D_{i,i}$ corresponding to the degree of node $i$ and $A$ is the adjacency matrix.
The node and edge encodings are initialized to the values of the $k$-RRWP values obtained with $M$, where the diagonal elements correspond to node encodings and off-diagonal elements correspond to edge encodings.

A downside to RRWP is the excessive GPU memory requirement for training and additional preprocessing.
For large $k$, a large number of new edges for the relative position encodings are created corresponding to the random walk probability from node $x_i$ to node $x_j$ even if $x_j$ is not a $k$-nearest-neighbor of $x_i$.
The absolute encoding is applied to the nodes, corresponding to the diagonal elements of the matrix $M$.
In this work, we do not include encodings beyond applying linear transformations to the incoming nodes and edges, which allows for significantly faster training. However, the full RRWP encoding scheme is available within GraphNeT.
Since the structural encodings are applied before the GRIT attention operators, alternative encoding schemes, such as random walk structural encoding (RWSE)~\cite{gps_transformer}, can be used without modifications to the model architecture.

The advantage of the \grit{} attention operation is that it attends to each node and edge globally, collecting long-range information that might be difficult for message-passing graph convolutions, such as EdgeConv~\cite{edgeconv}, which considers only the local neighborhood of nodes during convolution.
However, the trade-off is a larger network that may require significantly more training data and time compared to other networks.
Variations of the \grit{} attention operation, such as GRITSparse~\cite{graph_positional_encodings_benchmark}, have been introduced to reduce the quadratic complexity of attention by attending only to the local neighborhood of each node.
The \grit{} model presented here utilizes the same KNN graph-structured input as the \dynedge{} model, using six neighbors and pulse statistics for node definitions as defined in~\cref{subsec:common_techniques}.
The number of neighbors in the initial graph is a hyperparameter that can be tuned for specific use cases to reduce the overall computational burden.

\subsubsection*{Task Adaptations}
The implementation of the direction reconstruction model follows directly from that of the \dynedge{} model, where the neutrino direction is predicted and the $\texttt{vMF}_{\texttt{3D}}$ loss function is used during training.
For the energy reconstruction task, the neutrino energies are transformed with $\log_{10}(E)$, and the LogCosh loss function was used during training.
The binary classification task for tracks and cascades was implemented by predicting a single output probability of being a track, and the model was trained with the $\texttt{BinaryCrossEntropy}$ loss function as defined in~\cref{eq:bce}.
In the vertex position reconstruction task, the LogCosh loss function was used with a scaling factor of 0.05 to rescale inputs to $\mathcal{O}(1)$.
For the inelasticity reconstruction task, the final output values must be bounded between 0 and 1.
To do this, a sigmoid function is applied after the final layer of the inelasticity model.

\subsubsection*{Training Procedure}

The training procedure for the \grit{} models is similar to the procedure for the \dynedge{} model outlined in~\cref{subsec:dynedge} with minor modifications. 
Each task was trained using the complete datasets and therefore did not employ any morphology-based subsampling.
The initial learning rate was set to $5\times10^{-4}$ to improve training stability during the first epoch.
The time per batch varied considerably across datasets due to preprocessing and disk I/O, so the batch size and number of batches per epoch varied for each dataset. 
The smaller detector configurations trained significantly faster, as there were fewer pulses per event, resulting in faster preprocessing times.
The \texttt{ReduceLROnPlateau} learning rate scheduler with patience set to $3$ epochs and early stopping set to $10$ epochs.
We tabulated the relevant training parameters and the average number of epochs in Table~\ref{table:grit_training}.

\setlength{\tabcolsep}{4pt}
\begin{table}[h!]
  \centering
  \caption{\grit{} model training hyperparameters for each dataset.
  The average number of epochs is the number of training epochs averaged over the tasks for a given detector configuration.}
  \label{table:grit_training}
    \begin{tabular}{lccccc} 
    \hline
    \textbf{Dataset}  & \textbf{Batch size}  & \textbf{Batches/epoch} & \textbf{Avg. Num. epochs} \\
    \hline
    \hline
     Triangle & 256 & 18478 & 54 \\
     Cluster & 128 & 2048 & 46 \\
     % Flower S & 128 & 2048 & NaN \\
     Flower L & 128 & 18478 & 25 \\
     Flower XL & 128 & 2048 & 66 \\
     Hexagon Water & 128 & 2048 & 66 \\
     Hexagon Ice & 256 & 18478 & 65 \\
    \hline
    \hline
    \end{tabular}
\end{table}

\subsection{DeepIce}
\label{subsec:deepice}
\deepice{} is a transformer-based architecture presented in the Kaggle competition \textit{IceCube - Neutrinos in Deep Ice}~\cite{icecube_kaggle}. In the open-data competition hosted by the IceCube collaboration, participants competed to produce the most accurate direction reconstruction algorithms.
\deepice{} achieved second place, and its original model architecture and training procedure are described in Ref.~\cite{kaggle_3solutions}.

\begin{figure}[h]
    \centering
    \includegraphics[width=1.0\textwidth]{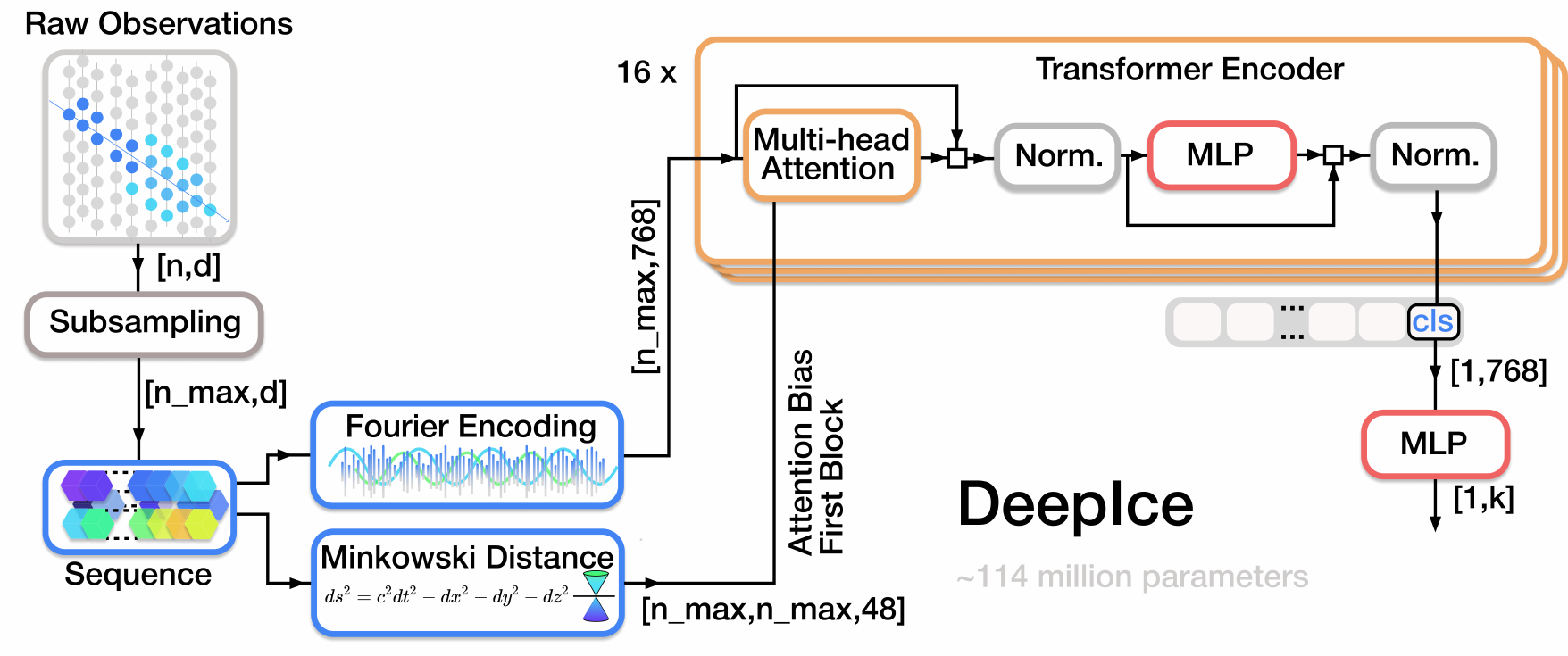}
    \caption{Illustration of the architecture of the  \deepice{} model \cite{icecube_kaggle} as used in this work.}
    \label{fig:deepice_architecture}
\end{figure}

The architecture of \deepice{} (\cref{fig:deepice_architecture}) is based on transformer layers. At its core, a transformer block uses attention mechanisms (first proposed in~\cite{attention_is_all_you_need}) to capture complex temporal and spatial dependencies in time-series data~\cite{transformers_time-series_survey}. To achieve this, \deepice{} employs two specialized encoders designed to preprocess the input data into a suitable subspace optimized for the self-attention mechanism.

Input pulse data is first standardized as described in~\cref{sec:standardization} and reshaped from a [n,d]-dimensional structure into a $[b,l,d]$ format, where $b$ represents the batch size and $l$ is the sequence length (pulses per event).
This reshaping organizes data into sequences, enabling unique positional information to each pulse within an event and ensuring the attention mechanism operates event-specifically.

The first encoder, the Fourier encoder, adapted from sinusoidal positional embeddings~\cite{attention_is_all_you_need}, encodes continuous input variables such as time, position and charge. Time and position are scaled by 4096, and charge by 1024, ensuring sufficient digitization resolution, dictated by the Fourier highest frequency, which corresponds to 1. 
For instance, with the normalization of $10^5ns$, the effective time resolution  can be computed $10^5 / 4096 \times 2\pi$, yielding approximately $3.9ns$. By mimicking the original implementation, this approach significantly improves model performance, as evidenced by results in~\cite{kaggle_3solutions}

The second encoder, referred to as the relative space-time interval bias, computes the Minkowski space-time interval $ds^2$ between pulse pairs $((i,j))$:
\begin{equation}\label{eq:minkowski_distance}
    (ds^2)_{ij} = -c^2 \cdot {(t_i - t_j)}^2 + {(x_i - x_j)}^2 + {(y_i - y_j)}^2 + {(z_i - z_j)}^2
\end{equation}
where $t_{i,j}$ and $(x, y, z)_{i,j}$ are temporal and spatial coordinates, and $c$ is the speed of light. The resulting $ds^2$ values are scaled by a factor of 1024 and processed using sinusoidal positional embeddings, followed by a linear layer, ensuring consistency between the dimensions of the input and output features.

Following the embedding process, the core of the \deepice{} consists of sixteen sequential transformer blocks. Each block includes a Multi-Head Attention (MHA) with 48 heads and an embedding dimension of 768. Additionally, each block includes a multilayer preceptron (MLP) layer, which is combined with its input via residual connections and subsequently normalized using \texttt{LayerNorm}.

A classification token $cls$ is initialized at the fifth transformer block, inspired by the \textit{BEiT} architecture~\cite{beit_v2}, this token aggregates prediction-relevant information in a single one-dimensional array. This array serves as a compact feature representation, which can be used later on to make the predictions.

The only modification made to the transformer block is the inclusion of an attention bias term, as proposed by the \textit{BeiT} architecture~\cite{beit_v2}. This proposal introduces an extension to self-attention that takes into account the pairwise relationships between input elements through the attention bias $a$, which in our case is the relative space-time interval bias $ds^2$.

\subsubsection*{Task Adaptations}
The direction task was the only one for which the model was trained. The same loss function as described in \cref{eq:vmf} is used without any further modification.

\subsubsection*{Training Procedure}
The \deepice{} model was traiend on the seven datasets listed in~\cref{table:datasets}. During training, events with more than 800 pulses were randomly sampled to a maximum sequence length of 800, balancing computational efficiency with performance. During inference, a maximum sequence length of 1500 were used.

Training was conducted using mixed precision arithmetic (FP16 and FP32) to improve computational efficiency and reduce memory usage without a significant loss of numerical accuracy.
The Adam optimizer with a weight decay of $0.01$ was employed during training~\cite{AdamWeightDecay}. 

A \texttt{OneCycleLR} learning rate scheduler~\cite{onecyclelr} was applied across all datasets, utilizing a cosine annealing strategy(~\cref{table:onecyclelr}). These values are consistent with those used in the original competition training~\cite{kaggle_3solutions}.

All models converged within 4 epochs, except for the \flowerl{}, which required 5 epochs. A patience of 2 epochs was used to determine convergence.

\setlength{\tabcolsep}{4pt}
\begin{table}[h!]
  \centering
  \caption{Overview of \texttt{OneCycleLR} learning rate policy variables for each training epoch}
  \label{table:onecyclelr}
    \begin{tabular}{lcc} 
    \hline
    \textbf{Epoch}  & \textbf{Max Learning Rate} & \textbf{Min Learning rate}\\
    \hline
    \hline
     \texttt{1} & $5 \times 10^{-4}$ & $\tfrac{1}{3} \times 10^{-5}$ \\
     \texttt{2} & $5 \times 10^{-4}$ & $2 \times 10^{-5}$ \\
     \texttt{3} & $1 \times 10^{-5}$ & $4 \times 10^{-7}$ \\
     \texttt{4} & $5 \times 10^{-6}$ & $\frac{1}{3} \times 10^{-7}$ \\
     \texttt{5} & $2 \times 10^{-6}$ & $2 \times 10^{-7}$ \\
    \hline
    \hline
    \end{tabular}
\end{table}

Training was conducted sequentially for each epoch, loading the weights from the previous epoch and setting the new learning rate scheduler each time. Gradient accumulation was used to achieve an effective batch size of 4096 while mitigating computational constraints. This technique accumulates gradients over several mini-batches before applying weight updates, simulating a larger batch size.

\end{document}